\documentclass[fleqn,usenatbib]{mnras}

\usepackage{newtxtext,newtxmath}

\usepackage[T1]{fontenc}
\usepackage{ae,aecompl}

\usepackage{graphicx}	
\usepackage{amsmath}	
\usepackage{booktabs}

\usepackage{pgf,tikz}
\usepackage{bm}
\usepackage{graphicx}
\usepackage{epstopdf}
\usepackage{float}

\usepackage{multirow}
\usepackage{hhline}
\usepackage{arydshln}

\newcommand{\ignore}[1]{}

\newcommand{\Me}{{\rm M}_{\oplus}}
\newcommand{\Msun}{{\rm M}_\odot}
\newcommand{\Rsun}{{\rm R}_\odot}

\newcommand{\Mp}{M_{\rm p}}

\newcommand{\ap}{a_{\rm p}}
\newcommand{\apn}{a_{\rm p,0}}

\newcommand{\hp}{h_{\rm p}}
\newcommand{\Mgap}{M_{\rm gas\,gap}}
\newcommand{\Mcrit}{M_{\rm crit}}
\newcommand{\Mciso}{M_{\rm c,iso}}
\newcommand{\Mcore}{M_{\rm core}}
\newcommand{\Mcoredot}{\dot{M}_{\rm core}}
\newcommand{\rp}{r_{\rm p}}
\newcommand{\Omegap}{\Omega_{\rm p}}
\newcommand{\Hp}{H_{\rm p}}
\newcommand{\tacc}{\tau_{\rm c,acc}}

\newcommand{\rheat}{r_{\rm HT}}
\newcommand{\Mdisk}{M_{\rm disk}}

\newcommand{\kappaenv}{\kappa_{\rm env}}
\newcommand{\fpl}{f_{\rm pl}}

\newcommand{\Ms}{M_\star}
\newcommand{\Rs}{R_\star}
\newcommand{\Ts}{T_\star}

\newcommand{\be}{\begin{equation}}
\newcommand{\ee}{\end{equation}}

\newcommand{\cs}{c_{\rm s}}
\newcommand{\xs}{x_{\rm s}}
\newcommand{\pnu}{{\rm p}_{\nu}}
\newcommand{\pnuzero}{{\rm p}_{\nu,0}}
\newcommand{\pchi}{{\rm p}_{\chi}}

\newcommand{\Fpnu}{F_{\pnu}}
\newcommand{\Fpchi}{F_{\pchi}}
\newcommand{\Gpnu}{G_{\pnu}}
\newcommand{\Gpchi}{G_{\pchi}}
\newcommand{\Kpnu}{K_{\pnu}}
\newcommand{\Kpchi}{K_{\pchi}}
\newcommand{\Fmax}{F_{\rm max}}
\newcommand{\tnu}{t_{\nu}}
\newcommand{\tlib}{t_{\rm lib}}
\newcommand{\dd}{{\rm d}}

\newcommand{\pnucorot}{{\rm p}_{\nu,{\rm corot}}}

\newcommand{\pchicorot}{{\rm p}_{\chi,{\rm corot}}}
\newcommand{\qnucorot}{{\rm q}_{\nu,{\rm corot}}}
\newcommand{\qchicorot}{{\rm q}_{\chi,{\rm corot}}}
\newcommand{\qdwcorot}{{\rm q}_{{\rm dw},{\rm corot}}}
\newcommand{\Kdwcorot}{{K}_{{\rm dw},{\rm corot}}}
\newcommand{\Mpnucorot}{{\rm M}_{{\rm p},\nu,{\rm corot}}}
\newcommand{\Mpchicorot}{{\rm M}_{{\rm p},\chi,{\rm corot}}}
\newcommand{\Mpdwcorot}{{\rm M}_{{\rm p},{\rm dw},{\rm corot}}}
\newcommand{\Mpcorot}{{\rm M}_{\rm p, corot}}
\newcommand{\Gamtot}{\Gamma_{\rm tot}}
\newcommand{\GamLR}{\Gamma_{\rm LR}}
\newcommand{\GamC}{\Gamma_{\rm C}}
\newcommand{\GamVHS}{\Gamma_{\rm VHS}}
\newcommand{\GamEHS}{\Gamma_{\rm EHS}}
\newcommand{\GamLVCT}{\Gamma_{\rm LVCT}}
\newcommand{\GamLECT}{\Gamma_{\rm LECT}}

\usepackage{hyperref} 
\hypersetup{
    colorlinks=true,
    linkcolor=blue,
    filecolor=magenta,      
    urlcolor=cyan,
}
\title[Bifurcation of Planet Formation Histories]{Turbulent Disk Viscosity and the Bifurcation of Planet Formation Histories}  
\author[Speedie et al. 2021]{
Jessica Speedie,$^{1, 2}$\thanks{jspeedie@uvic.ca (JS)}
Ralph E. Pudritz,$^{2}$\thanks{pudritz@physics.mcmaster.ca (REP)}
A. J. Cridland,$^{3,4}$
Farzana Meru,$^{5,6}$
Richard A. Booth$^{7,8}$
\\
$^{1}$ Department of Physics \& Astronomy, University of Victoria, Victoria, BC, V8P 1A1, Canada\\
$^{2}$ Department of Physics \& Astronomy, McMaster University,
Hamilton, ON, L8S 4M1, Canada\\
$^{3}$ Max-Planck-Institute for Extraterrestrial Physics (MPE), Giessenbachstr. 1, 85748 Garching, Germany\\
$^{4}$ Leiden Observatory, Leiden University, 2300 RA Leiden, The Netherlands\\
$^{5}$ Department of Physics, University of Warwick, Gibbet Hill Road, Coventry, CV4 7AL, UK\\
$^{6}$ Centre for Exoplanets and Habitability, University of Warwick, Gibbet Hill Road, Coventry CV4 7AL, UK\\
$^{7}$ Institute of Astronomy, University of Cambridge, Madingley Road, Cambridge, CB3 0HA, United Kingdom\\
$^{8}$ Astrophysics Group, Imperial College London, Blackett Laboratory, Prince Consort Road, London SW7 2AZ, United Kingdom\\
}

\date{Accepted XXX. Received YYY; in original form ZZZ}

\pubyear{2021}

\begin{document}
\label{firstpage}
\pagerange{\pageref{firstpage}--\pageref{lastpage}}
\maketitle

\begin{abstract} 
ALMA observations of dust ring/gap structures in a minority but growing sample of protoplanetary disks can be explained by the presence of planets at large disk radii - yet the origins of these planets remains debated. We perform planet formation simulations using a semi-analytic model of the HL Tau disk to follow the growth and migration of hundreds of planetary embryos initially distributed throughout the disk, assuming either a high or low turbulent $\alpha$ viscosity. We have discovered that there is a bifurcation in the migration history of forming planets as a consequence of varying the disk viscosity. In our high viscosity disks, inward migration prevails and yields compact planetary systems, tempered only by planet trapping at the water iceline around 5 au. In our lower viscosity models however, low mass planets can migrate outward to twice their initial orbital radii, driven by a radially extended region of strong outward-directed corotation torques located near the heat transition (where radiative heating of the disk by the star is comparable to viscous heating) - before eventually migrating inwards. We derive analytic expressions for the planet mass at which the corotation torque dominates, and find that this ``corotation mass'' scales as $\Mpcorot \sim \alpha^{2/3}$. If disk winds dominate the corotation torque, the corotation mass scales linearly with wind strength. We propose that the observed bifurcation in disk demographics into a majority of compact dust disks and a minority of extended ring/gap systems is a consequence of a distribution of viscosity across the disk population. 
\end{abstract}

\begin{keywords} 
protoplanetary discs -- planet-disc interactions -- planets and satellites: formation --  planets and satellites: physical evolution -- planets and satellites: individual: HL Tau
\end{keywords}


\section{Introduction}
\label{sec:introduction}


Protoplanetary disks, as the arenas for planet formation, are the mediators and record- keepers of the planet formation process. Over the last decade, our view of nearby protoplanetary disks has been revolutionized by high angular resolution images — most notably obtained by the Atacama Large Millimeter Array (ALMA). ALMA has resolved a diverse set of substructures present in a growing sample of protoplanetary disks, such as gaps, rings, spirals, clumps and crescents \cite[eg.][among many others]{van2013major, benisty2015asymmetric, wagner2015discovery, rapson2015peering, cieza2016alma, kudo2018spatially}. Of these structures, gaps and rings are the most common \citep{huang2018disk,long2018gapsringsALMA}. Probing the disk dust component, the DSHARP Survey \citep{andrews2016ringed,andrews2018disk} found gaps and rings ranging from circumstellar distances of 5 to more than 150 au. The HL Tau disk, in particular, exhibits dust gaps between $13-90$ au \citep{brogan20152014}.
  
A natural interpretation of these discoveries is that disk substructure is linked to the presence of young planets. Given that planets interact gravitationally with the disk in which they are embedded, substructures are thought to be caused by planets, at least in part. Many hydrodynamical simulations have shown that by judiciously choosing planet masses and orbital radii, one can successfully recreate the pattern of gaps and rings in HL Tau \cite[eg.][]{dong2015observational, dipierro2015planet, jin2016modeling}. While Jovian masses are required to open gaps in the gas, dust gaps can be created by much lower mass planets, possibly down to super-Earths or mini-Neptunes \citep{paardekooper2004dustgaps,paardekooper2006dust, rosotti2016minimum, long2018gapsringsALMA}.

Disks with gaps and rings, however, are in the minority of all disk systems known.  \citet{vandermarel2021-demographics} have analyzed the extant ALMA observations of all of the known protostellar disks, numbering over 700 disks. Specifically, of 692 disks analyzed in detail, only $ 6 \% $  are identified with structure resolved at 25 au scales.  These authors emphasize that the occurrence rate of disks with clear rings and gaps is similar to the occurrence rate of Jovian mass planets in the exoplanet population and displays the same dependence on stellar mass. 

Earlier work has already shown that high mass planets are rare at large disk scales. \citet{Fernandes2019} and \citet{Pascucci2019} used data from Kepler and RV surveys to show that giant planet occurrence rates peak at 2-3 au, around the position of snow lines, with about $4 - 6 \% $ at 3-7 au \citep{Wittenmyer2016}, and only $ 1 \% $ at  10-100 au. Similar trends beyond 10 au are also seen by the extensive Gemini GPIES, IR imaging survey  \citep{Nielsen2019}. The historic discovery of the pair of young forming massive planets in a transition disk around the T-Tauri star PDS 70 \cite[PDS 70b,c with 4-17 and 4-12 Jovian masses and orbital radii 20.6 and 34.5 au, respectively][]{Keppler2018, Haffert2019} indicates that in some cases, massive planets are indeed forming at disk radii exceeding 10 au.

It has been known for some time that disks with large gaps are more likely to be more massive \citep{Ercolano2017}.  More recently,  \citet{vandermarel2021-demographics} show that there is a good correlation of these ringed systems with more massive host stars (1.5-3 $\Msun$) in the survey, suggesting that ringed/gapped systems occur in more massive disks.   By comparing systems at different ages this work also showed that structured disks retain high dust masses up to at least 10 Myr, whereas the dust mass of compact, non-structured disks decreases over time.  

While this might imply that only a fraction of ring/gap disks are a consequence of giant planets \citep{Fernandes2019}, another possible explanation is that if such planets are indeed responsible for the production of the rings and gaps in disks at large disk radii during their formation, then they must have migrated back to 10 au or less before the final architectures of their planetary systems are established \citep{vandermarel2021-demographics}.

Our work is based on a careful analysis of planet-disk interaction wherein forming planets are subject to gravitational torques exerted by the surrounding gas disk which can significantly change their orbital location \cite[eg.][]{kley2012planet}. Type I migration, which pertains to low mass forming planets, depends on the competition between two kinds of torques arising from resonances between the planet and the disk \citep{goldreichtremaine1979}. Lindblad torques from waves launched at Lindblad resonances are fairly straight-forward \citep{tanaka2002torques3d} and are generally inward-directed. On the other hand, corotation torques depend on the viscosity \citep{masset2001corotation}, the radiative cooling efficiency \citep{kleycrida2008migrationrad} and the thermal, density and entopy gradients in orbits very nearly in co-rotation with the planet \citep{masset2006disk,ida2008toward,paardekooper2010torque, baruteaumasset2008corotation} -- and are typically outward-directed.  
 
The relative strength of these two competing torques  can lead to net inward or outward migration of planets in disks, depending on planet mass, disk viscosity, and local temperature gradients. There are also conditions where the torques are balanced.  These constitute planet traps, and can arise at sharp opacity transitions that occur at the water iceline, and dead zone boundaries where the turbulence amplitude rapidly changes due to a decrease in disk ionization. The thermal or viscous gradients in these cases favour a strong outward-directed and highly localized corotation torque that balances the inward Lindblad torque. A third, important instance is in the heat transition region of the disk where disk heating changes from viscous dissipation to irradiation by the central star. 

Viscous stresses due to disk turbulence are traditionally modelled with the $\alpha$ parameter \citep{shakura1973black}. This parameter can be determined observationally by measuring the amplitude of turbulence in the disks.  Most models of planet formation have used values in the range $\alpha=10^{-2}-10^{-3}$.

Given that co-rotation torques depend upon the turbulent viscosity of disks, it is natural to ask if all disks have similar levels of turbulence.  They do not. In fact, observations of line emission lacking turbulent broadening \citep{flaherty2015weakturbHD163296,flaherty2017turbulenceHD163296wDCO, flaherty2018turbulenceTWHya, flaherty2018turbulenceALMA, flaherty2020turbulenceDMTau}, small dust ring scale heights \citep{pinte2016HLTau}, sizes of protoplanetary disks \citep{Trapman2020observeddisksizes} and theoretical arguments based on low fragmentation velocities observed in laboratory experiments \citep{Pinilla2020fragilecollisions} increasingly point towards values of viscosity as low as $\alpha \simeq 10^{-4}$ in some systems.

In this paper, we compute how planets grow and migrate within  disk models whose detailed evolving astrochemistry is carefully followed.  We compute two different cases:  the conventional $\alpha=10^{-3}$, and a lower $\alpha=10^{-4}$. In relation to one another, we refer to these as \textbf{high viscosity} and \textbf{low viscosity} respectively. Our model allows for accurate calculation of thermal, density, and entropy gradients that are all important in computing the mass dependent, net torques on forming, migrating planets; both in magnitude and direction. In order to base our simulations for our general theoretical planet formation model as much as possible on real data, we adopted the conditions in the HL Tau protoplanetary disk \citep{cridland2019physics, cridland2019connecting}. We grow and evolve hundreds of planets, each in their own simulation, with initial orbital radii distributed throughout the disk. As a first step to computing the masses of these migrating planets, we adopt a very conservative estimate based on standard planetesimal accretion models.  

We find the remarkable result that there is a bifurcation in the migration behaviour of forming planets that is determined by the level of turbulence in their disks.  In the low turbulent viscosity ($\alpha=10^{-4}$ in our models) regime, strong outward-directed corotation torques beyond about 10 au, drive forming planets to the outer regions of the disk, where they are captured in the  heat transition trap.  They eventually reverse their migration and move inwards to smaller disk radii.  On the other hand, forming planets in the higher viscosity disks migrate inwards.  We confirm with analytical theory that the physics of this process can be well described by a new planetary mass scale which we call the co-rotation mass, $\Mpcorot = \Mpcorot(\alpha)$. It is the planet mass at which the outward-directed co-rotation torque achieves its maximum value, at some disk location. We argue that such low viscosity states are natural in more massive disks, whose higher column density will cut off the ionization of disks by external X-rays, and thus MRI induced turbulence within them.

This paper is structured as follows. In Section \ref{sec:model-and-sims}, we describe the theoretical formation model that we use to simulate planet formation and evolution in the HL Tau disk. Section \ref{sec:res:torqmaps} presents the first portion of our numerical planet formation results: the background torque landscape and key features within the torque maps. The second portion is presented in Section \ref{sec:res:planet-tracks}: the resulting planet migration tracks and final masses of the formed planet populations. In Section \ref{sec:res:insights-planetmigration}, we provide physical insight into our numerical results with analytic approximations and theory to derive expressions for the corotation mass. Finally, we discuss our results in Section \ref{sec:discussion} and conclude in Section \ref{sec:conclusions}.


\section{Model \& Simulations}
\label{sec:model-and-sims}

In this section, we describe the theoretical formation model that we use to grow and evolve planets in protoplanetary disks. We first establish the basic disk structure and dust evolution equations, and follow this with a description of the details of planet growth and migration.

\subsection{ Gas Disk Model}
\label{sec:diskmodel}

The gas disk model is based on the self-similar analytic model of \citet{chambers2009analytic}. Given a disk viscosity parameter $\alpha$, initial disk mass $M_0$, initial disk radius $s_0$ and protostar mass, radius and temperature ($\Ms$, $\Rs$, $\Ts$), the model computes (as a function of time) the disk accretion rate $\dot{\Ms}(t)$, and the disk surface density $\Sigma (t)$ and mid-plane temperature $T (t)$ radial profiles. We model HL Tau with a stellar mass of $\Ms = 1.2 \, \Msun$, a stellar radius of $\Rs = 3.0\, \Rsun$, and a temperature of $\Ts = 4395$ K \citep{white-hillenbrand-2004}.  

The disk is divided into two regions depending on the dominant disk heating mechanism: viscous dissipation or stellar irradiation. Viscous dissipation dominates in the inner regions where the disk's surface density is highest. The boundary between these two regions is known as the \textbf{heat transition (HT)}, $\rheat$, and is referred to extensively in this work. As a consequence, the disk surface density and temperature profiles take on a different power-law in each region:
\begin{equation}
\label{eqn:hlt-surface-density-profile}
    \frac{\Sigma }{\Sigma_0} (r) \propto 
    \begin{cases} 
      r^{-3/5} & r \ll \rheat \\
      r^{-15/14} & r \gg \rheat 
  \end{cases}
\end{equation}
\begin{equation}
\label{eqn:hlt-temperature-profile}
    \frac{T }{T_0} (r) \propto 
    \begin{cases} 
      r^{-9/10} & r \ll \rheat \\
      r^{-3/7} & r \gg \rheat \, ,
  \end{cases}
\end{equation}
where $\Sigma_0 = \Sigma_0 (t)$ and $T_0 = T_0 (t)$ depend on time through the evolving mass accretion rate. We assume that the mass accretion rate is constant over all disk radii, which requires: \begin{align}
    \dot{\Ms} = 3 \pi \nu \Sigma, 
    \label{eq:dm01}
\end{align}
where \begin{align}
    \nu = \alpha {\rm c_s} H
    \label{eq:dm01b}
\end{align} 
is the disk viscosity in the standard $\alpha$-disk paradigm \citep{shakura1973black}. \cite{chambers2009analytic} derived individual formulations for the different heating sources in the disk. 

Inward of $\rheat$ the disk is heated through viscous evolution. In the absence of a disk wind, the traditional assumption is that gravitational potential energy release at each radius is converted entirely into heat which is then radiated away as black body radiation from the disk.  As is well known, this rate of release is controlled by the accretion rate, which as seen above depends on both the viscosity and the column density. Thus, this region has an effective temperature of \citep{Ciesla2006visheating}:\begin{align}
    2\sigma T_{\rm eff}^4 = \frac{9\nu\Sigma\Omega^2}{4},
    \label{eq:dm02}
\end{align}
which results in the midplane temperature of :\begin{align}
    T_{\rm visc}^4 = \frac{3\kappa\Sigma}{8}T^4_{\rm eff} = \frac{27\kappa\nu\Sigma^2\Omega^2}{64\sigma},
    \label{eq:dm03}
\end{align}
where $\kappa$ is the (assumed constant) average dust opacity from the midplane to the disk surface. By combining Equations \ref{eq:dm01} and \ref{eq:dm03} one can (as \cite{chambers2009analytic} did) recover the radial dependence shown in Equation \ref{eqn:hlt-surface-density-profile} for $r \ll \rheat$ as well as the temporal evolution of $\dot{\Ms}$\, $\Sigma_0$, and $T_0$\footnote{For brevity we have largely neglected to show the mechanics of these derivations, and invite the reader to see \cite{chambers2009analytic} for details.}.

The second heating source, dominant for $r \gg \rheat$, is due to direct irradiation from the host star - not the release of gravitational potential energy of the accreting flow. In this case the midplane temperature profile use by \cite{chambers2009analytic} followed the model of \cite{chiang1997radiation}:\begin{align}
    T_{\rm irr} = T_{\rm rad}\left(\frac{r}{r_0}\right)^{-3/7},
    \label{eq:dm04}
\end{align}
where $r_0$ is the outer radius of the disk, and:\begin{align}
    T_{\rm rad} = \left(\frac{2}{7}\right)^{1/4}\left(\frac{\Ts}{T_c}\right)^{1/7}\left(\frac{\Rs}{r_0}\right)^{3/7} \Ts \, ,
    \label{eq:dm05}
\end{align}
where $T_c$ is a constant with units of Kelvin \citep[see][]{chambers2009analytic}. Notably, the temperature profile $T_{\rm rad}$ lacks a dependence on $\dot{\Ms}$ (and also time) and hence all of the temporal evolution of $\dot{\Ms}$ is encoded in the evolving gas surface density in regions of the disk that are mainly heated through direct irradiation.

We emphasize that $\rheat$ only prescribes a characteristic radial scale at which the disk's heating from viscous is comparable to that from direct irradiation. In reality, and as shown in numerical simulations \citep[for example, ][]{DAlessio2006diskmodel}, the temperature profile smoothly transitions from viscous-dominated to irradiation-dominated over a considerable range of disk radii. To account for this, we combine both temperature profiles (both computed over the whole radius range of the disk) by taking the sum of their energy distributions. In that case the midplane temperature becomes:

\begin{align}
    T^4 = T^4_{\rm visc} + T^4_{\rm irr}.
    \label{eq:dm06}
\end{align}

This is how the heat transition becomes an extended region spanning a range of disk radii (looking ahead to Fig. \ref{fig:3trap_definitions}). With this new temperature profile we update the gas surface density profile using their connection through Equation \ref{eq:dm01} for a given (current) value of $\dot{\Ms}$. We repeat this process for a wide range of time steps throughout the lifetime of the disk, from $t=10^5$ yr to $t=2\times 10^6$ yr, where we generate $\dot{\Ms}$ at each time step using the following modified power law \citep{chambers2009analytic}: 
\begin{equation}
    \dot{\Ms} =
    \begin{cases} 
       \frac{\dot{M}_0}{(1 + t/\tau_{\rm vis})^{19/16}}e^{-(t-\tau_{\rm init})/\tau_{\rm dep}} & t < t_1 \\
       \frac{\dot{M}_1}{\left[1 + (t-t_1)/\tau_{\rm rad}\right]^{20/13}} e^{-(t-\tau_{\rm init})/\tau_{\rm dep}} & t \ge t_1,
  \end{cases}
  \label{eq:dm07}
\end{equation}
where $t_1$ is the first time that $T_{\rm irr} > T_{\rm visc}$ at the outer edge of the disk and $\dot{M}_1$ and $M_1$ are the mass accretion rate and disk mass at $t_1$. If $T_{\rm irr} > T_{\rm visc}$ at $t = 0$ then $t_1=0$, $M_1 = M_0$, and $\dot{M_1} = \dot{M_0}$. The viscous accretion timescale in the viscously heated and radiative heating regimes are $\tau_{\rm vis} = 3M_0/16\dot{M}_0$ and $\tau_{\rm rad} = 7M_1/13\dot{M}_1$ respectively. The exponential term represents the impact of mass removal due to photoevaporation which lowers the overall mass accretion rate through the disk \citep{hasegawa2013planetary}. The parameter $\tau_{\rm init} = 10^5$ yr is the initial time at which we start our simulations and $\tau_{\rm dep} = 8$ Myr is the depletion time driven by photoevaporation. Our choice of a long depletion time is in line with a recent ALMA archival survey of local star forming regions which suggest that the average depletion time for protoplanetary disks (not in strong $F_{\rm UV}$ fields) is between 8-9 Myr \citep{Michel2021longliveddisks}. With this long depletion time we imply that the bulk of the mass accretion history is driven by viscous evolution.

As a final point, \cite{chambers2009analytic} assumed a constant $\kappa$ mainly to avoid complications due to a range of dust sizes, volatile abundances, and physical processes like vertical settling - all of which would negate an analytical derivation. An exception to this is at gas temperatures high enough to sublimate the dust, where an analytical derivation can again be obtained. Below this sublimation temperature ($\sim 1380$ K), $\kappa = \kappa_0 = 3 {\rm ~cm^2 g^{-1}}$, and is held constant across the disk. Above 1380 K silicate dust sublimates and it is assumed that $\kappa = \kappa_0(T/1380)^{-14}~{\rm cm^2g^{-1}}$ \citep{RudenPollack1991,Stepinski1998}. Here we wish to re-introduce some of these aforementioned complications to study how the freeze out of water can impact the local temperature structure of the disk and hence on planetary torques. 

The disk model of \cite{cridland2019physics} follows a straightforward path to incorporate variations in water ice abundances on the local temperature profile. The steps are:\begin{enumerate}
    \item Compute an initial temperature and surface density profile using $\kappa_0$ and the disk model of \cite{chambers2009analytic}.
    \item Compute the water ice distribution (see below).
    \item Compute new $\kappa(r)$ from ice distribution (see below).
    \item Re-compute viscous temperature profile from $\kappa(r)$ and original surface density using equation \ref{eq:dm03}.
    \item Re-compute new surface density with new temperature profile and equation \ref{eq:dm01}.
    \item Iterate steps iv and v until the functions converge.
\end{enumerate}

The astrochemistry of the disk dictates where opacity transitions such as ice lines will occur.  It also dictates how well ionized it is in a given X-ray background provided by the host star.  The ionzation, in turn, controls the coupling of the magnetic field to the disk and thus, whether MRI turbulence is present or not. This sets the extent of the dead zone.   These and other details of the initial conditions of the disk, and the value of disk viscosity in the dead zone and beyond, are discussed in Appendix \ref{sec:volatilechemicalmodels} - to which we refer the reader for details. 


\subsubsection{Disk winds}
As has been shown in a number of recent MHD disk simulations, MRI driven turbulence will be strongly damped in dense regions of the disk. The actual damping condition is reviewed in Appendix \ref{sec:volatilechemicalmodels}, where it is shown how we self consistently compute the extent of the dead zone using our detailed astrochemistry models.  Despite the virtual absence of turbulence in the dead zone, material must still accrete onto the central star. Disk winds have long been proposed as a main carrier of disk angular momentum \citep{Blandford1982, pudritz1983, Pelletier1992, ferreira1995}. Such winds efficiently extract a portion of the energy released in the accreting flow and carried off by the wind.  Recent non ideal MHD simulations \citep{Gressel2015,Bai2016} and observations \citep{Tabone2017,Tabone2020} have shown that disk winds are probably the major driver for global angular momentum transport, at least within the dead zone of protoplanetary disks.  These simulations have shown that MRI turbulence can be almost completely suppressed by Ohmic and ambipolar diffusion \citep{BaiStone2013, Gressel2015} while at the same time driving disk winds that dominate the angular momentum transport process.  \citet{BaiStone2013} found that $60 \%$ of the available energy is carried off by the MHD disk wind, the remainder being presumably dissipated as heat.

The loss of energy from the disk due to wind transport will reduce its temperature somewhat. As an estimate of this, we adopt the numerical results of \citet{BaiStone2013} to prescribe an efficiency for removing the energy in shearing flow by the wind as $\epsilon \simeq 0.60$, leaving a fraction of $(1-\epsilon)$ to be lost, presumably as heat which is then radiated away.  If we use this data, it allows one to estimate the disk temperature in this region as arising from the energy that is not carried off in the flow in this inner zone, but radiated away. Thus, $T_{\rm DZ} = (1-\epsilon)^{1/4} T_{\rm visc} \simeq 0.8T_{\rm visc} $.

The actual physical mechanism of disk heating have been examined in detailed recent global disk simulations \citep{gressel2020, wang2019}, that include irradiation, photodissociation and photoionization by X-ray and EUV photons, and the dissipation arising from Ohmic resistivity and ambipolar diffusion in the lightly ionized regions of the disk. The results show that turbulence is damped in wide regions of the disk by these magnetic dissipation effects. In the absence of turbulence on the midplane, Ohmic heating is $\eta_O J^2$,  where $J = (1/ 4 \pi) \nabla \times B $ is the current intensity and $\eta_O$ is the Ohmic magnetic diffusivity of the disk.  Numerical experiments show this to be small compared to heating by ambipolar diffusion (which depends on a magnetic field dependent diffusivity, $\eta_A$) that occurs preferentially above the disk plane. Even this non-ideal MHD heating mechanism may be small in the dead zone compared to heating by IR photons from the surface regions of the disk.

Given that the adjustment to the temperature in the dead zone region is not very large ($ \le 20 \% $), we elected to keep our underlying disk model simple by  keeping the $\alpha$ used in Equation \ref{eq:dm01b} constant at either $10^{-3}$ for the \textbf{high viscosity} model, or $10^{-4}$ for the \textbf{low viscosity} model. In the outer disk we assume that $\alpha_{\rm turb} = \alpha$ in the active region of the disk, which implies that turbulence is the primary source of angular momentum transport there. In the dead zone we reduce $\alpha_{\rm turb}$ by two orders of magnitude, but keep $\alpha$ constant. In this way we are assuming that disk winds have become the primary source of angular momentum transport in that region, and maintains a constant mass accretion rate throughout the disk. We note that a fully self-consistent treatment of both turbulence and disk winds has been performed for self-similar disk models \citep{chambers2019}, and further extended by \citet{2021-alessi-pudritz-inprep}. The latter shows that it is self-consistent to treat this as an effective $\alpha$ parameter that includes both turbulence and dis wind transport contributions.

\subsubsection{ Initial conditions }

As we are specifically targeting the young stellar system of HL Tau, we choose our initial disk gas mass and outer radius to best match the current estimates, assuming the age of HL Tau to be roughly 1 Myr. \cite{carrasco2016vla} report a dust mass range of $1-3 \times 10^{-3}$ $\Msun$ which, assuming the standard ISM gas-to-dust mass ratio of 100, results in a current gas mass of 0.1-0.3 $\Msun$. More recently \cite{ABooth2020HLTaumass} uses the rare (and optically thin) $^{13}{\rm C}^{17}{\rm O}$ isotopologue to estimate a total gas mass of $~\sim 0.2$ M$_\odot$ for HL Tau.

Therefore, we select initial disk gas masses so that, after the disk has evolved for 1 Myr, its integrated surface density profile works out to approximately 0.1 $\Msun$, 0.2 $\Msun$, and 0.3 $\Msun$. We refer to these as the \textbf{low-mass}, \textbf{medium-mass}, and \textbf{high-mass} models respectively. For the high viscosity ($\alpha = 10^{-3}$) cases, the initial disk gas masses that cover this range obtained from dust measurements are 0.14 $\Msun$, 0.28 $\Msun$, and 0.42 $\Msun$ \citep[as in ][]{cridland2019physics}; for the low viscosity ($\alpha = 10^{-4}$) cases, the intial disk gas masses are 0.11 $\Msun$, 0.22 $\Msun$, and 0.32 $\Msun$. Similarly, each of the models start with an initial disk radius of 91 au such that they evolve to have HL Tau's current $\sim 120$ au radius after 1 Myr of viscous evolution. Since each model starts with the same initial radius and different mass, their initial stellar mass accretion rate will be different. The six (6) formation scenarios are summarized in \hypertarget{hyp:tab:model-summary}{Table \ref{tab:model-summary}}. 

Figure \ref{fig:app:alldisks-masses-trunc93au} in Appendix \ref{app:numerical-results} plots the disk mass (integrated surface density profiles) inwards of 93 au over time. As these are quite high disk masses, we have calculated the Toomre-Q parameter to check the gravitational instability of these models (Figure \ref{fig:app:toomre}). The high- and medium-mass disks are unstable to gravitational collapse outwards of $\sim 30$ au initially, but this radius grows with time. By 1 Myr, all 6 models are stable over all radii. While the period of instability might impact the overall evolution of the mass accretion rate, it does not have a strong impact on our overall conclusions.

\begin{figure*}
	\includegraphics[width=18cm]{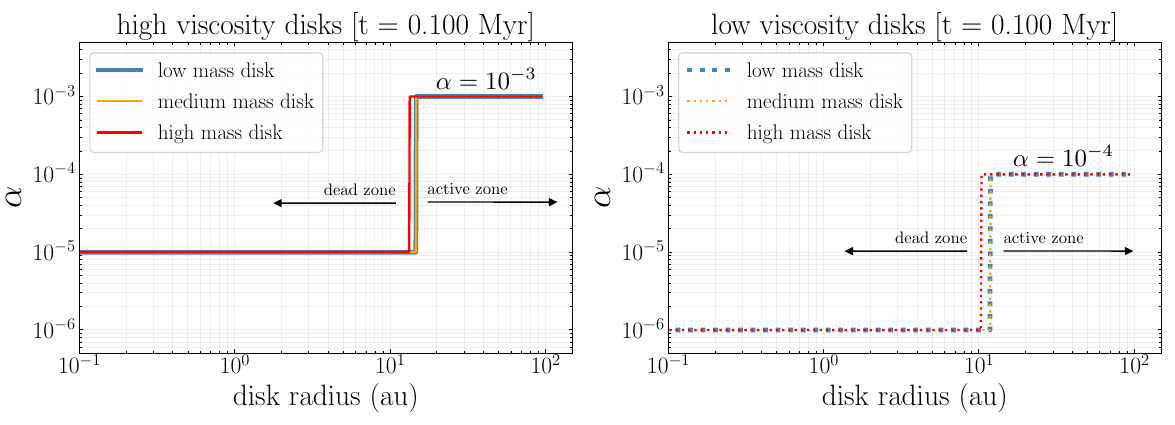}
	\includegraphics[width=18cm]{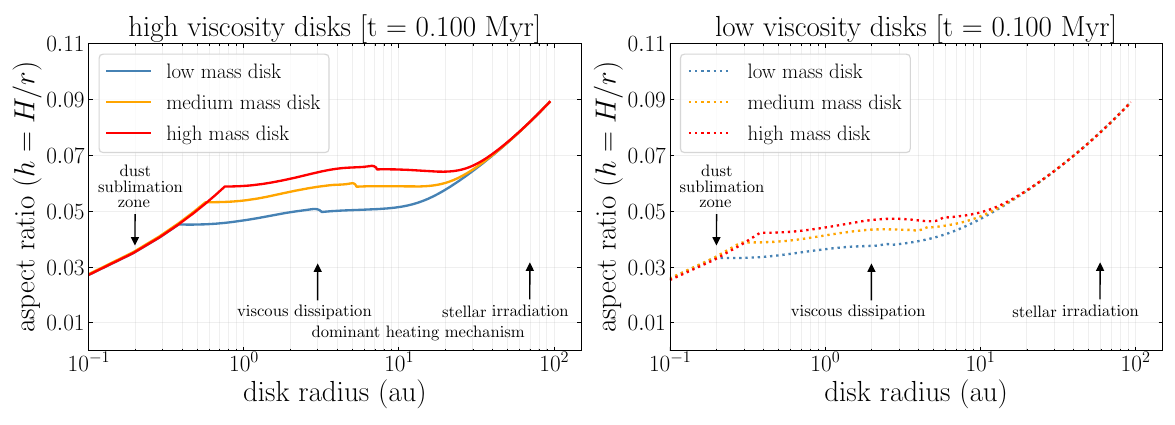}
	\includegraphics[width=18cm]{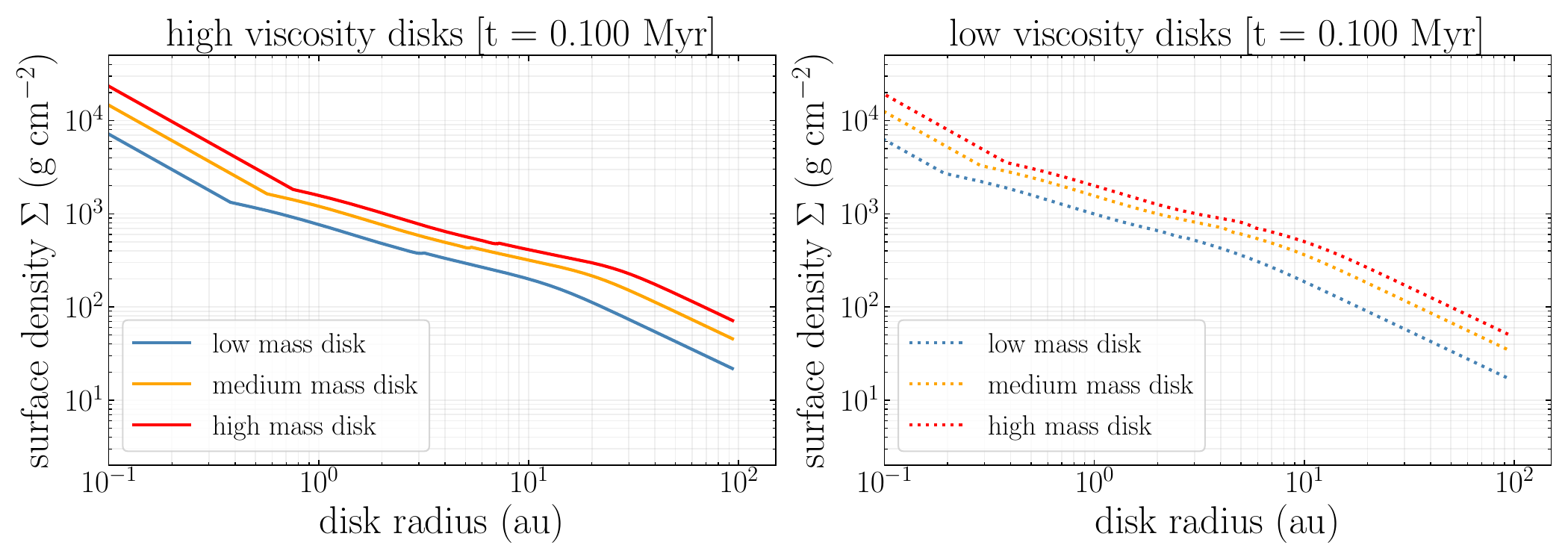}
    \caption{ \textbf{  \protect\hyperlink{hyp:fig:model_alphas_h_sigma}{Important properties of our planet formation models.}} Each panel shows the initial conditions for our 6 formation scenarios: three disk masses and two levels of viscosity. \textbf{Top row:} Radial profiles of the $\alpha$ viscosity parameter. These two panels effectively serve as the definition of what we refer to as ``high viscosity'' and ``low viscosity''. The MRI dead and active zones are labelled. \textbf{Middle row:} Profiles of the disk aspect ratio, $h=H/r$. We label the regions of the disk according to the dominant heating mechanism; the ``heat transition'' refers to the transition between viscously and radiatively heated regions. \textbf{Bottom row:} Profiles of the gas surface density, $\Sigma$. Table \ref{tab:model-summary} provides numerical values of these three quantities evaluated at $r=10$ au and $t=0.100$ Myr, as well as movies showing their full evolution over the course of our simulations.}
    \label{fig:model_alphas_h_sigma}
\end{figure*}

In \hypertarget{hyp:fig:model_alphas_h_sigma}{Figure \ref{fig:model_alphas_h_sigma}}, we plot radial profiles of three quantities for all six disks: the viscous $\alpha$ parameter, the disk aspect ratio and the gas surface density. The scale height plays several critical roles: in determining gas gap opening masses, the Hill sphere of the planet, and the saturation parameter which controls the magnitude of the corotation torque (see Sec. \ref{sec:res:insights-planetmigration}). The aspect ratio is determined by local hydrostatic balance between gas pressure and gravity at each disk radius, which for thin Keplerian disks $h \equiv H/r \ll 1$, with vertically isothermal gas profiles gives: \begin{equation}
    h(r) = \cs / v_{\rm Kep}(r)
\end{equation}
Since $\cs \propto T(r)^{1/2}$, it is the temperature behaviour of the disk, and its evolution, that determines this scale height. The inner, viscously heated region of the disk gradually cools as the column density decreases which causes $h(r)$ to decrease with time. The radiatively heated part of the disk however, is exposed to a constant stellar flux (we ignore stellar evolution in this work), and so $h(r)$ in the region beyond the heat transition retains a constant shape and value. The gas scale height is often modelled as a power law with disk radius of the form: $H \sim r^\beta$ \citep[for example ][]{Rich2021}. Under this form the gas scale height scales with $\beta = 1.05$ in the viscously heated region of the disk and $\beta = 1.29$ in the radiatively heated region of the disk. These are in line with scattered light observations of small dust grains and CO emission used to constrain the scale height through the disk \citep{Rich2021}.
  
\begin{table*}
\caption{ \textbf{ \protect\hyperlink{hyp:tab:model-summary}{Six planet formation scenarios:}} three disk masses and two levels of viscosity. Values shown here are at $r=10$ au and $t=0.100$ Myr (start of simulations). Disk mass stated is the integrated gas surface density inside 93 au, also at $t=0.100$ Myr. See Fig. \ref{fig:model_alphas_h_sigma} for radial profiles and Fig. \ref{fig:app:alldisks-masses-trunc93au} for disk mass over time. Within each scenario, we grow and evolve 100 planetary embryos (in separate simulations) with initial orbital radii logarithmically spaced between $r=0.2-94$ au, over almost 2 Myr ($t=0.100 - 1.888$ Myr). ``DZ'' means ``dead zone''.}
\begin{tabular}{@{}ccccccccccc@{}}
\toprule
            & \multicolumn{2}{c}{\textbf{$M_{\rm gas\,disk}$} ($M_{\odot}$)} & \multicolumn{4}{c}{\textbf{$\alpha$} parameter \protect\href{https://youtu.be/pi9fze1dO2A}{[link to movie]}}                          & \multicolumn{2}{c}{\textbf{$h=H/r$} \protect\href{https://youtu.be/vCE1fJFPenE}{[link to movie]}} & \multicolumn{2}{c}{\textbf{$\Sigma$ (${\rm g\,cm^{-2}}$)} \protect\href{https://youtu.be/Q7hxdWvqwp8}{[link to movie]}} \\ \midrule
            & \textbf{high $\alpha$}       & \textbf{low $\alpha$}        & \multicolumn{2}{c}{\textbf{high $\alpha$}} & \multicolumn{2}{c}{\textbf{low $\alpha$}} & \textbf{high $\alpha$}      & \textbf{low $\alpha$}      & \textbf{high $\alpha$}       & \textbf{low $\alpha$}        \\
            &      & & inside DZ    & outside DZ    & inside DZ    & outside DZ    &           &           &             &             \\ \midrule
\textbf{low-mass}    & 0.14 &  0.11    & $10^{-5}$       & $10^{-3}$        & $10^{-6}$       & $10^{-4}$       & 0.051     & 0.047     & 199         & 186        \\
\textbf{medium-mass} & 0.27 & 0.22    & $10^{-5}$       & $10^{-3}$        & $10^{-6}$       & $10^{-4}$       & 0.059     & 0.048     & 318         & 364        \\
\textbf{high-mass}   & 0.41 & 0.32    & $10^{-5}$       & $10^{-3}$        & $10^{-6}$       & $10^{-4}$       & 0.065     & 0.050     & 411         & 502        \\ \bottomrule
\end{tabular}
\label{tab:model-summary}
\end{table*}

\subsection{ Planet Growth }
\label{sec:formation-model}

Currently, two main theories for planet formation exist: formation through core accretion which we adopt here, and formation through gravitational instability. The core accretion scheme is further split into two mechanisms, depending on the size of the objects that accrete onto the planet: planetesimal accretion (10-100 km bodies) and pebble accretion (mm-cm bodies). We note that while accretion rates by pebbles are two orders of magnitude more rapid than accretion by planetesimals \citep{bitsch2015growth, johansen2017forming}, it still remains unclear whether pebble accretion can build large solid cores that are inferred in giant planets \citep{Brouwers2018,AliDib2020}. 

The discovery of rings in planetary disks has recently focused a great deal of attention on dust trapping by pressure bumps. This is a quickly developing new area of research that addresses possible planetesimal formation within the bumps \citep{Jiang2021, Carerra2021}.  We note that this could be an additional, or perhaps even major source of planetesimals under some conditions.   

{\it A conservative treatment of planetesimal accretion:} In this paper, we compute planet growth during migration using a conventional model for the source of planetesimals that ignores the back reaction of forming planets on disk structure. We do this for two reasons: deliberately in order to gauge how massive our planets can become in that picture; and practically because of the difficulty of including this self consistently in our already very demanding numerical simulations.  Specifically, we assume the standard planetesimal accretion paradigm of \citet{ida2004toward}. Furthermore, we do not include any dust evolution, nor the production of planetesimals from the underlying dust distribution. Instead, we assume that the availability of planetesimals follows the gas surface density, proportional to $\Sigma$ by a radially constant planetesimal-to-gas ratio $\fpl$:
\begin{equation}
    \label{eqn:dust-surface-density}
    \Sigma_{\rm pl} = \fpl \, \Sigma \, .
\end{equation}

 We now describe the three phases of planet growth and how we calculate a planet's accretion history if/when it gains enough mass to enter each subsequent phase. At these discrete times in a planet's formation history, we change the planetesimal-to-gas ratio $\fpl$ to approximately reflect changes in the dynamical effect of the growing planet on the surrounding population of plantesimals, and their probability of accretion onto the growing planet - which would otherwise be an intractable problem in our semi-analytic formalism. 

We initialize planetary cores with mass 
\begin{equation}
    \label{eq:Mcore}
    \Mcore = 0.01 \, \Me \, .
\end{equation}

The first phase consists of oligarchic growth. During this phase, the core grows by successive accretion of planetesimals that come close enough (specifically, within its 10 Hill radii feeding zone, $r_{\text{Hill}} = \ap \,[\Mp / 3\Ms]^{1/3}$). The heat generated by this accretion prevents any gas accretion onto the core. The core accretes at a rate
\begin{equation}
    \label{eq:Mcoredot}
    \Mcoredot = \frac{\Mp}{\tacc} \, ,
\end{equation}
where $\tacc$ is the core accretion timescale of \cite{ida2004toward}:
\begin{align}
    \tacc =& 1.2\times 10^5 {\rm yr} \left(\frac{\Sigma_{\rm pl}}{10 {\rm gcm}^{-2}}\right)^{-1}\left(\frac{a}{{\rm au}}\right)^{1/2} \left(\frac{\Mp}{M_\oplus}\right)^{1/3}\left(\frac{\Ms}{M_\odot}\right)^{-1/6}\nonumber\\
	\times & \left[\left(\frac{\Sigma}{2.4\times 10^3 {\rm gcm}^{-2}}\right)^{-1/5} \left(\frac{a}{{\rm au}}\right)^{1/20}\left(\frac{m}{10^{18} {\rm g}}\right)^{1/15}\right]^{2},
	\label{eq:acc_timescale}
\end{align}
where $\Ms$ and $m$ are the mass of the central star and incoming planetesimals (which we assume are all $10^{18}$g in mass). We assume that during oligarchic growth the young planet is too hot for gas to be collected and hence the accretion of solids is the only source of planetary growth. As such, the time derivative of the planet mass is strictly:\begin{align}
    \frac{{\rm d} M_{\rm p}}{{\rm d} t} = \Mcoredot.
    \label{eq:massevo}
\end{align}
At later phases, when gas accretion becomes the dominate source of mass evolution, equation \ref{eq:massevo} will also include a gas accretion term.

Throughout this first phase, we set $\fpl = 0.01$. Using the $\Sigma_{\rm pl} = \fpl \, \Sigma$ proportionality assumes very efficient ($\sim100\%$) planetesimal formation from the underlying dust density distribution. Indeed isolated streaming instability simulations \citep[eg.][]{Schafer17} rapidly (within $\sim 30$ orbits) convert all of the available dust to planetesimals. However, a given disk radius is \textit{not} isolated, since radial drift continually replenishes dust from larger radii. Therefore our assumption that $\fpl = f_{\rm dust,ISM} = 0.01$ represents a balance between higher expected dust-to-gas ratios driven by radial drift, and lower planetesimal formation efficiencies. 

This first phase of growth by solid accretion slows once the core depletes its $10$ $r_{\text{Hill}}$ feeding zone and reaches the core isolation mass \citep{ida2004toward}:
\begin{equation}
    \label{eqn:Mciso}
    \Mciso = 10 \, \Me \left( \frac{1}{10^{-6} \, \Me \, \text{yr}^{-1}} \frac{\text{d} \Mp }{\text{d}t} \right)^{1/4} \left( \frac{\kappa_{\text{env}}}{1\text{ cm}^2 \text{g}^{-1}} \right)^{0.3},
\end{equation}
where $\kappa_{\text{env}}$ is the opacity of the envelope \citep{mordasini2014grain}. 

The second phase marks the end of core formation and the beginning of the planet's atmospheric growth. Gas accretion begins slow, with timescales on $O(10^6)$ years, and as such the planet can migrate a significant amount during the initial growth of its proto-atmosphere. As it migrates it encounters a new population of planetesimals at different orbits, perturbing them potentially into its feeding zone. This results in a small (ie. very slow) increase in refractory mass in the planet as planetesimals are directly accreted into the growing atmosphere  \citep[see for example][]{NGPPS2020a}. To model this slow accretion of planetesimals we reduce $\fpl$ by a factor of 10 to be $\fpl=0.001$ and continue to allow growth via Equation \ref{eq:Mcoredot} (along with Equation \ref{eq:Mgasdot} below) to represent an evolved population of planetesimals that have been partially cleared by the migrating planet. 

The above decrease in $\fpl$ has the effect of changing the planetesimal accretion timescale from $O(10^5)$ years to $O(10^6)$ years. This longer timescale is reflective of the typical migration timescale for planets with a mass equal to the isolation mass in our disk model. In this way, our planetesimal accretion rate reflects the fact that the planet must move into a region of the disk that has previously untouched planetesimals in order to further its solid accretion.

The gas accretion is regulated in our model by the Kelvin-Helmholtz contraction time scale of the gas envelope $\tau_{\text{KH}}$:
\begin{equation}
    \dot{M}_{\rm gas} \simeq \frac{\Mp}{\tau_{\text{KH}}},
    \label{eq:Mgasdot}
\end{equation}
where \citep{hasegawa2013planetary}:
\begin{equation}
    \tau_{\text{KH}} \simeq 10^c \, \text{yr} \,  \left(\frac{\Mp}{\Me} \right)^{-d}.
\end{equation}

The parameters $c$ and $d$ depend on the opacity of the envelope $\kappaenv$; the values that have been identified by \citet{alessi2018formation} to best reproduce the observed mass-period relation of exoplanets are $c=9$, $d=3$ and $\kappaenv = 0.001$ cm$^{2}$ g$^{-1}$. Unstable gas accretion can occur if the planet becomes massive enough to lower the Kelvin-Helmholtz timescale to $\sim 10^5$ years \citep{cridland2016composition}.

The total mass evolution of the planet (equation \ref{eq:massevo}) becomes:\begin{align}
    \frac{{\rm d} M_{\rm p}}{{\rm d}t} = \Mcoredot + \dot{M}_{\rm gas},
    \label{eq:massevo2}
\end{align}
where $\Mcoredot$ follows from \ref{eq:Mcoredot} and the appropriate change of $f_{\rm pl}$. Note here we use the variable $\Mcoredot$ to stay consistent with our earlier notation, rather than explicitly stating that solid accretion results in solid delivery directly to the core. Once the proto-atmosphere is sufficiently massive most planetesimals no longer survive their trip to the core, instead evaporating in the gas \citep{Mordasini2016}. The distinction between a planetesimal reaching the core or not does not play a significant role in our model, as we are only interested in the overall mass evolution of the planet.

The third and final phase begins when a planet opens a gap in the local gas surroundings. It decouples from its surroundings and the gas accretion geometry changes \citep{szulagyi2014accretion}. The standard gap opening criterion is met if the torque exerted by the planet on the disk exceeds the disk's viscous torque, or equivalently, if the planet's Hill sphere exceeds the disk's pressure scale height:
\begin{equation}
    \label{eqn:Mgap}
    \Mgap = \Ms \min \Big( 3h^3 , \sqrt{40 \alpha h^5} \Big) \, ,
\end{equation}
where $h=H/r$ is the disk aspect ratio \citep{Lin1993,matsumura2006dead}.

In this third phase, both the gas and planetesimal accretion rates are once again modified. Due to the aforementioned change in gas accretion geometry we follow the gas accretion model of \cite{Cridland2018} to account for the interaction of the vertically flowing gas \citep[the so called `meridonial flow' of][]{Morbidelli2014,Teague2019} and the planet's internally generated magnetic field \citep[as proposed by][]{Batygin2018}. These interactions conspire to slow gas accretion\footnote{By a factor proportional to (nearly) the inverse of the planet's mass, see \cite{Cridland2018} for details.} and eventually lead to the termination of planetary growth. Along with the change in gas accretion geometry, the orbits of planetesimals potentially in the feeding zone of the planet become ever more  eccentric as the gas in the region is depleted, which further reduces the efficiency of their accretion onto a growing planet. To model this effect, we further reduce $\fpl$ by a factor of 1000 to be $\fpl=10^{-6}$ when the gap is opened. 

We note two important attributes of the formalism described in this section. Firstly, we are assuming that planetary core growth scales as $\Mcoredot \propto \Sigma_{\rm pl} \propto \Sigma$ (by Eqns. \ref{eq:Mcoredot}, \ref{eq:acc_timescale} and \ref{eqn:dust-surface-density}). We note that since the gas surface density $\Sigma$ drops over time according to Equation \ref{eq:dm01}, so too do our planet accretion rates. The length of time where planet core growth can be sustained thus depends on the disk viscosity $\alpha$ as well as the current position of the planet. Secondly, the $\fpl$ prescription negates any feedback or coupling between the growing planet and the disk (eg. increased availability of solids due to dust trapping in planet-induced pressure bumps). For these two reasons, we consider our planet accretion formalism to be conservative.

Figure \ref{fig:app:radius-vs-time} in Appendix \ref{app:numerical-results} shows the value of $\fpl$, and hence the growth stage, of each planet we form throughout their formation history. As our results will show, only planets that migrate to within $\sim 1$ au enter into the third and final growth phase. The potential planets relevant to the dust gaps/rings at large radii in HL Tau do not exceed the second phase by the end of our simulations.

\subsection{Planet Migration}

Planet migration is an inevitable consequence of planet-disk interaction. The key results for planet migration presented in this work pertain specifically to Type I migration. This is the regime of migration that all planets initially follow, until they have grown in mass enough to escape Type I disk torques by opening a gap in the gas (ie. by exceeding Eqn. \ref{eqn:Mgap}), at which point they transition into the Type II migration regime. In both regimes, the torque experienced by the planet depends on the planet's mass; in that sense, this section and the previous (Sec. \ref{sec:formation-model}) are intertwined.

The torque calculations performed in this work closely follow the method first developed in \citet{paardekooper2011torque}. The action of these torques was effectively visualized in what we call torque maps by \cite{coleman2014migration}. We have implemented both of these approaches in \citet{cridland2019physics}, and then applied them in our combined studies of  planet formation and migration in \cite{cridland2019connecting}. The mathematical details of Type I migration are deferred to our theoretical results in Section \ref{sec:res:insights-planetmigration}. First, we briefly summarize the basic physical concepts for the general reader. 

\subsubsection{Type I Migration} 
\label{subsec:typeI-migration}

Type I planet migration in a gaseous disk refers to a change in a planet's semi-major axis caused by the exchange of angular momentum between the planet and the disk in which it is embedded. Angular momentum is exchanged by gravitational torques. The torques that drive low mass planets in the Type I regime are of two kinds: \textbf{Lindblad torques} and \textbf{corotation torques}. 

Lindblad resonances between the disk gas and planet are located interior and exterior to the planet's orbit.  A wave flux is excited at each of these that carry off angular momentum through the disk \citep{goldreichtremaine1979, goldreich1980disk}. The net Lindblad torque, resulting from the difference of the inner and outer Lindblad torques, generally leads to inward planetary migration and outward tranport of the planet's angular momentum. 

The second kind of angular momentum exchange is not wavelike, and takes place in bands of planetary orbits lying very close to corotation with the planet. These corotation torques depend on a number of different kinds of physical processes such as disk viscosity (Sec. \ref{subsec:insights-Mpnucorot-Mpchicorot}), thermal diffusion (also Sec. \ref{subsec:insights-Mpnucorot-Mpchicorot}), and the action of disk winds (Sec. \ref{sec:discussion:diskwinds}), all of which affect the flow of angular momentum into this region \citep{paardekooper2011torque}.

The direction and magnitude of the total Type I torque exerted on a planet by the disk is the sum of the Linblad and total corotation torques. These depend not only on the planet's mass, but also on the local disk properties such as the gradients in local temperature and column density, and the adiabatic index of the gas.  These gradients vary across the disk, as for example in the disk heating which switches from viscous to stellar irradiation dominance as the disk evolves.  Thus, the total torque and its direction varies with the  planet's semi-major axis. This also implies that  planet-disk interaction and therefore planet migration is dynamic, changing as both the planet and the disk evolve.  
  
The central point here is that the direction of this total torque in Type I migration \textit{can} be outward, depending upon planet mass, disk viscosity and aspect ratio. We show this first in the simulations, and then in the continuation of this thread in Section \ref{subsec:insights-background}.
 
We note that the torque maps throughout this work are normalized to the reference torque \citep{tanaka2002torques3d}:
\begin{equation}
    \label{eqn:Gamma_0}
    \Gamma_0 = \Big( \frac{q}{\hp} \Big)^2 \, \Sigma_p \, \rp^4 \, \Omegap^2
\end{equation}
where $q = \Mp / \Ms$ and $\hp = \Hp / \rp$, and the index p denotes evaluation of the quantity at the position of the planet.  The quantity $ \Gamma_0 $ sets the magnitude of the net Lindblad torque that arises from the difference between the inner and outer Linblad torques \citep{Nelson2018}.

\subsubsection{Type II planet migration}
\label{sec:typeIImigration}

Returning to our formation model: If a gas gap is opened (Eqn. \ref{eqn:Mgap}), the planet clears its corotation region of gas and the strongest Lindblad node is evacuated. Type I migration is turned off and the planet transitions into Type II migration (which also corresponds to a change in the gas and planetesimal accretion rate, third phase in Sec. \ref{sec:formation-model}). Under this migration scheme the planet acts as an intermediary for angular momentum transport through the disk, moving its orbit to smaller radii. It does so on the viscous timescale:\begin{align}
    \frac{d\ap}{dt} = \frac{\ap}{t_\nu},
\end{align}
where $t_\nu = \ap^2/\nu$.

In the event that the planet's mass exceeds the total mass of the gas disk within its orbital radius, \begin{equation}
    \label{eqn:Mcrit}
    \Mcrit = \pi \Sigma \ap^2 \, ,
\end{equation}
then we lengthen the viscous timescale to $t^\prime_\nu = t_\nu(1 + \Mp / \Mcrit)$.

\section{Numerical Results: Migration and Torque Maps}
\label{sec:res:torqmaps}

\begin{figure*}
	\includegraphics[width=18cm]{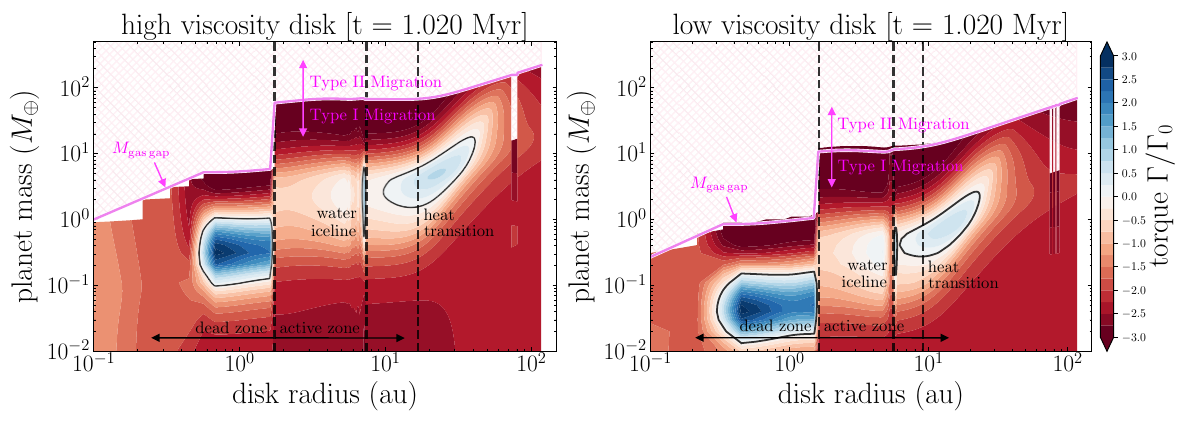}
    \caption{ \textbf{ \protect\hyperlink{hyp:fig:torque_maps}{Torque maps (migration maps) of evolving disks.}} [Shown for our high-mass disk]. The disk exerts a torque on a forming planet as a function of the planet's mass and semi-major axis. We show this torque at one demonstrative time snapshot, $t=1.020$ Myr. For comparison, the high viscosity (\textbf{left panel}) and low viscosity (\textbf{right panel}) models are shown side by side. Negative torque $\Gamma/\Gamma_0 < 0$ (red) corresponds to inward migration, and positive torque $\Gamma/\Gamma_0 > 0$ (blue) to outward migration. At planet masses above the gas gap-opening mass (pink line; Eqn. \ref{eqn:Mgap}), the planet is free to move under Type-II migration (pink cross-hatching). Three main features of the torque maps, corresponding to contours of zero net torque, are labelled (the dead zone, water iceline and heat transition; see Fig. \ref{fig:3trap_definitions} for formal definitions). Those three contours of zero net torque also appear in Fig. \ref{fig:madiagram_alldisks}. The distinct notches in these diagrams that occur near 100 au are minor spurious numerical artifacts arising from the resolution used in the astrochemisty simulations.}
    \label{fig:torque_maps}
\end{figure*}

In this section, we focus purely on the background disk and the features that dictate planet migration. Section \ref{sec:res:planet-tracks} presents the resulting planetary evolution tracks. 

As described in Section \ref{sec:model-and-sims} and summarized in Table \ref{tab:model-summary}, we explore six planet formation scenarios. Of particular interest to this work is the comparison of evolution outcomes between two levels of viscosity, set by the $\alpha$-parameter to $\alpha = 10^{-3}$ or $\alpha = 10^{-4}$ over the bulk of the disk. In relation to one another, we refer to these as \textbf{high} and \textbf{low viscosity}, respectively. Three initial disk masses (\textbf{low-mass}, \textbf{medium-mass}, \textbf{high-mass}) are chosen to bracket the observationally constrained values for the HL Tau disk. Throughout the following sections, all figures show our high-mass disk models unless otherwise stated. See Table \ref{tab:movies} for movies of the the low- and medium-mass disk model results.

In \hypertarget{hyp:fig:torque_maps}{Figure \ref{fig:torque_maps}} we calculate the total torque that would be exerted on a planet of any given mass and semi-major axis and display it at an intermediate time snapshot in our simulations. We call this landscape a ``torque map''. The colourbar indicates the magnitude and direction of the total torque. Inward-directed torque ($\Gamma/\Gamma_0 < 0$, shown in red) works to decrease a planet's semi-major axis, and outward-directed torque ($\Gamma/\Gamma_0 > 0$, shown in blue) works to increase it. Where these two opposing forces cancel ($\Gamma/\Gamma_0 = 0$, shown in white) are special locations within the disk known as planet traps, which we discuss more below. As in Figure \ref{fig:model_alphas_h_sigma}, we show the high viscosity case on the left, and low viscosity on the right. The features of the torque maps are as follows.

\textit{Type I/II migration regimes.} The upper boundary of the torque maps is outlined by the gas gap-opening mass $M_{\rm gas\, gap}$ (Eqn. \ref{eqn:Mgap}, shown in pink). Beyond this mass, a planet is locally detached from the gas and therefore free of the torques associated with Type I migration. Instead, it follows Type II migration (Sec. \ref{sec:typeIImigration}), moving slowly inwards on a viscous timescale of millions of years as the gas is slowly accreted onto the star, or evaporated. We show the Type II regime with pink cross-hatching. Note that torques exerted by the planet on the disk always work to open a gap, while viscous flow competes to diffusively smoothen the resulting surface density gradients and fill the gap back in.

\textit{Dead and active zones.} At the radial location between the dead zone and active zone, the upper boundary of the torque maps jumps suddenly by a factor of 10. This is a consequence of two things: (a) the dependence of the gas gap-opening mass on viscosity, $M_{\rm gas\, gap}\propto \alpha^{1/2}$, and (b) the step in viscosity between the two zones (see top row of Fig. \ref{fig:model_alphas_h_sigma}). In both our high and low viscosity models, $\alpha$ increases by 2 orders of magnitude in going outward from the dead into the active zone (either from $10^{-5}$ to $10^{-3}$, or from $10^{-6}$ to $10^{-4}$).

\textit{The scaling effect of viscosity.} How the level of viscosity affects the torque maps and hence planetary evolution is a key theme of this work and we discuss it extensively in Section \ref{sec:res:insights-planetmigration}. For now, we note that the factor of 10 decrease in $\alpha$ between the high and low viscosity models results in a global ``downward'' shift in the torque maps, such that all of the torque map features that determine planet migration occur at lower planet masses. 

\textit{Time evolution.} The torque landscape evolves over time \cite[gradually, owing to our 8 Myr depletion time;][]{Michel2021longliveddisks} in two ways. Firstly, the outermost radius increases as the disk spreads viscously outwards. Secondly, features in the torque maps move slowly inwards over time. \citet{coleman2016formation,coleman2016giant} also observed this behaviour in their simulations. It is a consequence of the gradual reduction of the disk's surface density due to viscous evolution  \citep{cridland2019physics}. The evolution can be seen more clearly in the movie provided for Fig. \ref{fig:planet_tracks} in Sec. \ref{sec:res:planet-tracks}. 

In particular, the gradual reduction of the column density due to viscous evolution, eventually followed by photoevaporation, means that the disk becomes more easily ionized by external radiation (X-rays and FUV) in its outer regions. The dead zone therefore shrinks - its outer boundary moves inwards as it becomes possible to sustain magnetized turbulence in the expanding region of lower column density gas beyond. The shrinking of the viscously heated inner disk region inevitably also moves the heat transition inwards. As such, the planet trap that is associated with the position of the heat transition (see below) also moves radially inwards with time. 

\begin{figure}
	\includegraphics[width=\columnwidth]{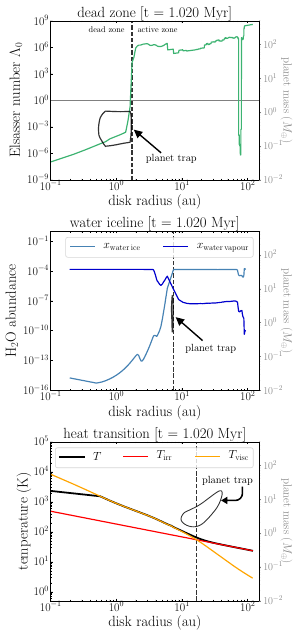}
    \caption{ \textbf{  \protect\hyperlink{hyp:fig:3trap_definitions}{Formal definitions of three planet traps.}} [Demonstrated for our high viscosity, high-mass disk at $t=1.020$ Myr]. In all panels, the black contours correspond to the contours of zero net torque in Fig. \ref{fig:torque_maps} (left panel) with height given by the right-hand y-axis. Vertical dashed black lines indicate the specific radius of each associated physical change, which in the case of the dead zone and water iceline, coincide with the radial location of planet trapping. \textbf{Top panel:} The outer edge of the dead zone is the radius where the ohmic Elsasser number (see Eqn. \ref{eq:elsasser} and App. \ref{sec:volatilechemicalmodels}) exceeds unity. \textbf{Middle panel:} The water iceline occurs where the abundances of water vapour and water ice are equal ($r_{\rm IL}$, App. \ref{sec:volatilechemicalmodels}). \textbf{Bottom panel:} The heat transition is an extended radial region over which the disk temperature profile (black line, Eqn. \ref{eq:dm06}) transitions from the viscously-heated profile (orange, Eqn. \ref{eq:dm03}) to the radiatively-heated one (red, Eqn. \ref{eq:dm04}). The associated planet trap is \textit{not} at $\rheat$ (vertical dashed black line; Eqns. \ref{eqn:hlt-surface-density-profile} \& \ref{eqn:hlt-temperature-profile}). }
    \label{fig:3trap_definitions}
\end{figure}

\textit{Contours of zero net torque.} In both panels of the torque maps in Figure \ref{fig:torque_maps}, we outline three occasions of zero net torque in black contours. The shape and location of the contours depends on the underlying disk chemistry, temperature and density gradients, disk viscosity, and planet mass. Each contour is associated with the radial location of a planet trap, which are in turn associated with a change in some physical quantity within the disk. The precise radius at which the disk undergoes one of these changes is indicated in Figure \ref{fig:torque_maps} with a vertical dashed line and labelled. 

\textit{Planet traps.} In essence, a planet trap is a location of zero net torque (ie. a point along a contour), bounded from the inside by outward-directed torque, and bounded from the outside with inward-directed torque. Planet traps are convergent in the sense that no matter what edge of the trap a planet starts on, it will be pulled towards the trap and kept there. The same is true in the opposite sense for a planet that starts inward of a trap (ie. in a blue zone). 
The existence of these traps is computed self consistently from the disk properties and using the torque formulae in Section \ref{subsec:insights-background}. The three types of traps that we have found are expected on general grounds \citep{hasegawa2011dust}, as discussed below. 

In \hypertarget{hyp:fig:3trap_definitions}{Figure \ref{fig:3trap_definitions}}, we dedicate three panels to each of the three contours of zero net torque and show explicitly the associated physical change. The latter closely coincides with the radial location of planet trapping for the dead zone and water iceline, but not the heat transition. The vertical dashed lines are the same ones shown in Figure \ref{fig:torque_maps}. We discuss them each in turn.

\textbf{The dead zone.} The transition between the dead zone and active zone, as described in Appendix \ref{sec:volatilechemicalmodels}, occurs where the ohmic Elsasser number (Eqn. \ref{eq:elsasser}) exceeds unity, shown by the light grey horizontal line. At $t=1.020$ Myr, this happens around $r\approx1.5$ au in our high-mass models. 

\textbf{The water iceline.} The water iceline (also known as the water snowline) is defined as the disk radius at which water vapour ($x_{\rm water\, vapour}$) is equally abundant as water ice ($x_{\rm water\, ice}$). The radius of the water iceline is roughly $r\approx 7$ au at $t=1.020$ Myr.

In the case of the dead zone and water iceline traps, the outer edges of the corresponding contours constitute a clear planet trap at a sharp radial location which is constant over a certain range of planet masses for which trapping is effective. The radial localization is because there is a very sharp radial change in the turbulence gradient (for the dead zone), and opacity gradient (for the water iceline). The former is a consequence of the rapid quenching of MRI with disk radius (ie. increased screening of X-rays), and the latter due to the sharp phase transition that defines the vapour to solid transition for water. 

\textbf{The heat transition.} The heat transition describes a radially extended region where the disk goes from being heated predominantly by viscous dissipation to predominantly by stellar irradiation. As described above Equation \ref{eqn:hlt-temperature-profile}, the midplane temperature profile follows the power-law $T\propto r^{-9/10}$ inside the heat transition, and $T\propto r^{-3/7}$ outside. The radius where viscous- and radiatively-heated temperature profiles formally intersect is $r_{\rm HT} \approx 17$ au (at $t=1.020$ Myr). The midplane temperature profile transitions between the two smoothly and, importantly, gradually, via Equation \ref{eq:dm06}. This yields a contour of zero net torque that itself spans a large range of values in radius, extending outwards to tens of au (the outermost radius depending on disk mass), and turning upwards in planet mass. The region of blue outward-directed torque enclosed by this contour of zero net torque plays a key role in determining the radial evolution of planets formed in our low viscosity disks -- a point we will return to many times in the rest of this work.

\section{Numerical Results: Planet Evolution Tracks}
\label{sec:res:planet-tracks}

\begin{figure*}
	\includegraphics[width=18cm]{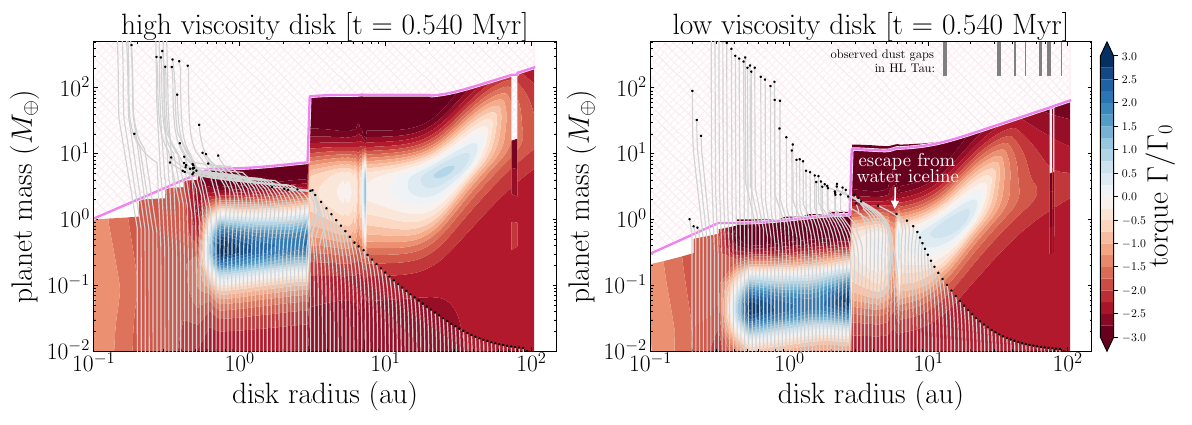}
	\includegraphics[width=18cm]{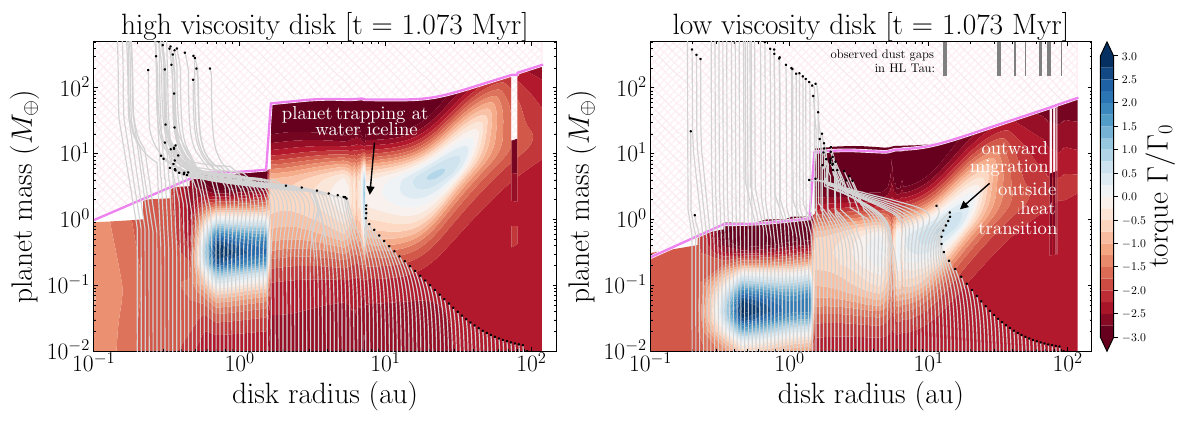}
	\includegraphics[width=18cm]{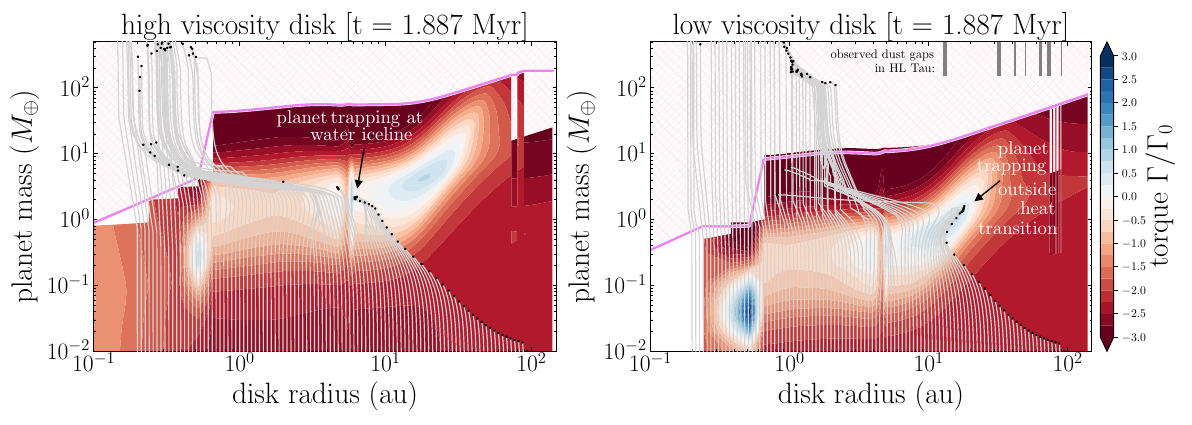}
    \caption{ \textbf{ \protect\hyperlink{hyp:fig:planet_tracks}{Planet evolution tracks.}} [Shown for our high-mass disk]. Like Fig. \ref{fig:torque_maps}, but overlaid in grey lines are the growth and migration histories of 100 planets, with black dots indicating a single planet's location at time: \textbf{(top row)} $t=0.540$ Myr, showing planets in the low viscosity disk escaping from the water iceline trap; \textbf{(middle row)} $t=1.073$ Myr, showing trapping of planets at the water iceline in the high viscosity disk and outward migration of planets in the low viscosity disk; and  \textbf{(bottom row}) $t=1.888$ Myr, the end of our simulations. Each planet is formed in its own separate simulation, so there is no interaction between planets. In the low viscosity disk panels, we mark the radial locations of the observed gaps in the dust distribution of the HL Tau disk \protect\cite[Table 2 of ][]{brogan20152014}. Each grey tick corresponds to a dust gap, with thicker lines indicating those gaps whose locations are more well constrained. \textbf{ \protect\href{https://youtu.be/r47qs4BTxxU}{ [Link to movie.] }} }
    \label{fig:planet_tracks} 
\end{figure*}

Having described the background torque landscape and the features that dictate planet migration, we now present the resulting planetary evolution tracks.

In each of our six planet formation scenarios, we grow and evolve 100 planets, each with a different initial orbital radius between $0.2-93$ au, distributed logarithmically. Planetary cores are initiated with a mass of $0.01$ $\Me$, and grow by our conservative planetesimal core accretion formalism.  Each planet is formed in its own separate simulation, so there is no interaction between them.

In \hypertarget{hyp:fig:planet_tracks}{Figure \ref{fig:planet_tracks}}, we show the formation histories of the planets we form in our high-mass HL Tau disk models for two levels of viscosity. Overlaid atop the torque maps in grey lines are planet evolution tracks, showing how a planet has grown in mass due to accretion of material from the disk at each radius, and migrated over time as determined by the disk torques. Each black dot at the end of a planet track indicates that planet's current mass and semi-major axis. We provide three demonstrative time snapshots (top, middle and bottom rows) and label the key events happening at those times. The evolution of planetary trajectories in our torque maps can be better appreciated when viewed as a movie, for which we provide a link in the caption of Figure \ref{fig:planet_tracks} as well as Table \ref{tab:movies}. We'll begin by describing the effect of viscosity on planet growth, followed by its effect on Type I planet migration. 

As previously described (see Sec. \ref{sec:formation-model}), forming planets accrete an amount of solid material directly proportional to the local gas surface density at their orbital radius ($\Mcoredot \propto \fpl \, \Sigma $). This point can be most easily seen if the reader pauses the movie during the first few frames -- after a few$\times 10^4$ years of planet growth, but before too much of the disk has accreted onto the central star. At those early times, the distribution of planet masses falls off with radius following the gas surface density profile. Planet growth is easy at small disk radii, and difficult at large disk radii.

A second and crucial point pertains to planet growth over time. By Equations \ref{eq:dm01} \& \ref{eq:dm01b}, the lower viscosity disk loses its mass to the star at a lower rate. Therefore, the planet-building solids are retained in the lower viscosity disks for longer, enabling more sustained planet growth as time goes on. We note that this difference between viscosities is made subtler by our long depletion time, $\tau_{\rm dep} = 8$ Myr (Eqn. \ref{eq:dm07}). 

The differences in planet migration between the high and low viscosity models arise from differences in the resulting torque landscapes. As mentioned briefly in Section \ref{sec:res:torqmaps}, lowering the disk viscosity lowers the planet mass at which key features in the torque maps occur -- most notably, the extended region of outward-directed torque associated with the heat transition. We dive into the theory to explain why this happens in Section \ref{sec:res:insights-planetmigration}, and focus now on the outcomes.

The result foreshadowed in Section \ref{sec:res:torqmaps} is presented in the middle right panel of Figure \ref{fig:planet_tracks}. Planets in the lower viscosity disk ($\alpha=10^{-4}$) initialized near the heat transition (as close in as $\approx 10$ au) are captured by the extended region of outward torque and migrate outwards to large disk radii early on in their evolution (beginning around $t\approx 0.750$ Myr). At around $t=1.5$ Myr, this population of planets reach their maximum orbital radii, $\approx 20$ au. They encounter inward-directed torques at their exterior and their outward migration is halted. Trapped at this location of zero net torque associated with the heat transition, they go on to slowly migrate inward as the disk viscously evolves. To connect this result to the observed dust gaps in the HL Tau disk, we mark the gaps' radial locations on the low viscosity panels of Figure \ref{fig:planet_tracks} \cite[occuring between roughly 10 and 90 au; Table 2 of][]{brogan20152014}. 

While they are trapped, this population of planets continues to grow in mass but only slightly. This cessation of further growth is a direct consequence of our conservative accretion prescription - namely, that the solids fraction available for accretion onto the planet in the form of planetesimals is a constant $\fpl=0.01$ (Sec. \ref{sec:formation-model}), or even reduced to $\fpl=0.001$ in a couple cases (top right panel, Fig. \ref{fig:app:radius-vs-time}). The bottom right panel of Figure \ref{fig:planet_tracks} shows these planets still trapped at $t=1.887$ Myr, the end of the simulation. While the white regions of near zero torque extend to higher masses and orbital radii, our accretion model limits them from being driven to these larger mass and radial scales. 

In the higher viscosity disk, inward migration dominates, tempered effectively only by planet trapping at the water iceline. This is highlighted in the left panels of Figure \ref{fig:planet_tracks}. The planets trapped at the water iceline do not escape before the end of the simulation.  This is a consequence of the higher escape masses that this trap has in comparison to the low viscosity case.  One can observe this difference directly by noting that planets near the iceline in our low viscosity disk grow in mass reaching the value necessary to escape ($\approx 1 \, \Me$ in the high-mass disk) around $t=0.5$ Myr. 

Turning to the innermost regions ($r \lesssim 1$ au) of both the high and low viscosity disks, the high column density makes for rapid planet accretion and we see planets quickly exceeding the gas gap-opening threshold. They enter the Type II migration regime, having interacted with the outward-directed torque feature within the dead zone very little if at all.

\begin{figure*}
	\includegraphics[width=8cm]{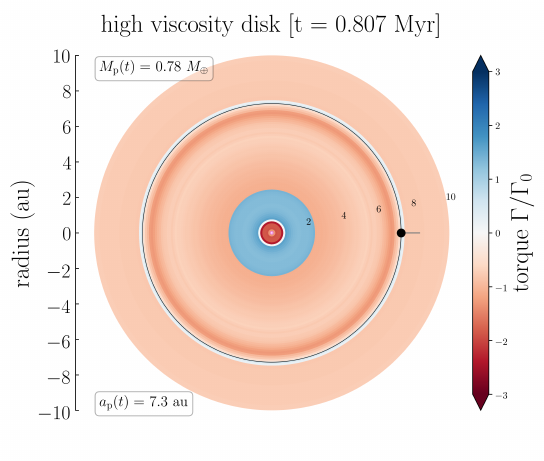}
	\includegraphics[width=8cm]{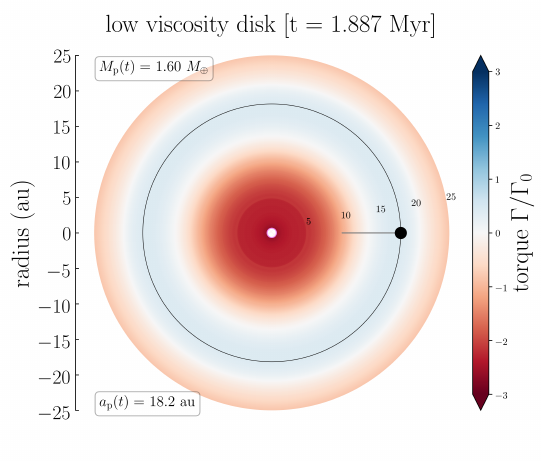}
    \caption{ \textbf{  \protect\hyperlink{hyp:fig:torqdisk_visualization}{Planet-trap interaction: single planet examples.}} [Shown for our high-mass disk]. An alternative visualization, best understood through a movie, of how disk torques determine planet migration. In each panel, we follow the evolution of a single, demonstrative planet. As a function of time, we take a horizontal slice through the torque map (Fig. \ref{fig:torque_maps} or \ref{fig:planet_tracks}) at the planet's current mass to yield a radial torque profile. We tile this profile azimuthally to form a face-on disk. The planet's current semi-major axis is shown as a thin black circle, and its current mass (also listed in the top left corner) is proportional to the size of the black dot. \textbf{Left panel:} The formation history of a planetary embryo in the high viscosity, high-mass disk, with initial semi-major axis $a_{\rm p,0}=8.3$ au. The planet is trapped at the water iceline and does not escape before the end of the simulation. \textbf{ \protect\href{https://youtu.be/sYTi1LZYjY0}{[Link to movie.]} \textbf{Right panel:}} The formation history of a planetary embryo in the low viscosity, high-mass disk, with initial semi-major axis $a_{\rm p,0}=10.7$ au. Around $t=0.700$ Myr, the planet migrates outward to almost twice its initial orbital distance, where it gets trapped for the rest of the simulation. \textbf{ \protect\href{https://youtu.be/qQnr9R4a7yI}{[Link to movie.]}}}
    \label{fig:torqdisk_visualization}
\end{figure*}

\begin{table*}
\caption{Collection of movies illustrating the key results of this work.}
\begin{tabular}{@{}c p{0.25\linewidth} p{0.5\linewidth}  c@{}}
\toprule
\textbf{Figure} & \textbf{Model}    & \textbf{Description} & \textbf{Link} \\ \midrule
\ref{fig:planet_tracks}      & high-mass, high \& low viscosity & Planet evolution tracks atop torque maps (all planets)    & \protect\href{https://youtu.be/r47qs4BTxxU}{[click]}               \\
      & medium-mass, high \& low viscosity &     & \protect\href{https://youtu.be/IA72QOti2m0}{[click]}               \\
      & low-mass, high \& low viscosity &     & \protect\href{https://youtu.be/oT4Td6gCqqQ}{[click]}               \\\midrule
\ref{fig:torqdisk_visualization} (left)      & high-mass, high viscosity         & Planet trapping at the water iceline (single planet)    & \protect\href{https://youtu.be/sYTi1LZYjY0}{[click]}              \\
\ref{fig:torqdisk_visualization} (right)      & high-mass, low viscosity          & Outward planet migration by the heat transition (single planet)     & \protect\href{https://youtu.be/qQnr9R4a7yI}{[click]}              \\\midrule
\ref{fig:single-madiagram-alldisks}      & 3 disk masses, low viscosity          & Outward planet migration by the heat transition (all planets)    & \protect\href{https://youtu.be/ihiTKngTfp4}{[click]}              \\ \midrule
\ref{fig:corot_lindblad_breakdown}      & high-mass, low viscosity          & Outward planet migration \& decomposition of disk torques (single planet)    & \protect\href{https://youtu.be/Rs8F7fn_31I}{[click]}              \\ \bottomrule
\end{tabular}
\label{tab:movies}
\end{table*}

Before leaving Figure \ref{fig:planet_tracks}, we note that all of the models shown produce Hot Jupiter planets. In previous previous papers, \citep{alessi2020formation, alessi2018formation} population synthesis studies showed that distributions of disk masses and lifetimes could explain the broad structure of planetary populations in models where Type I migration was drastically reduced by means of the various planet traps discussed here. In this paper, we focus on a disk model for these extended systems that features both fairly massive as well as long lived disks.  These conditions are exactly right for producing close in planets.  The rarity of Hot Jupiters is in turn a reflection of the relative scarcity of massive, long lived disks in the disk populations around young stars.  

As an additional aid for visualizing planet interaction with disk torques, we provide \hypertarget{hyp:fig:planet-trap-interaction}{Figure \ref{fig:torqdisk_visualization}} and two accompanying movies. We select a single planet from each of the high and low viscosity disks and follow their evolution over time. As each planet's mass changes, we take a horizontal slice through the torque map at that mass to create a 1D profile of the torque as a function of radius. We tile this profile azimuthally to create the image of an axisymmetric face-on disk -- the torque landscape as seen by the planet. 

In the left panel of Figure \ref{fig:torqdisk_visualization}, we highlight the interaction between a demonstrative planet (initial orbital radius $8.3$ au) and the water iceline in the high viscosity, high-mass disk. We place the planet (arbitrarily) at 3 o'clock, represented by a black dot whose size is proportional to the planet's mass. A grey line again records the planet's migration history, and a black circle indicates the planet's semi-major axis. At the time snapshot shown in the figure, the planet is trapped at the water iceline, where it grows in mass and migrates inward at a rate dictated by the trap for the rest of the simulation.

In the right panel of Figure \ref{fig:torqdisk_visualization}, we highlight the outward migration of a planet interacting with the extended region of outward torque associated with the heat transition in the low viscosity, high-mass disk. The time snapshot shown ($t=1.887$ Myr) corresponds roughly to when this planet reaches its largest orbital radius. With a linear radial scale, it is more apparent how much of the disk is dominated by this blue region when the planet is at the right mass.

In both of these planet-trap interaction examples, the planets do not reach a high enough mass to escape the influence of the trap. We discuss this \textbf{trap escape mass} in more detail in Section \ref{sec:res:insights-planetmigration}. (Looking ahead to that section, this mass is a function of time and disk mass, and in the case of the water iceline it is well described by Eqn. \ref{eqn:Mpnucorot}). For now we note that in our models, it takes at least $3-5 \, \Me$ to escape the water iceline, and at least $\approx 1.5-2.3 \, \Me$ to migrate up and over the heat transition trap, depending on the disk mass (see Table \ref{tab:populations}). \citet{cridland2019connecting} find similar trap escape masses for their water iceline (their Fig. 5).

\begin{figure*}
	\includegraphics[width=18cm]{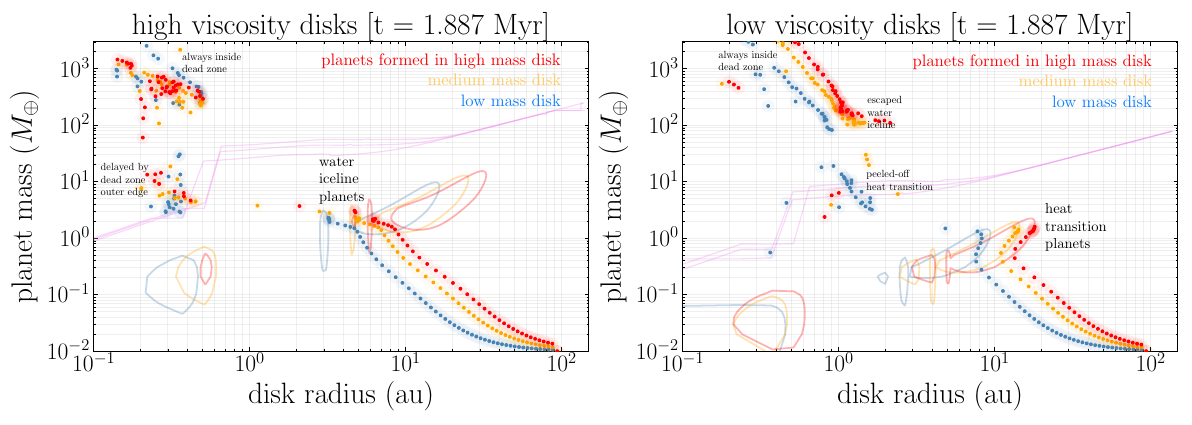}
	\caption{ \textbf{  \protect\hyperlink{hyp:fig:madiagram_alldisks}{Planet masses and semi-major axes at the age of HL Tau.}} Results from all six of our formation scenarios: high-mass (red), medium-mass (orange) and low-mass (blue) HL Tau disks, at two levels of viscosity (broadly, $\alpha=10^{-3}$ and $\alpha=10^{-4}$; see top panel of Fig. \ref{fig:model_alphas_h_sigma}). Each dot (planet) is surrounded by a faded circle simply to give the reader a sense of number density. As always, results for the high viscosity disks are shown on the left, and low viscosity on the right. For reference, and to put the planets' locations in the context of the torque maps, we overlay the gas gap-opening mass in pink (Eqn \ref{eqn:Mgap}), and the three contours of zero net torque in colours corresponding to disk mass (see also Fig. \ref{fig:torque_maps} \& \ref{fig:planet_tracks}). We identify planets trapped at the water iceline in the high viscosity disks as ``water iceline planets,'' and those trapped outside the heat transition in the low viscosity disks as ``heat transition planets'' (see Fig \ref{fig:single-madiagram-alldisks}'s movie).}
    \label{fig:madiagram_alldisks} 
\end{figure*}

\begin{figure}
	\includegraphics[width=\columnwidth]{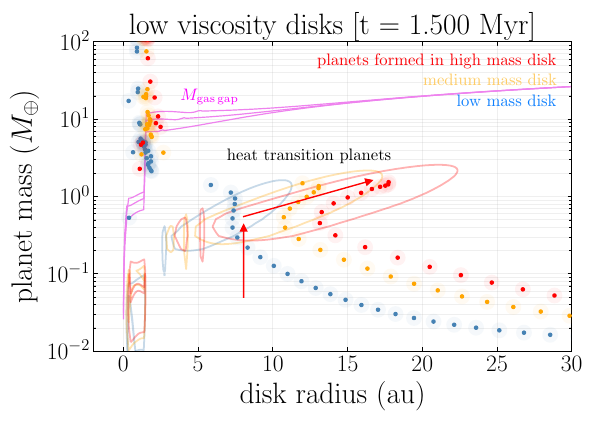}
	\caption{ \textbf{  \protect\hyperlink{hyp:fig:single-madiagram-alldisks}{Heat transition planets and the mechanism for outward migration.}} [Low viscosity disks]. A zoom-in on the right panel of Figure \ref{fig:madiagram_alldisks}. The x-axis is now in linear scale, showing how large a fraction of the disk is spanned by the contour of zero net torque associated with the heat transition (lightly coloured contours). Inside of these contours, the total torque is outward-directed, and outside, it is inward-directed. We indicate schematically the general migration trend that results from interaction with this zero-torque contour with red arrows, using the high mass disk as an example. \protect\href{https://youtu.be/ihiTKngTfp4}{\textbf{[Link to movie]}}}
    \label{fig:single-madiagram-alldisks}
\end{figure} 

\begin{table*}
\caption{Total mass, core mass and orbital radius of a select planet in each of the trapped populations at $t=1.888$ Myr (see Fig. \ref{fig:madiagram_alldisks}). In the bottom row we provide each trap's ``escape mass'': the planet mass corresponding to the zero net torque contour peak. Planets would leave the trap and migrate inwards again if their mass grew to exceed this threshold.}
\centering
\begin{tabular}{ccccccc} \toprule
        & \multicolumn{3}{c}{\textbf{Water iceline planets} (high viscosity disk)} & \multicolumn{3}{c}{\textbf{Heat transition planets} (low viscosity disk)}  \\
        & high-mass & medium-mass & low-mass        & high-mass & medium-mass      & low-mass      \\ \midrule
 $\Mp$    & 2.22 $\Me$    & 2.12 $\Me$     & 2.11 $\Me$    & 1.61 $\Me$ & 1.42 $\Me$ &  1.34 $\Me$  \\
$\Mcore$ & 2.10 $\Me$    & 2.03 $\Me$   & 2.02 $\Me$    & 1.55 $\Me$  & 1.40 $\Me$ & 1.32 $\Me$ \\
 $\ap$ & 6.39 au    & 5.02 au   & 3.22 au    & 18.1 au & 14.3 au &  7.78 au \\ \midrule 
 trap escape mass & 5.0 $\Me$    & 3.9 $\Me$     & 3.0 $\Me$    & 2.3 $\Me$ & 2.0 $\Me$ & 1.5 $\Me$        \\ \bottomrule 
\end{tabular}
\label{tab:populations}
\end{table*}

Up to now, we have featured results only from our high-mass disk models. In Table \ref{tab:movies}, we provide movie analogues to Figure \ref{fig:planet_tracks} for the low- and medium-mass disks. In \hypertarget{hyp:fig:madiagram_alldisks}{Figure \ref{fig:madiagram_alldisks}}, we present the results of all six planet formation scenarios in the form of a mass vs. semi-major axis diagram (M-a diagram). As usual, the high viscosity case is shown on the left, and the lower viscosity case on the right. We colour the planets according to the mass of the disk in which they formed. 

As we have argued, understanding the underlying torque map features is crucial for interpreting points on an M-a diagram, and so we include the three contours of zero net torque in the background. For the same purpose, upper edge of the torque map (the gas gap-opening mass, Eqn. \ref{eqn:Mgap}) is shown in pink. 

In the context of the underlying torque map features, we describe the groupings of planets in each panel according to the torque feature that most influences their evolution history and label them on Figure \ref{fig:madiagram_alldisks}. The ``water iceline planets'' and ``heat transition planets'' are those trapped at the water iceline and contour of zero net torque associated with the heat transition, respectively. Masses and semi-major axes for a representative planet in each population are given in Table \ref{tab:populations}, as well as the escape mass for each trap.

\textbf{Water iceline planets.} The high viscosity disks form a population of planets that spend a significant fraction of their formation history trapped at the water iceline. This trap occurs at larger distances from the star for higher disk masses, simply because the temperature profile at the location of the iceline $T_{\rm visc} \propto \Sigma$, and the higher mass disks have higher surface density (see bottom panel Fig. \ref{fig:model_alphas_h_sigma}). At $t=1.888$ Myr, the position of the ice line, $r_{\rm IL}$, is between $3-6$ au.  The iceline planets in the low mass disks come closer to escaping the trap than those in the higher mass disks because the escape mass is lower (see the dependence of Eqn. \ref{eqn:Mpnucorot} on $h$, which is lower for a lower $\Mdisk$).

\textbf{Heat transition planets.} The low viscosity disks form a population of planets with masses between $1-2\, \Me$ and with orbital radii $8-20$ au. To highlight this result, \hypertarget{hyp:fig:single-madiagram-alldisks}{Figure \ref{fig:single-madiagram-alldisks}} provides a closer look at the masses and semi-major axes of the heat transition planets in the low vs high viscosity disks, as well as the shape and extent of the zero-torque contour associated with the heat transition.  

Examining Figure \ref{fig:single-madiagram-alldisks}'s movie, planet interaction with the heat transition's torque feature unfolds, broadly speaking, as follows. Early in these heat transition planets' formation history, they encounter the low mass end of the zero-torque contour associated with the heat transition. As they grow in mass (represented by the ``up'' arrow), the outward-directed torque inside the contour boundary forces them to migrate outwards (indicated by the ``right'' arrow). If their growth is too rapid, they reach the ``top'' of the contour, peeling off and migrating inwards. Otherwise, they continue to travel ``along the blue band'' inside the zero-torque contour to its outermost radius, where they are trapped until they can acquire the mass needed to grow ``up and over'' the trap and escape. The ultimate fate of the heat transition planets is therefore to grow more massive and likely migrate inward. As we will see in greater detail in the theory section to follow, the contour shape turns upwards in the outer regions because $h=H/r$ increases with radius in the outer, radiatively heating dominated region of the disk, and it is this shape that aids planet trapping outside the heat transition.

\begin{figure}
	\includegraphics[width=\columnwidth]{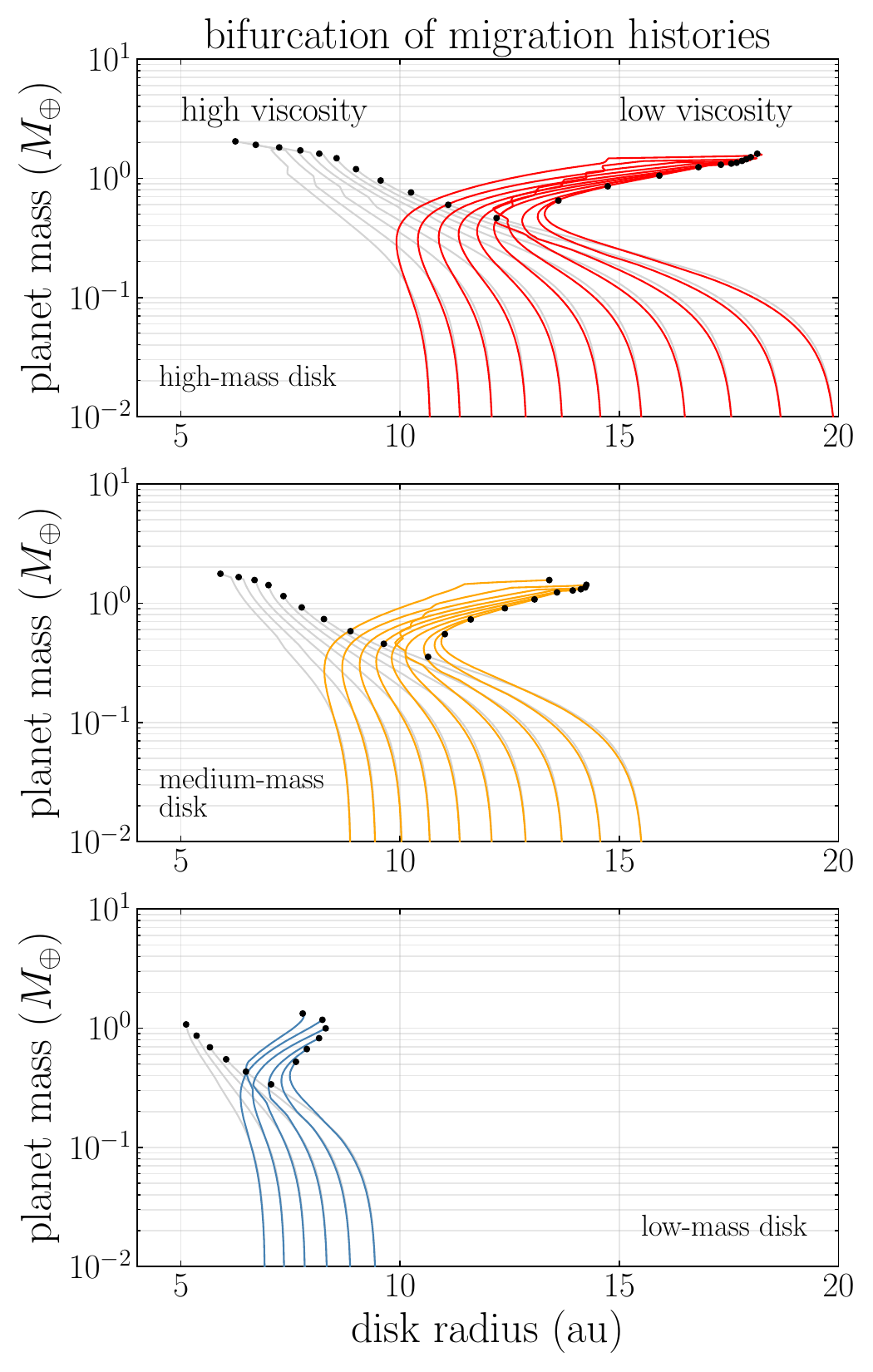}
	\includegraphics[width=\columnwidth]{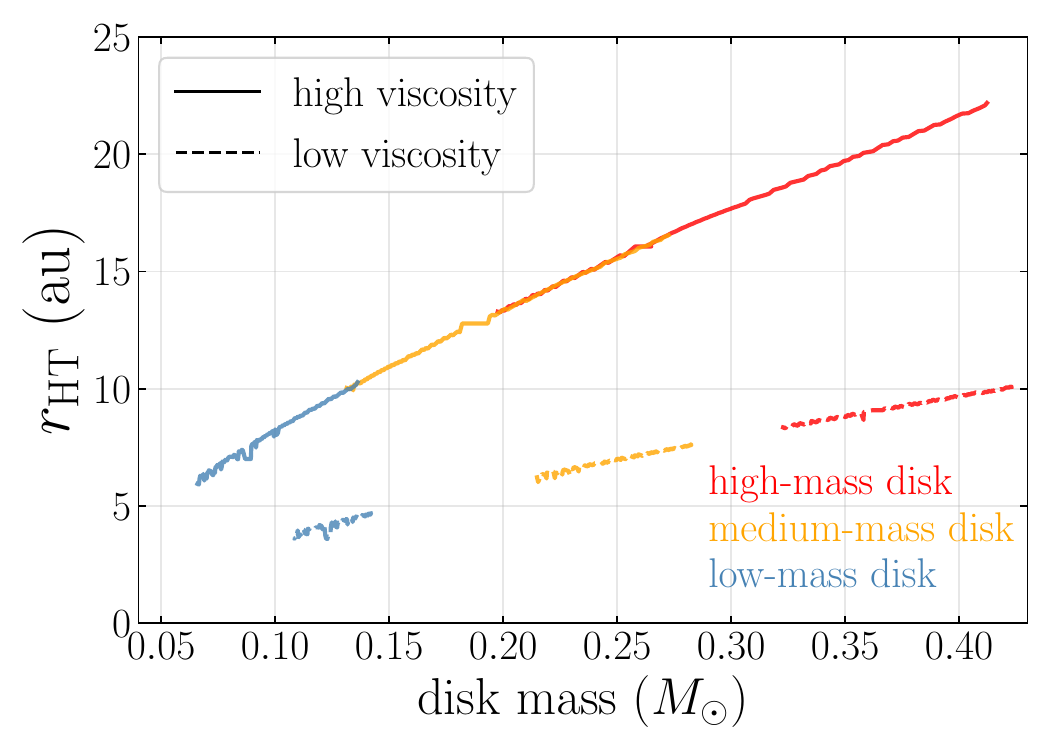}
	\caption{ \textbf{  \protect\hyperlink{hyp:fig:bifurcation-diskmass}{Bifurcation of migration histories.}} \textbf{Top three panels:} Planet tracks of all the planets that migrate outward in the low viscosity disks (red, orange and blue lines), and planet tracks of the planets initialized at the same locations in the high viscosity disks, which migrate inwards (grey lines). Black dots are their masses and orbital radii at $t=1.88$ Myr. \textbf{Bottom panel:} The radial evolution of $\rheat$, the formal location of the heat transition (vertical dashed black line in bottom panel of Fig. \ref{fig:3trap_definitions}), as a function of disk mass over time in each of our 6 planet formation scenarios.  }
    \label{fig:bifurcation-diskmass}
\end{figure}


Figure \ref{fig:single-madiagram-alldisks} also shows that there is some variation in the degree of outward migration at different disk masses. In \hypertarget{hyp:fig:bifurcation-diskmass}{Figure \ref{fig:bifurcation-diskmass}} we investigate this variation by overplotting the migration trajectories of planets initialized in the same radial locations but in the two different viscosity disks, for each disk mass. The degree of bifurcation in the migration histories between the two viscosities is most pronounced in the high-mass disk. As shown in the bottom panel of Figure \ref{fig:bifurcation-diskmass}, the radial location of the heat transition is farther outward in higher mass disks. Thus, the disk viscosity is the fundamental quantity responsible for the bifurcation, and the disk mass controls the degree or extent to which the migration outcomes are different. Low viscosity, massive disks make the best case for extensive outward migration.

We note that disk mass could affect the gas gap opening criterion that we have used \citep{Lin1993}, which is based solely on torque balance effects,  to predict when Type II migration will set in.  \citet{Malik2015} have shown that an additional constraint must be considered - namely - that the  gap the crossing time of a planet be longer than the gap opening time.  As our low mass planets don't get into the Neptune mass regime, this is unlikely to be significant here.  

This concludes our presentation of planet formation simulation results. \textbf{To summarize:} A lowered viscosity shifts the features of the torque map that determine planet migration ``downward'', to occur at lower planet masses. As such, the disk viscosity dictates the planetary mass range at which outward planet migration can occur. In our low viscosity disks, planetary embryos initialized near the heat transition reach this mass threshold and are propelled to twice their initial orbital radii -- an extended planetary system. The threshold is too high in our high viscosity disks, and inward migration takes hold, resulting in compact planetary systems, tempered only by planet trapping at the water iceline around $\sim 5$ au. The theory section to follow provides a rigorous theoretical perspective that explains these results and leads to simple scaling laws for the nature of cororation masses and planet formation in disks of different viscosities.


\section{Theoretical Results: Planet Formation at Large Disk Radii}
\label{sec:res:insights-planetmigration}

\begin{figure*}
	\includegraphics[width=18cm]{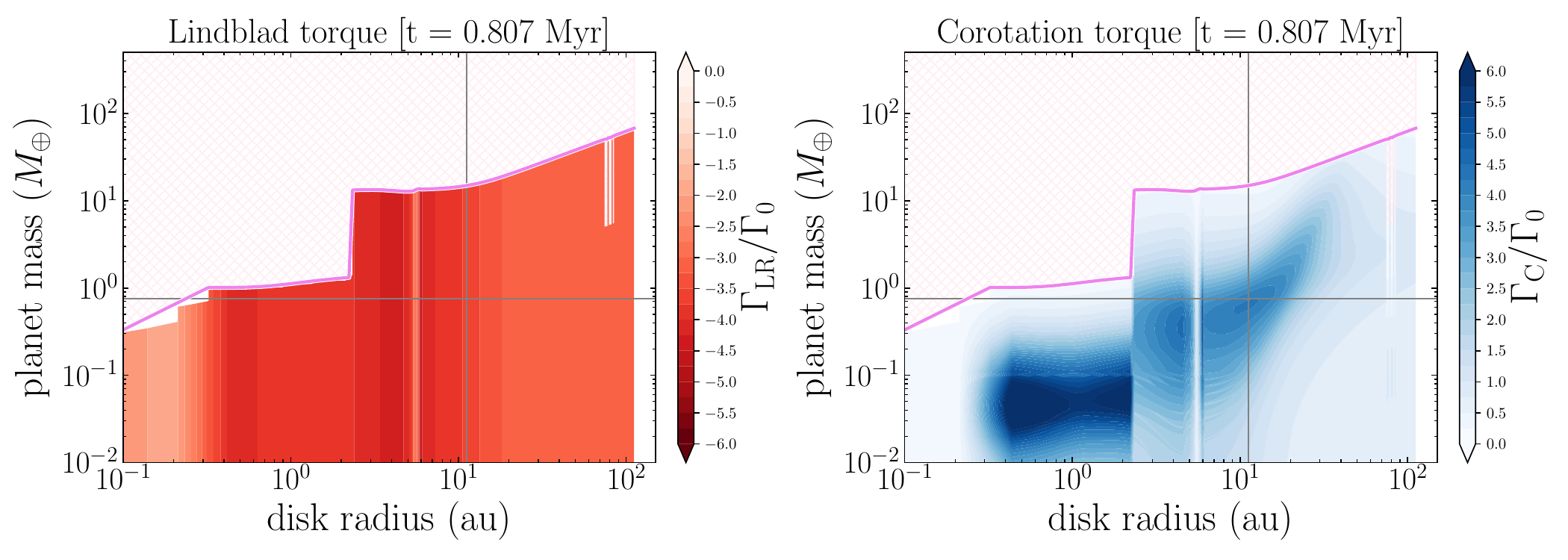}
	\includegraphics[width=17cm]{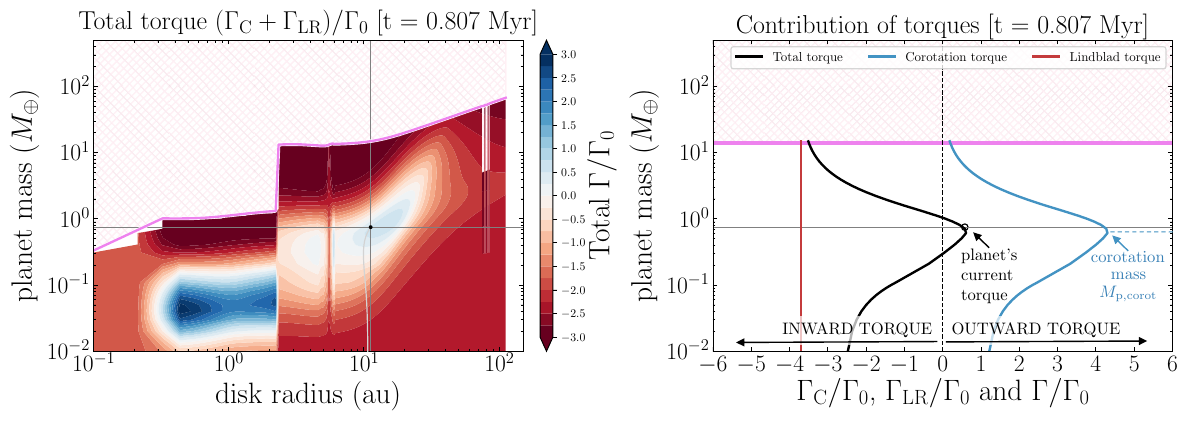}
    \caption{ \textbf{  \protect\hyperlink{hyp:fig:corot_lindblad_breakdown}{Decomposing the total torque and visualizing the planet mass of maximum outward torque.}} [Low viscosity, high-mass disk]. The total torque exerted by the disk on a planet is the sum of the outward-directed corotation torque  and inward-directed net Lindblad torque. To illustrate why planets in the low viscosity disks are propelled outward, we take a vertical slice through the Lindblad torque map (\textbf{top left panel}), the corotation torque map (\textbf{top right panel}), and total torque map (\textbf{bottom left panel}) at the planet's current orbital radius to yield three torque profiles as functions of planet mass. We plot these profiles in the \textbf{bottom right panel}. The reader familiar with \citet{paardekooper2011torque} will recognize them as rotated versions of their Figure 6. Where the total torque profile (black) intersects with the horizontal line (the planet's current mass) is the torque this planet experiences. (It is the same planet as in the right panel of Fig. \ref{fig:torqdisk_visualization}, with $a_{\rm p,0}=10.7$ au). The vertical dashed line at $\Gamma=0$ delineates negative inward torques from positive outward ones. \textbf{ \protect\href{https://youtu.be/Rs8F7fn_31I}{[Link to movie.]}}}
    \label{fig:corot_lindblad_breakdown}
\end{figure*}

The physics of planet-disk interaction by gravitational torques is subtle. The previous two sections have shown our numerical results: the torque maps and evolutionary tracks of planets in the full, non-linear context of planet-disk interactions. The purpose of this section is to pair our numerical results with the physical insight derived from analytical approximations of the full torque theory.

\subsection{Theoretical Background}
\label{subsec:insights-background}

Here we pick up on the thread we started in Section \ref{subsec:typeI-migration}.
The torque calculations and maps are based on the full set of equations originally from Eqns 50-53 in \cite{paardekooper2011torque}, and summarized below.

In \hypertarget{hyp:fig:corot_lindblad_breakdown}{Figure \ref{fig:corot_lindblad_breakdown}}, we show how these torques actually work using a clear visualization.  In particular, we take another look at the outward migration of a demonstrative planet in the low viscosity, high-mass disk. This is the same planet also presented in the right panel of Figure \ref{fig:torqdisk_visualization}. Early on in its formation, the planet interacts with the extended region of outward torque associated with the heat transition and is propelled from its initial $\apn=10.7$ au to $\ap=18.2$ au. We discuss each of the four panels of Figure \ref{fig:corot_lindblad_breakdown} in turn, starting at the bottom left and going clock-wise.

\textbf{Total Type I torque ($\Gamtot$).}  As described in \cite{paardekooper2011torque}, the total Type I torque exerted by a gaseous disk on an embedded planet is the sum of two physical processes: torques at Lindblad resonances \citep{goldreichtremaine1979, goldreichtremaine1980}, and corotation torques (co-orbital and horseshoe torques). For a planet with zero eccentricity and inclination, this simply means:
\begin{align}
	\label{eq:torq01}
    \Gamtot &= \GamLR + \GamC \, ,
\end{align}
where $\GamLR$ is the Lindblad torque, and $\GamC$ constitutes the corotation torques. In the top two panels of Figure \ref{fig:corot_lindblad_breakdown}, we decompose the total torque into its two competing components: the Lindblad torques (top left panel, red), and corotation torques (top right panel, blue).

\textbf{Lindblad torques ($\GamLR$).} Lindblad torques arise from waves at the locations of Lindblad resonances throughout the disk, both interior and exterior to the planet's orbit ($r_{\rm Lind} = [1 \pm 1/m]^{2/3} \, \ap$, where $m\geq2$ is an integer). The waves generate spiral arms that either carry angular momentum away from (outer wave) or deposit it onto (inner wave) the planet. The direction of the net Lindblad torque can therefore be inward or outward, depending on the interplay between the gradient of the column density and that of the disk temperature: 
$ \gamma_{\rm eff} \GamLR / \Gamma_o  =  (-2.5 - 1.7 \beta_{\rm T} + 0.1 \alpha_{\Sigma} )  $, where $(-\beta_{\rm T})$ is the power law index of the temperature on disk radius, and $ (- \alpha_{\Sigma}) $  the index for the column density. The effective adiabatic index of the gas is $\gamma_{\rm eff}$  \cite[see equations 46,47 in][]{paardekooper2011torque}.  Using the power law indices for the temperature and column density regimes summarized in our equations \ref{eqn:hlt-surface-density-profile} and \ref{eqn:hlt-temperature-profile}, we compute that $\gamma_{\rm eff} \GamLR / \Gamma_o = -3.79; - 2.85$ for the disk radii inside and outside the heat transition, respectively.  The negative values indicate inward directed torques.  In other words, the Lindblad torque is always directed inward (i.e. red in our torque maps) throughout our entire disk model and for all disk masses. The top left panel of Figure \ref{fig:corot_lindblad_breakdown} confirms this analytic result -- in our models the net Lindblad torque is always directed inwards. Outward planetary movement in our disks therefore depends entirely on the physics of the corotation torque.

\textbf{Corotation torques ($\GamC$).} Corotation torques result from gravitational perturbations to the gas close to the planet - inside its co-orbital and (closed) horseshoe region. The co-orbital component is linear and depends on the gradient of vortensity, $(\nabla \times \vec{v})/ \,\Sigma$. The horseshoe component is non-linear. Gas undergoing horseshoe orbits gains and loses angular momentum over the cycle, and if the entropy of the gas decreases with disk radius (non-adiabatically), the resulting azimuthal asymmetry in density causes outward planet migration \citep{paardekooper2010}. 

Specifically, the corotation torques are comprised of vorticity and entropy-related horseshoe drag torques ($\GamVHS$, $\GamEHS$) and linear vorticity and entropy-related corotation torques ($\GamLVCT$, $\GamLECT$) as follows:
\begin{align}
	\label{eq:GamC}
    \GamC &= \left[\, \GamVHS \Fpnu \Gpnu + \GamEHS \Fpnu \Fpchi \sqrt{ \Gpnu \Gpchi } \, \right]  \nonumber\\
    &+  \GamLVCT \left( 1 - \Kpnu \right) + \GamLECT \sqrt{ \left( 1 - \Kpnu \right) \left( 1 - \Kpchi \right) } \, 
\end{align}
These latter three functions $F$, $G$ and $K$, discussed further in Sec. \ref{subsec:insights-Mpnucorot-Mpchicorot}, are the amplitudes of the combined Lindblad and corotation torques that measure the saturation of the torques, and depend on a saturation parameter, to be discussed below.  Only the $F$ amplitude varies significantly over parameter ranges of interest, and its peak value will determine where the corotation torque hits a maximum.  

The top right panel of Figure \ref{fig:corot_lindblad_breakdown} shows that the corotation torques in our models are directed outward, and their strength depends on both planet mass and radius. Features in the corotation torque map give rise to features in the total torque map (ie.positions and shapes associated with the dead zone, ice line, and heat transition traps).

\textbf{Contribution of torques.} In the bottom right panel of Figure \ref{fig:corot_lindblad_breakdown}, we take a vertical slice in planet mass through each of the three torque maps ($\Gamtot$, $\GamLR$, $\GamC$) at the planet's current orbital radius. The result is three curves of torque amplitude as a function of planet mass, where planet mass is on the y-axis to match the torque map panels. The horizontal grey line indicates the planet's current mass, and where it intersects with each of the torque curves indicates the value of that torque the planet is currently experiencing. We provide a movie in Table \ref{tab:movies}.

This panel clearly shows that the Lindblad torque profile is constant in planet mass, and strongly negative (inward-directed). The corotation torque is positive (outward-directed); its dependence on planet mass dictates the mass dependence of the total torque profile, and makes outward planet migration possible. At the time snapshot shown, the planet is undergoing outward migration simply because it is at the right mass to reach peak corotation torque. 

In the derivations that follow, we refer to this planet mass of maximum outward torque ($\GamC$) as the \textbf{corotation mass}, or $\Mpcorot$. In other words, 
\begin{equation}
    \GamC (\Mpcorot, \,r) = \max{ \GamC (r) }\, .
\end{equation}
We label the corotation mass in the bottom right panel of Figure \ref{fig:corot_lindblad_breakdown}. Despite using quite a different underlying disk model, our corotation torque profiles strongly resemble those in Figure 6 of \citet{paardekooper2011torque}. Starting from the framework laid out in that same work, we derive an analytic recipe for the planet mass of maximum outward torque depending on the local properties of the disk.

\subsection{The viscous \& thermal corotation masses}
\label{subsec:insights-Mpnucorot-Mpchicorot}

The amplitude or magnitude of the corotation torque is prone to saturation over time. In the absence of replenishing processes, the torque modifies the angular momentum of gas in the horseshoe region (which is not connected to other orbits within the disk) and destroys the vortensity and entropy gradients that bring it to life. This occurs over the libration timescale, $\tlib$ \citep{paardekooperpap2009a}:
\begin{equation}
    \label{eq:tlib}
    \tlib = \frac{4}{3 \, \xs} \frac{2 \pi}{\Omegap}\, , 
\end{equation}
where $\xs$ is the dimensionless radial half width of the horseshoe band of orbits around the planet's orbital radius, $\rp$, and $\Omegap$ is the planet's orbital angular frequency. We note that angular momentum exchange can occur continuously at the Lindblad resonances, and that $\GamLR$ is not subject to saturation.

The width of the corotation region has been analyzed in great detail, by means of fits to 3D numerical simulations \citep{paardekooperpap2009b}. These authors used the \texttt{FARGO} code, and introduced a numerical softening parameter $b$ in computing gravitational potential of the planet. Their simulation results could be well matched using a value of $(b/h) = 0.4$, where $h=H/r$. They find that 
\begin{equation}
    \label{eq:xs}
	 \xs = C(b/h) \, \cdot \, \sqrt{\frac{q}{h}}\, , 
\end{equation}
where  $q = \Mp / \Ms$ is the ratio of the planet's mass to that of its host star. The coefficient $C$ is a function $C = C(b/h) \simeq 1 $ that can be written as a power law around the value $(b/h) = 0.4$ such that  $C = (1.1 / \gamma_{\rm eff}^{1/4})\, \cdot \, [0.4 / (b/h)]^{-1/4} $, where the effective adiabatic index of the gas is $\gamma_{\rm eff}$.

In order to circumvent saturation and sustain the corotation torque, the vortensity and entropy gradients within the horseshoe region need to be restored more quickly than $\tlib$. The amplitude of $\GamC$ thus implicitly depends on processes that transport angular momentum within the disk. For non-isothermal disks, such as the models we use here, the replenishing processes necessary to maintain the corotation torque are (1) thermal diffusion, $\chi$ (Eqn. \ref{eq:chi}) and (2) viscosity, $\nu$ \citep{paardekooper2008}.

The degree to which the corotation torque is saturated is described by saturation parameters: $\pchi$, associated with thermal diffusion, and $\pnu$, associated with viscosity. These parameters are the subscripts of the $F$, $G$ and $K$ functions that appear in the equation for the corotation torque $\GamC$ (Eqn. \ref{eq:GamC}), where for example $F_{\rm p}$ is shorthand for $F(\rm p)$. The numerical results of \cite{paardekooper2011torque} show that the $G$ and $K$ functions vary only slightly, and so of the four terms, the first two terms involving $\Fpnu$ and $\Fpchi$ (inside square brackets in Eqn. \ref{eq:GamC}) will contribute the most to the functional form or variation of $\GamC$. The form of these functions is reproduced in Figure \ref{fig:corot_lindblad_breakdown}, lower right panel, where the maxima in $F_{\rm p}$ are there shown as a function of planet mass (on the y-axis), for $\alpha=10^{-4}$. 

Thus, the basic point is that we take the amplitude of the corotation torque $\GamC$ to be mainly determined by:
\begin{equation}
    \label{eq:Fpnu-Fpchi}
    \Fpnu = F(\pnu) \, {\rm and}\, \Fpchi =  F(\pchi) \, ,
\end{equation}
each an identical function of two saturation parameters - one due to viscous diffusion, and one due to thermal diffusion. To find $\Mpcorot$, our task is to find the planet mass of peak $\Fpnu$ and $\Fpchi$. We do so by locating the value of $\pnu$ for which $\Fpnu$ takes its maximum \cite[relying on Fig. 6 of][]{paardekooper2011torque}, and then translating this $\pnu$ into a planet mass: $\Mpnucorot$. We repeat the process for $\pnu$ and $\Fpnu$ to find $\Mpchicorot$, and then combine the two into a net $\Mpcorot$ (Sec. \ref{subsec:insights-Mpcorot}).

We first focus on the effects of viscous diffusion, $F(\pnu)$. In ideal disk models, both the vortensity and entropy are conserved along stream lines of the fluid. Consider the case where the entropy decreases with radius. Fluid in orbits just beyond the planet's orbit are in colder gas, those inside the orbit have hotter gas. A fluid element just outside will make a U-turn as it approaches the planet, and goes on to enter the inner, slightly hotter region. In order that pressure balance be preserved, a density increase must occur in the colder fluid. Similarly, there is a density drop in the hotter region and this density difference results in a torque that pushes the planet outward. This density bump must be maintained, however, in the face of decoherence brought on by phase mixing.  In this process, because the horseshoe libration period is different for different orbits in the region, the density jump can be quickly smeared out and disappears. If the viscous time scale of the gas $\tnu$ is comparable to this libration time however, then the surface density and vortensity gradients can be maintained against these looses, and the torque remains unsaturated \cite[see review,][]{Nelson2018}.

The viscous diffusion time scale of gas across a region of width $\xs$ at the planet's orbital radius $\rp$ is
\begin{equation}
    \label{eq:tnu}
    \tnu = \frac{(\xs \, \rp)^2}{\nu} \, ,
\end{equation} 
where $\nu=\alpha \cs H$ is the viscosity of the disk. The associated saturation parameter has been computed numerically for non-isothermal disks and is expressed in \cite{paardekooper2011torque}:
\begin{equation}
    \label{eq:pnu}
    \pnu = \frac{2}{3} \sqrt{ \frac{\xs\, \tnu \, \Omegap}{2 \pi}} \, = \frac{2}{3} \sqrt{ \frac{\xs^3 \, \rp^2 \, \Omegap }{2 \pi \nu} } \, ,
\end{equation}
where the second equality comes from substituting $\tnu$ (Eqn. \ref{eq:tnu}).  The physical meaning of the viscous saturation parameter $\pnu$ is that it is a direct expression of the ratio of  $\tnu$ to $\tlib$, which is readily derived by using the expressions for $\tlib$ and Equation \ref{eq:tlib}
\begin{equation}
    \label{eq:pnu-ratio}
    \pnu = \frac{4}{3 \sqrt{3}} \, \sqrt{ \frac{\tnu}{\tlib}} \, ,
\end{equation}
and which is natural given that the libration timescale $\tlib$ needs to be compensated for in order that the corotation torque remain unsaturated.

Continuing from the second equality in Equation \ref{eq:pnu}, substituting $\nu=\alpha \cs \Hp$ and noting that for thin disks, $\Omegap / \cs = 1/\Hp $, we see that the saturation parameter depends explicitly upon $\alpha$ and $\hp = \Hp / \rp$ at the planet's orbital radius:  
\begin{equation}
    \label{eq:pnu-planet}
	 \pnu = \frac{2}{3 \sqrt{2\pi}} \, \alpha^{-1/2} \, \xs^{3/2} \, \hp^{-1} \, .
\end{equation}

Combining this with Equation \ref{eq:xs}, we obtain an expression for the saturation parameter $\pnu$ that depends on the properties of the full disk, namely the viscosity $\alpha$ and the aspect ratio $h$, as well as the planet's mass, as measured by the mass ratio $q$:
\begin{equation}
    \label{eq:pnu-fulldisk}
	\pnu = \pnuzero \cdot \frac{q^{3/4} }{ \alpha^{1/2} h^{7/4}}  \, ,
\end{equation}
where we define the coefficient $\pnuzero$ as 
\begin{equation} 
    \label{eq:pnu0}
	\pnuzero =  \frac{2 }{ 3 \sqrt{2\pi}} C^{3/2} = 0.266 \, ,
\end{equation}
the numerical value given corresponding to the typical case where $C \simeq 1$.

If viscous diffusion is the only diffusive mechanism putting angular momentum into the corotation region, then the maximum outward directed corotation torque will occur for a value of $\pnu$ where $\Fpnu = \Fmax$; ie. where $ \dd \Fpnu / \dd \pnu = 0$. We denote the value of $\pnu$ where $\Fpnu$ takes its maximum as $\pnucorot$. The numerical results of \cite{paardekooper2011torque} (Figure 6) show that 
\begin{equation}
    \label{eq:pnucorot}
	\pnucorot \simeq 0.35 \, .
\end{equation}

The physical insight offered by Equations \ref{eq:pnu-fulldisk} \& \ref{eq:pnucorot} is as follows. As seen in Equation \ref{eq:pnu-ratio}, the level of saturation described by $\pnu$ is a question of timescales. At small mass ratios $q$, when $\pnu \ll 1$, the corotation torque is in the weak (linear) regime, and migration will be dominated by the inward Lindblad torque. As the mass ratio $q$ grows such that the viscous timescale is \textit{comparable} to the libration timescale, the corresponding saturation parameter $\pnu$ is \textit{of order} unity (ie. $\pnu = \pnucorot \sim 1$). Here, the corotation torque is at maximum strength and we can have outward planet migration. With further mass $q$ increase, the planet moves into the non-linear, $\pnu \gg 1$ regime, wherein the corotation torque saturates, and again, the Lindblad torque pushes the planet inwards. Therefore, it is when the timescale of a replenishing (diffusive) process is comparable to the libration timescale that is crucial for outward planet migration. While we have just described this for the case where viscosity is the replenishing process, it applies broadly to other processes that transport heat and angular momentum (eg. thermal diffusion, see below; disk winds, see Sec. \ref{sec:discussion:diskwinds}).

Setting Equation \ref{eq:pnu-fulldisk} equal to $\pnucorot$ and rearranging for $q$, we may define a value of the mass ratio $\qnucorot$ for which a planet can undergo the strongest, outward directed corotation torque. We see that it depends upon the disk viscosity and gas aspect ratio as:
\begin{equation}
    \label{eq:qnucorot}
	 \qnucorot = \Big( \frac{\pnucorot }{ \pnuzero } \Big)^{4/3} \alpha^{2/3} h^{7/3} = 1.44\, \alpha^{2/3} h^{7/3} \, .
\end{equation}
Many disk models suppose typical conditions for these disk parameters as $\alpha = 10^{-3} $ and $h = 0.05$, for which we find $\qnucorot = 1.33 \times 10^{-5}$. For a solar mass star (note that $\Ms = 1.2 \, \Msun$ in our HL Tau models), this corresponds to a planetary mass of 
\begin{equation}
	\label{eqn:Mpnucorot}
	\boxed{ \Mpnucorot = 4.43\, \Big( \frac{\alpha }{ 10^{-3}} \Big)^{2/3} \, \Big( \frac{h }{ 0.05} \Big)^{7/3} \, \Me \, . }
\end{equation}
We refer to $\Mpnucorot$ as the \textbf{viscous corotation mass}.

\begin{figure*}
	\includegraphics[width=8cm]{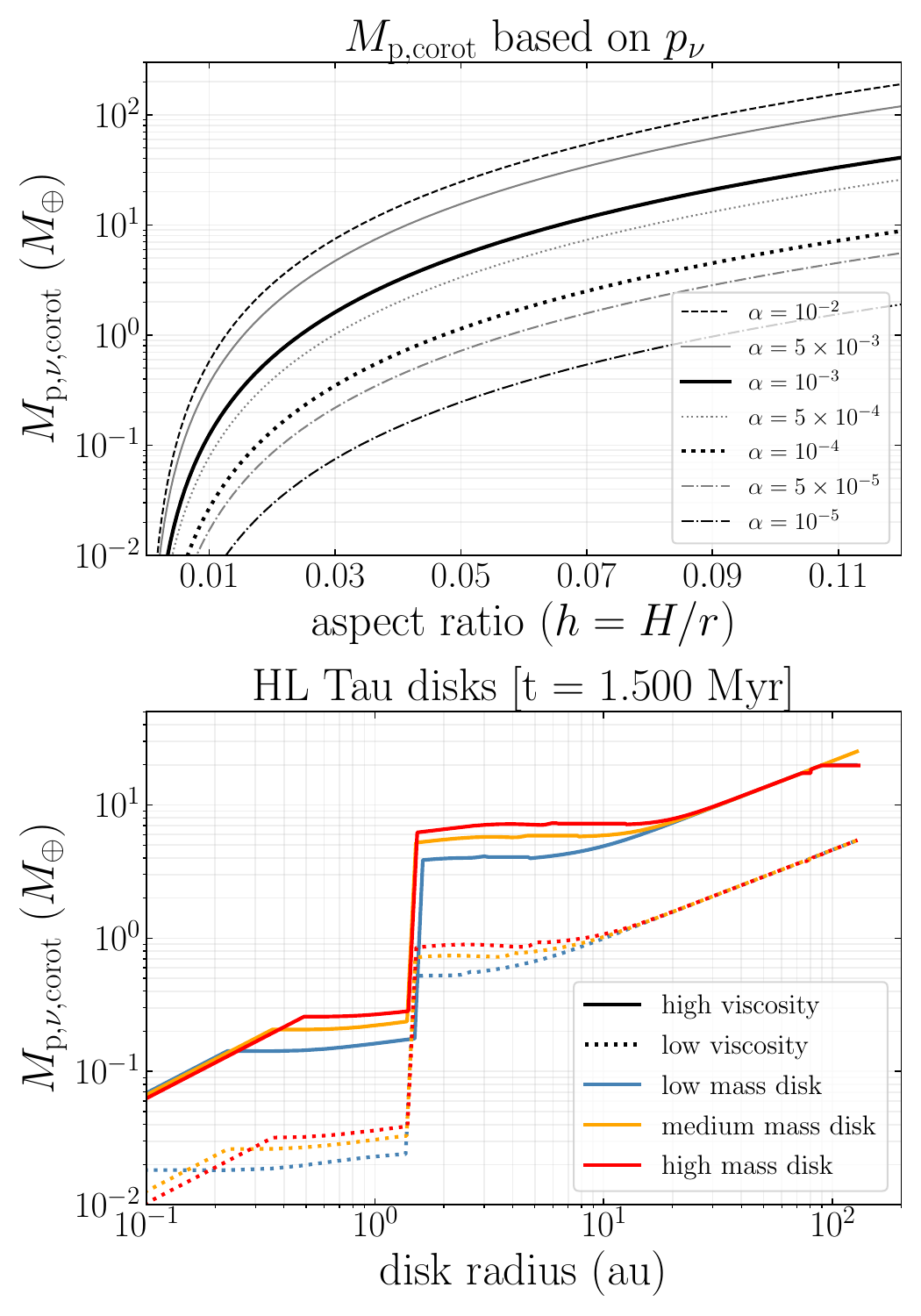}
	\includegraphics[width=8cm]{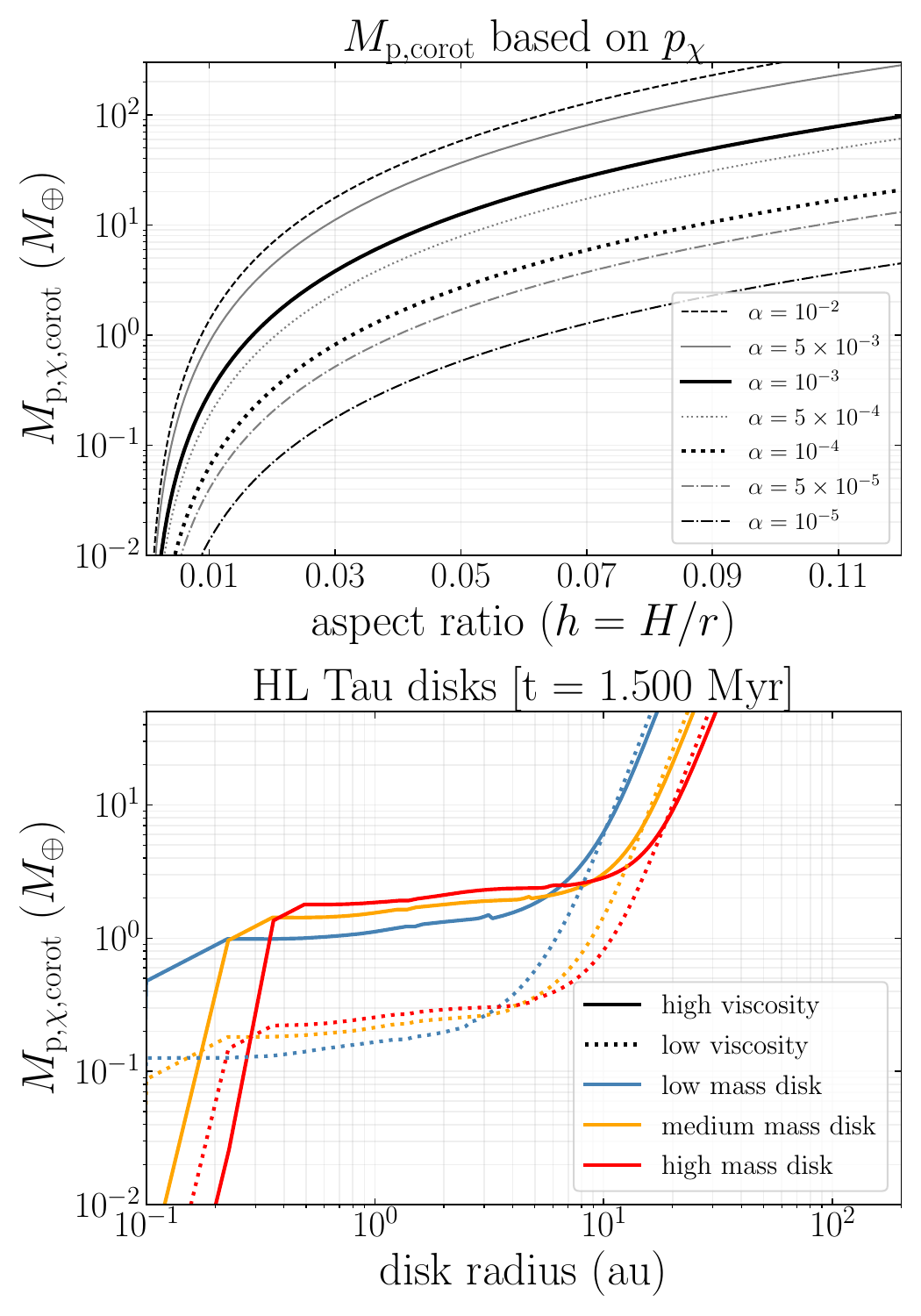}
    \caption{ \textbf{  \protect\hyperlink{hyp:fig:Mp_corot_curves}{Planet masses of maximum outward torque: viscous corotation mass ($\Mpnucorot$, Eqn. \ref{eqn:Mpnucorot}) \& thermal corotation mass ($\Mpchicorot$, Eqn. \ref{eq:Mpchicorot-units}).}} The form of our two analytic approximations for the planet mass of maximum outward torque derived purely from the viscous saturation parameter, $p_{\nu}$ (\textbf{left column}) and from the thermal saturation parameter, $p_{\chi}$ (\textbf{right column}). In the \textbf{top row}, we plot the curves as a function of the disk aspect ratio ($h=H/r$) for a range of $\alpha$ parameters ($\alpha=10^{-5}-10^{-2}$). The values of $\alpha$ explored in this work are the thick black solid and dotted lines. In the \textbf{bottom row}, we plot the curves as a function of disk radius through the radial dependence of $h$ and $\alpha$ contained in our HL Tau models (see top and middle rows of Fig. \ref{fig:model_alphas_h_sigma}), for all three model disk masses and two levels of viscosity. A lower $\alpha$ results in a lower planet mass of maximum outward torque, because both $\Mpnucorot$ and $\Mpchicorot$ scale as $\alpha^{2/3}$.}
    \label{fig:Mp_corot_curves}
\end{figure*}

With these ideas outlined, we now turn to thermal diffusion effects, $F(\pchi)$, to find the \textbf{thermal corotation mass}, $\Mpchicorot$. The physical discussion follows exactly the same course as the preceeding viscous diffusion arguments. The thermal diffusion saturation parameter $\pchi$ is related to that due to viscosity by 
\begin{equation}
    \label{eq:pchi}
    \pchi = \pnu \cdot \frac{3 }{ 2}\, \sqrt{\frac{\nu }{ \chi}} \, ,
\end{equation}
where $\chi$ is the thermal diffusivity:
\begin{equation}
    \label{eq:chi}
    \chi = \frac{4 \gamma (\gamma -1) \, \sigma T^4}{3 \kappa \rho^2 H^2 \Omega^2} \, .
\end{equation}
Here, $\gamma$ is the adiabatic exponent, $\sigma$ is the Stefan-Boltzmann constant and $\kappa$ is the opacity.

In fluid mechanics, the ratio of these two diffusivities is known as the Prandtl number: 
\begin{equation}
    \label{eq:prandtl}
    \Pr = \frac{ \nu }{ \chi }\, ,
\end{equation}
It describes the relative importance of viscous versus thermal diffusion. The Prandtl number plays an important role in the corotation torque. For reference, we provide \hypertarget{hyp:fig:app:prandtl_vs_radius}{Figure \ref{fig:app:prandtl_vs_radius}} in Appendix \ref{app:insights-planetmigration}, which shows how the thermal diffusivity $\chi=\chi(r)$, the level of viscosity $\nu=\nu(r)$ and therefore the Prandtl number $\Pr=\Pr(r)$ varies across disk radius in our models.

We again refer to the numerical results of \citet{paardekooper2011torque} (their Fig. 6) for the value of the saturation parameter at which the corotation torque amplitude takes its maximum. While they do not provide the pure function $F(\pchi)=\Fpchi$, they do show the product $\Fpchi \cdot \Fpnu$. As this is a product of near-Gaussians, the amplitude peak is hardly affected, and like $\pnucorot$, 
\begin{equation}
    \label{eq:pchicorot}
    \pchicorot \simeq 0.35 \, .
\end{equation}

From here, we substitute ${\rm Pr}$ (Eqn. \ref{eq:prandtl}) and $\pnu$ (Eqn. \ref{eq:pnu-fulldisk}) into Equation \ref{eq:pchi}:
\begin{equation}
    \pchi = \frac{3 }{ 2} \, \pnuzero \, \sqrt{\Pr}\,  \, q^{3/4} \, \alpha^{-1/2} \, h^{-7/4}\, .
\end{equation}
Setting $\pchi=\pchicorot$ so that $q=\qchicorot$, re-arranging for $\qchicorot$, noting that $\pnucorot=\pchicorot$ and absorbing terms in common to Equation \ref{eq:qnucorot} into $\qnucorot$, we find the thermal corotation mass ratio:
\begin{equation}
    \label{eq:qchicorot}
    \qchicorot = \Big( \frac{2 }{ 3} \Big)^{4/3} \,  \Big( {\Pr} \Big)^{-2/3}\,  \qnucorot \, .
\end{equation}
Equivalently, in terms of the viscous corotation mass, we find the thermal corotation mass:
\begin{equation}
    \label{eq:Mpchicorot}
    \Mpchicorot = \Big( \frac{2 }{ 3} \Big)^{4/3} \,  \Big({\Pr}\Big)^{-2/3}\,  \Mpnucorot \, .
\end{equation}
Putting this in physical units (again $\Ms = \Msun$), we have, finally:
\begin{equation}
    \label{eq:Mpchicorot-units}
    \boxed{ \Mpchicorot = 2.58 \,  \Big( {\Pr} \Big)^{-2/3}  \Big( \frac{\alpha }{ 10^{-3} } \Big)^{2/3} \, \Big( \frac{h }{ 0.05} \Big)^{7/3}\, \Me \, . }
\end{equation}
Comparing this to Equation \ref{eqn:Mpnucorot}, we see that the thermal corotation mass differs from the viscous corotation mass by a constant and a Prantl number prefactor, $\Pr^{-2/3}$.

In \hypertarget{hyp:fig:Mp_corot_curves}{Figure \ref{fig:Mp_corot_curves}}, we compare the behaviour of our two corotation masses -- $\Mpcorot$ based on $\pnu$ in the left column (the viscous corotation mass, $\Mpnucorot$, Eqn. \ref{eqn:Mpnucorot}) and $\Mpcorot$ based on $\pchi$ (the thermal corotation mass, $\Mpchicorot$, Eqn. \ref{eq:Mpchicorot-units}) in the right column. A planet of mass equal to either corotation mass will experience the maximum possible magnitude of corotation (outward) torque under the competing saturating influence of either viscous or thermal diffusion. The top two rows show $\Mpnucorot$ and $\Mpchicorot$ as a function of disk aspect ratio $h$, for a range of disk viscosities: $\alpha=10^{-5} - 10^{-2}$. Note that in the right panel, $\Mpchicorot$ was calculated assuming $\Pr = 1$. The bottom panels in Figure \ref{fig:Mp_corot_curves} show the behaviour of the two corotation masses in the context of our models at a certain time snapshot ($t=1.5$ Myr,  also chosen for Fig. \ref{fig:prandtl_fits}). The variation of the disk aspect ratio is folded into the radial dependence as $h=h(r)$ (see middle panel of Fig. \ref{fig:model_alphas_h_sigma}). The effect of lowering the disk viscosity by a factor of 10 is clear from comparing the solid ($\alpha=10^{-3}$) and dashed ($\alpha=10^{-4}$) lines. Both corotation masses scale as $\Mpcorot \propto \alpha^{2/3}$, and so the planet mass of maximum outward torque is lower for lower disk viscosities.

\subsection{The net corotation mass}
\label{subsec:insights-Mpcorot}

\begin{figure*}
	\includegraphics[width=18cm]{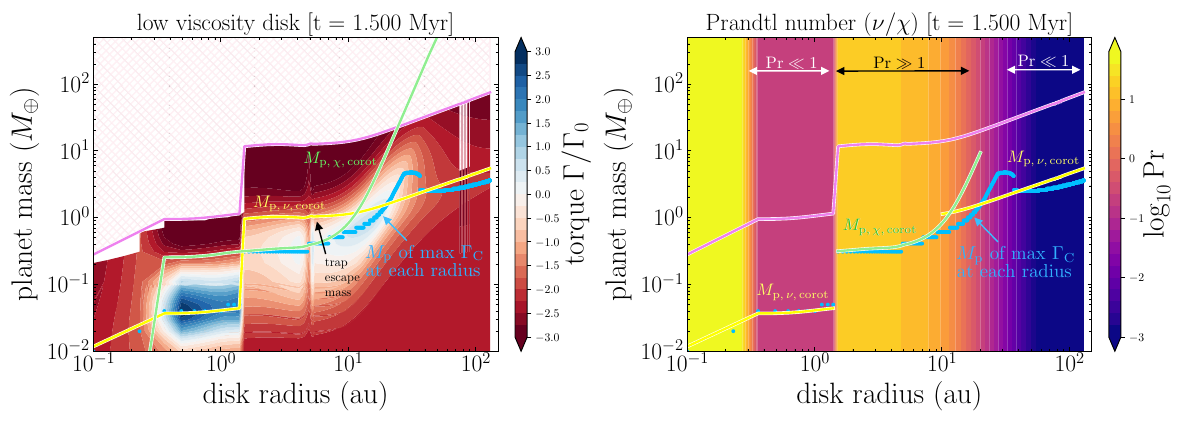}
    \caption{ \textbf{ \protect\hyperlink{hyp:fig:prandtl_fits}{The planet mass of maximum outward torque in the limits of $Pr \ll 1$ and $Pr \gg 1$.}} [Low viscosity, high-mass disk]. Here, we illustrate how well our two analytic approximations for the planet mass of maximum outward torque agree with the numerically calculated values from our torque maps. \textbf{Left panel:} Atop the total torque map of the high-mass, low viscosity disk, we show the numerically calculated planet mass of maximum outward torque at each radius (blue dots), the viscous $\Mpnucorot$ based on $\pnu$ (yellow line), and the thermal $\Mpchicorot$ based on $\pchi$ (light green line). \textbf{Right panel:} Same as the left panel, but overlaid on a map of Prandtl number (${\rm Pr}={\rm Pr}(r)=\nu / \chi$, see right panel of Fig. \ref{fig:app:prandtl_vs_radius}). The $M_{\rm p,\, \nu\, corot}$ curve predicts what planet masses are required for outward migration in regions where $Pr \ll 1$, and $M_{\rm p,\, \chi\, corot}$ does the same where $Pr \gg 1$. When each curve is \textit{not} describing $M_{\rm p}$ of maximum $\Gamma_{\rm C}$, it is delineating the boundary to inward migration (eg. $\Mpnucorot$ describes the water iceline escape mass). See left panel of Fig. \ref{fig:app:prandtl_vs_radius} for the same curves overlaid on purely the corotation torque map.}
    \label{fig:prandtl_fits}
\end{figure*}

\textit{Which corotation mass applies to which regions of the disk?} The equivalent question is: What is the relative importance of viscous versus thermal diffusion in different regions in the disk? This is the utility of the Prandtl number (Eqn. \ref{eq:prandtl}). 

In the inner region defined by the dead zone, the viscosity is very low ($\alpha=10^{-5}$ or $10^{-6}$). The heat generated in that region is therefore carried out by thermal diffusion, so in the dead zone $r \lesssim 1$ au we expect $\chi \gg \nu$ and $\Pr \ll 1$. In the outer parts of the disk where radiative heating dominates viscous heating, here too thermal diffusion must be important to maintain the necessary cooling of the disk, so for $r \gtrsim \rheat$ we anticipate that $\Pr \ll 1$.

At intermediate disk radii, (the region just beyond the dead zone and out to where radiative heating begins to dominate), viscous heating and thermal diffusion can be more comparable. In the models of \cite{paardekooper2011torque}, $\Pr$ does not exceed unity. In our models however, we find that the Prandtl number reaches high values $\Pr \gg 1$ (see Fig. \ref{fig:app:prandtl_vs_radius}; $\max \Pr \simeq 20$). We discuss reasons for this difference in Appendix \ref{app:insights-planetmigration}. Nonetheless, we can expect that at intermediate radii our planet formation results arise from the corotation mass that depends on the thermal diffusivity rather than the viscosity. 

In the variational analysis to follow, we will see that these physical ideas are indeed born out by the theoretical analysis and the numerical results. When $ \Pr \ll 1$, the applicable corotation mass is the viscous corotation mass $\Mpnucorot$, while in the $\Pr \gg 1$ regime, the important corotation mass is that derived from thermal diffusion dependent saturation, $\Mpchicorot$. Note that while the Prandtl number does inform us of the relative importance of $\chi$ and $\nu$, this does not translate into a description of the relative importance of the corresponding saturation parameters (ie. $\Pr \neq \pnu / \pchi$).

As discussed above Equation \ref{eq:Fpnu-Fpchi}, based on the numerical results of \cite{paardekooper2011torque} we consider that the variation of the corotation torque $\GamC$ is dominated by the two terms involving the $\Fpnu$ and $\Fpchi$:
\begin{equation}
    \label{eq:varGamC}
    \GamC \simeq \GamVHS \Fpnu \Gpnu \, + \, \GamEHS \Fpnu \Fpchi \sqrt{ \Gpnu \Gpchi } \, ,
\end{equation}
which we can re-write as
\begin{equation}
    \GamC  \simeq  [ A + B \, \Fpchi ] \, \Fpnu  
\end{equation}
to indicate the slowly varying multipliers $A$ and $B$.

Finding the peak corotation mass is equivalent to finding where $\delta \GamC = 0$: 
\begin{equation}
    \delta \GamC  \simeq \delta ( [ A + B \, \Fpchi ] \, \Fpnu ) = 0 \, .
\end{equation}
Carrying out the variation, we have
\begin{equation}
    \label{eq:var-terms}
    [A + B \Fpchi ]\frac{ d \Fpnu }{ \dd \pnu} \delta \pnu  + B\, \Fpnu \frac{\dd \Fpchi }{ \dd \pchi } \delta \pchi = 0
\end{equation}
Recalling the relation between $\pnu$ and $\pchi$ (Eqn. \ref{eq:pchi}), we see that $\delta \pchi \propto {\Pr}^{1/2} \, \delta \pnu $. Thus, the 2nd term (depending on $\delta \pchi$) in the above equation scales as ${\Pr}^{1/2}$ with respect to the first term. In the limit of small Prandtl numbers, as $ {\Pr} \ll 1$, the 1st term dominates and Equation \ref{eq:var-terms} is satisfied if
\begin{equation}
   \frac{ \dd \Fpnu }{ \dd \pnu } = 0 \, .
\end{equation}
In other words, the amplitude of the corotation torque $\GamC$ is dictated by the saturation influence of viscosity $\pnu$, and the planet mass of maximum outward torque will be given by $\Mpnucorot$.  

In the limit of large Prandtl numbers $ {\Pr} \gg 1$, the 2nd term dominates Equation \ref{eq:var-terms} and we require 
\begin{equation}
   \frac{ \dd \Fpchi}{ \dd \pchi } = 0 \, .
\end{equation}
Here, the amplitude of the corotation torque $\GamC$ is dictated by the saturation influence of thermal diffusion $\pchi$, and the planet mass of maximum outward torque will be given by $\Mpchicorot$.

With these results in hand, can describe the corotation mass function throughout the whole disk.  At each disk radius, we first check to find the value of the Prandtl number there. The net corotation mass is then defined by an approximating piece-wise formula:
\begin{equation}
    \label{eq:piecewise-Mpcorot}
    \boxed{\Mpcorot = 
         \begin{cases}  
               \Mpnucorot \, , & \Pr \ll 1 \\
               \Mpchicorot \, , & \Pr \gg 1
         \end{cases}}
\end{equation} 
where $\Mpnucorot$ is given by Equation \ref{eqn:Mpnucorot} and $\Mpchicorot$ by Equation \ref{eq:Mpchicorot-units}. We note that, if we had included additional processes in our models that transport heat or angular momentum (eg. disk winds, see Sec. \ref{sec:discussion:diskwinds}), we would need to incorporate them into the processes of finding the net $\Mpcorot$ too.

In \hypertarget{hyp:fig:prandtl_fits}{Figure \ref{fig:prandtl_fits}} we assess the performance of our analytical approximations. In the left panels, we plot $\Mpnucorot$ and $\Mpchicorot$ overlaid on maps of the total torque. The small blue dots show the numerical values of $\Mpcorot$ at each radius, computed directly from the torque map. In the right panels, we plot the piece-wise definition of $\Mpcorot$ (Eqn. \ref{eq:piecewise-Mpcorot}) overlaid on a map of Prandtl number to highlight the regimes wherein each corotation mass applies (note $\Pr$ is independent of planet mass -- we tile the $\Pr$ profile for visual purposes). 

As expected, $\Mpnucorot$ matches in the $\Pr \ll 1$ regime (inside the dead zone and outside the heat transition), and $\Mpchicorot$ matches in the $\Pr \gg 1$ regime (intermediate disk radii). One can interpolate between these two regimes for the case that $ \Pr \simeq 1$. 

Finally, we note an additional property of the viscous corotation mass $\Mpnucorot$: at intermediate disk radii (outside the dead zone but inside the heat transition), $\Mpnucorot$ matches the mass at which planets trapped at the water iceline can escape. As an explanation for this, note in Figure \ref{fig:prandtl_fits} that the inward directed Linblad torque at the radial position of the ice line has a very sharp onset (ie. dark red region) just above this mass scale.  Below it, the net torque is near zero - and above, the planet must be pushed rapidly inwards away from the iceline.  Following this logic, we see that at larger radii, $\Mpchicorot$ demarcates a similarly sharp transition above the heat tranistion trap region, which we predict will also behave as the escape mass condition.  These results inform us that planets, upon increasing their mass sufficiently to escape their traps, will quickly reverse course and be driven in towards small disk radii by the Linblad torques that await them - assuming masses are all in the Type I regime. If planets achieve enough mass to open a gas gap (above the pink line), slow, inward directed Type II migration will ensue.

\subsection{Extension: Corotation mass for MHD disk wind driven accretion}
\label{sec:discussion:diskwinds}

Our analysis so far has not addressed the effect of disk winds on planetary migration and formation, but it is expected to be significant. Work by \cite{Hasegawa2017} shows that disk winds may be important in understanding both the high accretion rates as well as low levels of turbulence (and hence rapid dust settling) in the HL Tau system \citep{ueda2021, Rich2021}. 

In general, any mechanism that transports angular momentum through or out of the disk will have an effect on the corotation torque, and hence on how planets will migrate.  The analysis always comes down to a basic comparison of two timescales, namely,  the timescale for replenishing the angular momentum in the corotation region by whatever is the dominant mechanism for angular momentum transport and the libration timescale which smears out the fluctuation and eliminates the corotation torque. Disk winds result in inward mass advection (inflow)  as the disk's angular momentum is removed by the wind.  This is a laminar not turbulent flow, so the action is not viscous in nature. Disk winds set up a radial inflow speed $v_{\rm r}$ whose  magnitude is related to the strength of the outflow.  In this situation the replenishment of angular momentum in the corotation region occurs on a timescale
\begin{equation}
t_{\rm w} = \frac{\xs \, \rp \, }{  v_{\rm r} }
\end{equation} 
The radial inflow speed driven by a disk wind can be written as 
\begin{equation}
v_{\rm r} = -2 \, r \, \frac{\dot \Sigma_{\rm w} }{ \Sigma } (\lambda - 1)   
\end{equation} 
where $\dot \Sigma_{\rm w}$ is the mass outflow rate per unit area from the disk surface.  The high efficiencies of MHD disk winds arise from their long "lever arms" which is measured by $\lambda = (r_{\rm A}(r_{\rm o}) / r_{\rm o})^2$ where $r_{\rm A}$ is the Alfven radius on the field line out to which the gas is accelerated, from the footpoint of the field line at a radius $r_{\rm o}$ on the disk. In general this is a slowly varying function of $r_o$ \citep{Pelletier1992}.  For self similar disks as modeled by  \citet{Blandford1982},  $\lambda = {\rm const}$.  

\cite{kimmig2020winds} have simulated planetary migration in 2D disks in which angular momentum transport is mediated by disk winds.   These were not explicitly MHD simulations, rather, the wind torque prescription was described analytically. Lindblad torque effects were not included.  They discovered that both inward and outward migration of planets could take place depending upon a parameter $K$ that is the ratio of the time for disk gas to advect through the corotation region under the action of the disk wind, $t_{\rm w}$, to the libration timescale: 
\begin{equation}
    K = \frac{t_{\rm w} }{ t_{\rm lib}}
\end{equation}
In their analysis, the radial advection speed is $v_{\rm r}= (\lambda - 1)(b/\pi) \Omega \, \rp $, where 
\begin{equation}
\label{eqn:disk-winds-b}
 b = \frac{\dot \Sigma_{\rm w} }{ \Sigma} \, \frac{2 \pi }{ \Omega}   
\end{equation} becomes a parameter that measures the wind mass loss rate from the disk.  We will call $K$ the ``wind saturation parameter,'' in accord with the definitions of the viscous and thermal saturation parameters $\pnu$ and $\pchi$ (Sec. \ref{subsec:insights-Mpnucorot-Mpchicorot}). From the results above one derives \citep{kimmig2020winds}:
\begin{equation}
    \label{eqn:kimmig_found}
    K = \frac{3 }{ 4}\frac{q }{ h} \frac{1 }{ ({\lambda -1)} b} \, .
\end{equation}

More detailed physics is required to actually compute the mass loss rate of the wind, but theoretical results and numerical simulations, now supported by observations \citep{Watson2016}, show that 
\begin{equation}
   \frac{\dot M_{\rm w} }{ \dot M_{\rm a}} = \lambda^{-1} \simeq 0.1 
\end{equation} 
where $\dot M_{\rm a}$ is the accretion rate driven through the disk. Thus, in what follows, we take $\lambda \simeq 10$ for the theoretically derived and observationally verified value.

In order to estimate likely values of the wind mass loss rate for observed jets and disk winds, note that with the total mass loss rate from the disk scaling as $\dot M_{\rm w} = 2 \pi r^2 \Sigma_{\rm w}$, and $\dot M_{\rm a} = 2 \pi \Sigma v_{\rm r} r$, and assuming thin disks where $h=\cs/\Omega r$, then  Eqn. \ref{eqn:disk-winds-b} can be written as 
\begin{equation}
\label{eqn:disk-winds-b-again}
    b = \frac{\dot M_{\rm w} }{ \dot M_{\rm a}} \frac{v_{\rm r} }{ \cs} h \, =\,  0.5 \times 10^{-5} \, \Big( \frac{ \dot M_{\rm w} / \dot M_{\rm a} }{ 10^{-1}  }\Big) \, \Big( \frac{v_{\rm r}/\cs }{ 10^{-3}} \Big) \, \Big( \frac{h }{ {0.05}} \Big)
\end{equation}
where we have normalized the inflow rate at an effective rate of $ v_{\rm r}/ \cs \simeq 10^{-3}$. 

The simulation results of \citet{kimmig2020winds} show that outward planetary migration occurs for values $K \simeq 10$. This is the hallmark of an unsaturated corotation torque, as the authors note, and the parameter $K$ plays exactly the same kind of role as the saturation parameters $\pnu$ and $\pchi$ that we have already discussed (Sec. \ref{subsec:insights-Mpnucorot-Mpchicorot}). 

From this result, we can define an analogous corotation mass when wind driving is the dominant angular momentum transport mechanism in the disk. If the maximum outward corotation torque occurs at $\Kdwcorot = 10$, then we can solve the corotation mass scale of a planet under the action of disk winds, $\Mpdwcorot$, by rearranging Eqn. \ref{eqn:kimmig_found}, giving:
\begin{equation}
    \qdwcorot =  \frac{40 }{ 3} \, \frac{(K / \Kdwcorot) }{ 10}\,  (\lambda - 1) \, b \, h  \, .
\end{equation}

Thus we find the intriguing result that for a fiducial disk wind, the disk wind cororation mass ratio is comparable to an Earth mass: $ \qdwcorot = 3 \times 10^{-5} \, (b / 0.5 \times 10^{-5}) \, (h / 0.05)$.  Writing out the mass scale for a planet migrating due to a disk wind around a solar mass star:
\begin{equation}
    \label{eq:Mpdwcorot}
    \boxed{ \Mpdwcorot = 1.13 \, \Big( \frac{(K / \Kdwcorot) }{ 10} \Big) \, \Big(  \frac{b }{ 0.5 \times 10^{-5}}\Big) \, \Big( \frac{h }{ 0.05} \Big) \, \Me }\, 
\end{equation}
where we have taken $\lambda = 10$.

This result is in accord with the \cite{kimmig2020winds} simulations.  They found that to drive a Jupiter mass outward, they needed a value of $b = 10^{-2}$.  Using the longer lever arm from wind observations, we see that the corotation mass corresponding to Jupiter is $b = 1.4 \times 10^{-3}$.  However, examining the estimate for the value of $b$ given in Eqn,  \ref{eqn:disk-winds-b-again}, this would require an unrealistically high mass loss rate from the disk, of roughly $\dot M_{\rm w} / \dot M_{\rm a} \simeq 10$ which is far too high for winds observed in disk systems \cite{Watson2016}.  Another possibility is that the radial inflow speed becomes much larger, and would have to reach $\alpha_{\rm w} \simeq 10^{-1}$.  This would imply accretion rates through the disk that are at least an order of magnitude larger than the observations would suggest. 

There are still only very few works that address the full physics of MHD disk winds in low viscosity regimes. Recently \cite{mcnally2020} have completed fully 3D, non-ideal simulations of corotation torques and their effects on planetary motion.  An important new physical process arises for magnetized gas in 3D, namely magnetic buoyancy. These 3D simulations investigate magnetic buoyancy effects in the corotation region.  While in 2D outward motion is indeed driven by MHD torques \citep{mcnally2017}, in 3D magnetic buoyancy alters local vortensity gradients. This leads to a dominant {\it negative} corotation torque; ie, low mass planets move even more rapidly inwards than without MHD effects.  Arguably the main caveat to this as indicated by the authors, is that heating of the corotation region by the accreting planet alters the temperature gradients in such a way as to strongly reduce the buoyancy of the gas.  This remains to be investigated in future 3D MHD studies. 

We note that disk winds could contribute to early outward, and then later inward directed torques.  Here's how.  As already discussed, jet and disk observations clearly show that wind loss and accretion rates are strongly coupled. At the earliest phases of disk formation, the disk accretion rate are the highest, and decrease with time. The wind mass rates will follow this trend. This implies that disk winds early in the life of a disk have the greatest potential to drive forming planets outwards.   Later, as the wind levels drop below thresholds derived above, the wind torque effects will reverse and push the planets inwards.  This mimics the behaviour we have outlined for the low viscosity case. Disk winds will have a more prevalant role in low viscosity disks, so we expect winds to reinforce the migration behaviour due to pure viscous disks.    


\section{Discussion}
\label{sec:discussion}

Our simulations and calculations lead to the important new insight that forming planetary systems can either be concentrated in the inner regions of the disk or dispersed to large disk radii before moving inwards again.  This bifurcation of planetary evolution depends on the effective viscosity of the host protoplanetary disk which dictates the corotation mass (ie. the threshold planet mass for outward migration beyond the heat transition).  We have shown that in low $\alpha$ disks, the corotation mass scale is markedly lower than in the high viscosity case. Forming planets in such disks are caught up in regions of outward co-rotation torque and pushed outwards.  These eventually move slowly inwards again as the disk viscosity diminishes and the heat transition trap moves slowly inwards dragging the planetary cores with them. 

A direct consequence of this finding is that the observational disk demographics in \citet{vandermarel2021-demographics} can be explained as a manifestation of the underlying distribution of turbulent viscosities across the whole protostellar disk population. As a natural consequence of our results, we expect that observed rings and gaps arise in those disks with low viscosity. We argue, below, that both observations of turbulent intensities in disks, and the broad range of disk masses imply a range of turbulent viscosities across disk populations.   

As for the masses attained by planets at these larger radii: our accretion models adopt a conservative picture of mass  accretion.  These end up forming planets in the 1-2 Earth mass range.  This is likely significantly below the mass needed for planets to form dust gaps \citep{rosotti2016minimum}.  We saw, however, that with more mass accretion planets have potential trajectories in the torque maps to take them to larger disk radii.  These planets are being starved of mass in this conventional mass accretion picture.  We discuss both this and the point above in greater detail, below.

The corotation mass idea allows us a clear picture of how this bifurcation mechanism works.   The corotation mass scale in disks with higher viscosities ($\alpha=10^{-3}$), is high enough that forming planets are caught up by the inward directed Lindblad torque and migrate inwards.  Their rapid type I inward is arrested by either ice lines, or dead zone traps which lie in the inner regions (less than 10 au) of the disk.   These traps occur at inner disk regions, not readily resolved by ALMA.  The outer regions of such disks are rapidly depleted in cores that could grow into more massive super-Earths or Jupiters and planet formation occurs in the inner disk regions. So compact dust disks and initial planetary configurations are predicted by our model.

In low viscosity disks ($\alpha=10^{-4}$ ) on the other hand, growing planetary embryos beyond 10 au are intercepted by the corotation torques whose corotation mass scales are much lower. In this situation, they are quickly pushed out to twice their starting radius, to a maximum of 20 au (but with possibilities to much further out, as indicated).  For our particular choice of accretion model, these planets are unlikely sufficient in mass to be able to open a dust gap (see below).  This result agrees with other works that the standard models of planetesimal growth don't provide planets with sufficient mass in the outer regions of disks \citep{johansen-bitch-2019-planetesimal}.  

Another important aspect of our model is that planets pushed out to large disk radii will gradually return to the inner regions of the disk.  This is also due to the nature of the corotation torque.  As the mass of the planet grows, this torque saturates.  If a planet becomes massive enough to decouple from the heat transition trap, it will migrate inward, pushed along by the Lindblad torque, until it becomes sufficiently massive to open a gap in the gas and slow down into the Type II migration mode.  In many cases, the planet can remain trapped in the extended heat transition trap, in which case it gradually moves inward as disk viscosity clears out the disk over a few million years. Ultimately, the disk lifetime is determined by the final rapid gas loss due to photoevaporation \citep{Ercolano2017}, which is also contained in our models.  In disks with short lifetimes,  photoevaporation will starve forming Jupiters of gas, leaving a dominant collection of SuperEarths \citep{hasegawa2013planetary, cridland2016composition, hasegawa2016super, alessi2020formation}.  In either case, planets that have migrated out to large disk radii will return to the inner disk regions while the disk is still present.  This is in excellent agreement with the picture painted by \citet{vandermarel2021-demographics} which hinges on the assumption that ring/gap-creating planets will eventually migrate inward again.  On another note, outward migration of Super-Jupiter planets has also been reported in simulations of low viscosity disks \citep{Dempsey2021}. 

Let us return to the question of the distribution of turbulent intensities in disks. The data are still scarce, but a variety of measurements exist, based on observations of mass accretion rates in T-Tauri protostellar disks \citep{Hartmann1998}, direct measurements of velocity dispersion in disks \citep{ flaherty2018turbulenceALMA, teague2018}, and constraints arising from the sharpness of dust ring features in ALMA observations \citep{dullemond2018}.  More recently, \citet{ueda2021} in their study of differential dust settling on the SED and polarization, constrain turbulent $\alpha$ to very low values ($\lesssim 10^{-5}$). In a related vein, \citet{doi2021} estimate the dust scale height from ALMA dust continuum image of HD 163296 and find $\alpha \simeq 10^{-4} - 3 \times 10^{-3} $ in two specific rings.
These  indicate that values for the turbulent intensity lie in the range $10^{-4} \le \alpha \le 10^{-2}$.  

Given this albeit still limited range of direct measurements of turbulence in disks, it is natural to think that turbulence amplitudes should vary across disk populations. Stars form in star clusters, so there are bound to be large variations in the ionizing radiation environments of disks across such regions (eg. being in the proximity of a massive star forming in a cluster, vs forming in a quiet "suburb" of the star cluster).   Since the masses of protoplanetary disks are linked to the masses of stars which ultimately form within them, disks will have a considerable range in mass and column density.  More massive disks, having higher column densities, are more highly screened from X-ray ionization, resulting in lower disk  ionization and hence, MRI turbulence.  All of these local conditions will determine the degree of the disk ionization by stellar X-rays \citep{matsumura2006dead},  on the level of MRI turbulence within them, and hence on the character of their forming planetary system.

\subsection{Confronting the Observations}
\label{sec:discussion:confrontingobservations}

Our results address several key observational constraints on planet formation:

\begin{itemize}
    \item {\bf Planets at large disk radii are not associated with opacity transitions}. One of the main challenges to standard models of planet formation is the importance of ice line formation. Our model shows that while ice lines are important traps for planet formation in higher viscosity disks, this is not so obvious for low viscosity disks.  Low mass forming planets in low viscosity disks are pushed out to large disk radii at a very early time, where they are trapped in extended heat transition regions. Such traps have no connection to opacity transitions or ice lines, and originate in the rather broad disk region in which the transition between predominantly viscous to radiative heating of the disk occurs. Unlike ice lines, or dead zone traps, planets associated with the heat transition could open gaps over a wide range of potential disk radii. Where this region occurs in the disk will depend on the strength of the turbulence, as we have seen in  Fig. \ref{fig:planet_tracks}. 
    
    These results are in excellent accord with observations that the gaps structures in HL Tau are not associated with opacity transitions, such as water, or other icelines \citep{van2018rings}.  If these gaps are indeed the result of planet formation, then these planets cannot be associated with such transitions, as our model clearly shows.  
    
    We note that planet formation at higher disk viscosities in our models is more closely connected with ice lines and opacity transitions, which are not currently well resolved by ALMA observations.
    
    \item {\bf Low observed frequency of massive planets beyond 10 au}. As discussed in the Introduction, occurrence rates for giant planets falls off markedly beyond 10 au.  
  Our model shows that planets in low viscosity disks reside at large disk radii temporarily.  They move steadily inwards, either by being tethered to their inward moving heat transition traps, or by decoupling with such traps and subsequent migration driven by Lindblad torques until they open a gap in the gas and undergo slow Type II movement.  Given that, in our picture, most disks may have higher turbulent viscosity, those systems will undergo strong Lindblad torques that rapidly clear out forming embryos from the outer regions of the disk.  As a third argument, we note that the stochastic "kicks" that turbulence gives to planets especially when they are of low mass, will increase their inward migration rates over those found in quiescent disks \citep{Baruteau2011}.  It follows that the loss of planetary cores will be much less severe in the lower viscosity disks, again reinforcing our results. 
    
    \item {\bf Low disk viscosity and pebble formation.} A completely independent line of evidence supporting low amplitude turbulence  ($\alpha \simeq 10^{-5}-10^{-4}$) arises upon considering the growth of pebbles \citep{Pinilla2020}.  With isotropic turbulence, pebble formation can happen in disks when the gas and dust diffusion has these low values. Otherwise, high turbulence limits pebble formation because of the increased grain fragmentation that will occur. Pebble accretion may be important in quickly generating giant planet cores \citep{bitsch2018pebble}.  Taken together with our own results, we see that Jovian planet formation could also occur in low viscosity disks.   It seems clear that pure pebble accretion quickly produces $\sim 10$ Earth mass planets if the dust is large and turbulence is weak \citep{ormel2017springer}. Whether or not this process extends to higher masses is possible \cite[eg. in combination with planet-planet collisions,][]{wimarsson2020} but is a topic for further investigation. 
    
    \item {\bf Low disk viscosity and disk winds.} The observation of low turbulence levels in a disk is a consequence of non-ideal MHD effects which damp out turbulence and drive angular momentum loss by MHD disk winds \citep{BaiStone2013,Gressel2015}.  Outflows are associated with protostellar disks and the formation of stars of all masses \cite[eg. review][]{Pudritzray2019}. Therefore our low viscosity models are in accord with the finding that ambipolar diffusion damps out the MRI turbulence to low levels leaving primarily the disk wind to extract the disk angular momentum. The energetic jet \citep{krist2008} observed in $H_{\alpha}$ and forbidden lines by HST, and the associated molecular outflow observed by ALMA \citep{klaassen2016} from HL Tau are the means by which this disk angular momentum is carried away. Disk winds will reinforce this trend in the migration of forming planets, as we have argued. 
\end{itemize}

\subsection{Dust settling and dust gap opening}  
\label{sec:discussion:dustgapopening}

Our results for the corotation masses at large disk radii indicate that in our conservative mass accretion model, planet masses lie in the range  1-2 $\Me $. This falls short of the predictions of simulations which suggest a dust gap opening mass \citep{Lambrechts2014, rosotti2016minimum} of 
\begin{equation}
    \label{eqn:Mdustgap_rosotti}
    M_{\rm dust \, gap} \simeq 15  \Big( \frac{h }{ 0.05} \Big)^3 \Me \, ,
\end{equation}
for $\alpha=10^{-3}$. More recently, \citet{dong2018multiple} carried out hydrodynamical simulations at very low viscosities ($\alpha = 10^{-6} - 10^{-5}$) and found that a planet as low in mass as Mars ($\Mp = 0.1 \, \Me$) could open multiple (4) dust gaps in the disk inner regions where the gas disk aspect ratio is low ($h=0.02$). On the other hand, if the aspect was high ($h=0.08$), a planet of mass $\Mp = 34 \, \Me$ was needed to open multiple (3) dust gaps. In our simulations, the gas disk aspect ratio $h$ at large disk radii where radiative heating dominates is independent of viscosity and has typical values of  $h$ between 0.05-0.06 out to 30 au, and then climbing to 0.09 at 100 au.   Does the high local values of $h$ mean that the heat transition planets computed in this work won't open dust gaps after all? 

There are several caveats to the calculation of dust gap opening that need to be addressed. Simulations have not included the result of dust settling, nor an in depth examination of the effect that different turbulent amplitudes will have on the result \citep{rosotti2016minimum}. Our own analysis also assumes complete mixing of the dust and gas.  This ignores the fact that dust particles will rather quickly settle to scale heights that depends on both their size (best measured by their Stokes number ${\rm St}$ in the Epstein drag regime) and the turbulence in the disk \citep{birnstiel2010, ueda2021, doi2021}. For Stokes numbers ${\rm St} \ll 1$, the dust scale height is related to the gas pressure scale height as
\begin{equation}
    H_{\rm d} = H \, \sqrt{\alpha / {\rm St}} \, .
\end{equation} 
The reduced dust scale height compared to that of the gas has recently been observed and quantified for a few systems \citep{Rich2021}.

Another point is that our planet formation calculations assume that it is planetesimal accretion that dominates the accretion of rocky planetary cores beyond an Earth mass or so.  The issue of whether or not pebble accretion builds up to Jovian planet mass cores is still under investigation.  However, we do note that low viscosity disks may be more favourable to their formation away from the ice line and out at larger disk radii.  Detailed 3D simulations indicate that strong turbulence suppresses pebble accretion \citep{Ormel2018}, but further investigation is needed. Perhaps of the greatest importance is the question of how much dust can be trapped in the pressure bumps that these forming planets raise.  That is a question we leave to forthcoming paper. 

Thus, in our view, it is still an open question as to whether or not dust gap opening needs to have a minimum of Neptune mass planets to open the dust gaps (as Eqn. \ref{eqn:Mdustgap_rosotti} suggests).  


\section{Summary}
\label{sec:summary}


We summarize the flow of technical steps and innovations that we have developed in this work for the convenience of the general reader. 

$\bullet$ For two levels of disk viscosity (high, $\alpha=10^{-3}$ and low, $\alpha=10^{-4}$), we first calculate maps of the total torque that would be exerted by the disk on a planet of any mass and semi-major axis for each time step in our simulations (see Fig. \ref{fig:torque_maps}). In these torque maps, we identify three key features (three contours of zero net torque enclosing regions of outward-directed torque embedded in an otherwise inward-directed torque landscape) that influence planetary migration in their respective regions of the disk. The torque features are common between the high and low viscosity disks, \textit{but occur uniformly at lower planet masses in the lower viscosity case}. 

$\bullet$ Each of the three zero net torque contours are associated with (but not necessarily co-located with) a change in a physical quantity within the disk and a planet trap (see Fig. \ref{fig:3trap_definitions}):
\begin{enumerate}
    \item The \textbf{dead zone}, where the ohmic Elsasser number exceeds unity (Eqn. \ref{eq:elsasser} and App. \ref{sec:volatilechemicalmodels}).
    \item The \textbf{water iceline}, where the abundances of water vapour and ice are equal (App. \ref{sec:volatilechemicalmodels}).
    \item The \textbf{heat transition}, where the disk transitions from being mainly viscously heated to mainly radiatively heated (Eqns. \ref{eqn:hlt-surface-density-profile}, \ref{eqn:hlt-temperature-profile} \& \ref{eq:dm06} and Sec. \ref{sec:diskmodel}).
\end{enumerate}

$\bullet$ In our high-mass, high viscosity disk at $t=1.02$ Myr (loosely the age of the HL Tau system) for example, the outer edge of the dead zone is co-located with its trap at $\sim 1.5$ au, the water iceline and its trap occur at $\sim 7$ au, and the contour enclosing outward-directed torque associated with the heat transition extends from 10 to 40 au (Fig. \ref{fig:3trap_definitions}).

$\bullet$ We then compute planet evolution tracks, which involve planet growth and migration under local disk conditions. These tracks are the lines of planet mass and semi-major axis over time superimposed on the evolving torque maps. This procedure allows us to clearly see how the torque landscape affects planet growth and direction of migration (Fig. \ref{fig:planet_tracks} and the movies in Table \ref{tab:movies}).

$\bullet$ It is through the extensive sets of such simulations that we discovered the migration bifurcation phenomenon. To develop a theoretical model that explains the behaviour, we derived analytic expressions, based on generalize torque theory,  for the planet mass at which the (outward-directed) corotation torque is at maximum strength. This we refer to as the \textbf{corotation mass} ($\Mpcorot$, Eqn. \ref{eq:piecewise-Mpcorot}). As the corotation torque is prone to saturation over time by (in our models) either viscous or thermal diffusion, we first derive the viscous and thermal corotation masses ($\Mpnucorot$, Eqn. \ref{eqn:Mpnucorot} and $\Mpchicorot$, Eqn. \ref{eq:Mpchicorot-units}); we then identify which regions of the disk to which they each apply based on the Prandtl number (Sec. \ref{subsec:insights-Mpcorot}).  

$\bullet$  Building on \citet{kimmig2020winds} who demonstrated that disk winds can also lead to outward migration, we show that there is a \textbf{disk wind corotation mass}, $\Mpdwcorot$ (Eqn. \ref{eq:Mpdwcorot}) at which the outward corotation torques are at a maximum (Sec. \ref{sec:discussion:diskwinds}). 

\section{Conclusions}
\label{sec:conclusions}

We have shown, by means of sophisticated astrochemistry and planet formation simulations backed by a detailed theoretical framework, that there is a bifurcation between the migration histories of planets forming in disks of high viscosity and those forming in disks of low viscosity. Co-rotation torques in low viscosity disks push planets outward to large disk radii before they return to the inner regions of the disk, whereas planets in high viscosity disks only migrate inwards.
From this we infer that extended, structured protoplanetary disk systems with dust gaps/rings observed at 10s of au are likely disks that have these low viscosities, while the majority of systems with more compact dust components have higher viscosity. 

Our specific conclusions are:

$\bullet$ The torque features associated with the water iceline and especially the heat transition significantly affect the migration of planets that interact with them:
\begin{enumerate}
    \item In the \textbf{low viscosity} models, a collection of planetary embryos are captured by the extended region of outward torque associated with the heat transition and are forced to migrate outwards to larger disk radii early in their evolution (starting around $t\approx 0.750$ Myr; see right panels of Fig. \ref{fig:planet_tracks} and the movies in Table \ref{tab:movies}). While some of these planets peel off the outward migration band before reaching the trap (see Fig. \ref{fig:single-madiagram-alldisks}'s movie), those that reach the largest orbital radii are trapped there. These \textbf{heat transition planets} have masses within $1-2\, \Me$ and orbital radii ranging $8-20$ au (depending on disk mass, see Fig. \ref{fig:single-madiagram-alldisks}), roughly twice as far from the star as where they were seeded. With further accretion, they would reach the trap escape mass ($1.5-2.3\, \Me$; Table \ref{tab:populations}) and migrate inwards again.
    \item In the \textbf{high viscosity} models, planet trapping at the water iceline is the main safeguard to otherwise unimpeded inward migration (see left panels of Fig. \ref{fig:planet_tracks} and the movies in Table \ref{tab:movies}). These \textbf{water iceline planets} do not accumulate enough mass to escape the water iceline trap and continue migrating inward before our simulations end. The mass needed to escape this trap decreases with time and increases with disk mass, ranging $3-5\, \Me$ (Table \ref{tab:populations}). 
\end{enumerate}

$\bullet$ In both cases of thermal or viscous diffusion in a planet's co-rotation region, the corotation mass scales as $\Mpcorot \propto \alpha^{2/3}$, explaining how a lower $\alpha$ results in the strongest corotation torque reaching lower planet masses. Our analytic expressions describe our numerical results well (Fig. \ref{fig:prandtl_fits}).

$\bullet$ The extent of outward planet migration in the low viscosity disks depends on the disk mass.  The most extensive bifurcation and outward migration under the influence of the co-rotation torque occurs for the most massive disks.  
    
$\bullet$ Analogous to how raising the turbulent $\alpha$ raises the viscous and thermal corotation masses, increasing the disk wind mass outflow rate through the parameter $b$ increases the disk wind corotation mass, $\Mpdwcorot \propto b$. Stronger winds in the early phases of disk evolution, especially in the case of low turbulent viscosity, should contribute to strong outward corotation torques.  As winds die off during later disk evolution, the direction of the wind torque will reverse, helping push planets inwards. Using realistic disk wind strengths that align with outflow observations suggests a value for $\Mpdwcorot$ in the range of Earth to mid super-Earth masses.

\medskip
\noindent
\textbf{Our work has several major implications:}

$\bullet$ The statistical analysis of hundreds of protoplanetary disks shows that the dust component of disks falls into two populations: (1) radially compact, with no (resolved) structure, or (2) radially extended, with ample resolved gap/ring structure \citep{vandermarel2021-demographics}. The inevitability of dust radial drift suggests that compact disks simply contain no mechanism for trapping and retaining their dust at large radii \cite[eg. pressure bumps,][]{whipple1972}, whereas structured/extended disks do. One possible source of pressure bumps is planets - which then requires a mechanism by which planets can be present at large radii early in their formation history.

$\bullet$ If turbulence is primarily due to the MRI instability, then we argue that the most massive disks \cite[which are the structured/extended disks of][]{vandermarel2021-demographics} will naturally have low values of MRI disk turbulence. This is due to increased shielding from ionizing radiation.  As the most massive disks are more common around more massive stars, there should be an associated trend of lower disk viscosity in disks around more massive stars.  We conclude that low values of disk turbulent viscosity ($\alpha=10^{-4}$) could be a characteristic of the extended, gap/ring protoplanetary systems.

The models presented here do not take into account the back reaction of the forming planets on the disk structure (eg. increased availability of solids due to dust trapping in planet-induced pressure bumps) - and this, we think, is why our heat transition planet masses are relatively low. We address this question in a future paper. 

In the meantime, we note that our theory can be tested in principle. The key is in identifying where the heat transition occurs in observed disks. Since viscous heating rates and hence the extent of the viscous region in disks depends upon the turbulent amplitude $\alpha$, observations will need to measure the turbulence strength at various disk radii to compare with the external heating rate due to the host star. 




\section*{Acknowledgements}

We thank the anonymous referee for a thoughtful report and constructive comments that helped clarify the manuscript. We thank Nienke van der Marel for interesting discussions and comments on an early version of this manuscript; and Ruobing Dong, Giovanni Rosotti and Lisa Patrascu for comments. REP and FM thank the organizers of the inspiring {\it New Horizons in Planetary Systems} conference, May 13-17 in Victoria, B.C, where our discussions planted the seeds for this paper. JS completed a large part of this work as part of an Honours Integrated Science Thesis at McMaster with REP. Additional support for JS is provided by an NSERC CGS-M award at the University of Victoria.  REP is supported by a Discovery Grant from the National Science and Engineering Research Council of Canada (NSERC). FM acknowledges support from The Royal Society Dorothy Hodgkin Fellowship.  RAB acknowledges support from the STFC consolidated grant ST/S000623/1 and funding from the European Research Council (ERC) under the European Union’s Horizon 2020 research and innovation programmes PEVAP (grant agreement number 853022) and DUSTBUSTERS (grant agreement number 823823).


\section*{Data Availability}

The data underlying this article will be shared on reasonable request to the corresponding authors.


\bibliographystyle{mnras}
\bibliography{paper1_v3} 


\appendix

\bigskip
\section{Volatile Chemical Models}
\label{sec:volatilechemicalmodels}

As in \cite{cridland2019physics}, we are interested in the relation between discontinuities in disk properties like the water iceline and dead zone on the absolute value and direction of migration torques. For this, we use a novel approach of combining disk physics with the astrochemistry in protoplanetary disks. We start with the (evolving) temperature and density profiles described above, given the initial conditions. 

As the disk evolves we compute the astrochemical evolution of H, He, C, N, O, S-bearing species, along with the ionization of Mg and Fe. The relevant elemental abundances (for water chemistry) are C/H = 1.0$\times 10^{-4}$, O/H = 2.5$\times 10^{-4}$. The ionization rate due to cosmic rays is $\chi_{\rm CR} = 10^{-17}$/s, and the chemistry is initialized as a set of molecules determined from cold cloud models. The model is based on a modified \citep[in][]{Cridland2019popandchem} version of the Michigan chemical code \citep[see for example,][]{Fogel2011chem}.

With the astrochemical results we have the radial distribution of water ice throughout the disk evolution. These distributions allow the model to compute the location and evolution of the water iceline, as well as the gradient of the water ice abundance across the ice line. We then use the method of \cite{Miyake1993} to compute the total dust opacity as a function of the water ice mass fraction. This calculation results in a change in the dust opacity across the water iceline - which is the physical change that is expected to cause planet trapping at the ice line \citep{hasegawa2013planetary}. For brevity, see \cite{cridland2019physics} (their Sec. 4) for an outline of the specific method used in computing the dust opacity.

The result of this calculation is a small, smooth reduction in the gas temperature across the ice line. The dust opacity and the gas surface density both depend on the gas temperature, so we update these properties and iterate our method until changes in the temperature profile are small. The small reduction in temperature that results is accompanied with a small increase in the surface density of gas in order to maintain a constant mass accretion rate through the ice line. 

\ignore{These changes conspire to strengthen the corotation torque across the water iceline, producing a positive torque which produces the water iceline trap at its outer edge.}

Along with the ice abundance, we compute the electron fraction from our chemical model. The electron fraction dictates how strongly the gas is linked the the disk magnetic field, and hence how susceptible the disk is to the magnetorotational instability (MRI). We assume that the disk turbulence is solely driven by the MRI hence the \textit{turbulent} $\alpha_{\rm turb}$ is directly liked to the electron fraction. We reduce the turbulent $\alpha_{\rm turb}$ by two orders of magnitude when the Ohmic Elsasser number drops below one. The Ohmic Elsasser number is: \begin{equation}
    \Lambda_O \equiv \frac{\rm c_s^2}{\eta_O\Omega}
    \label{eq:elsasser}
\end{equation} 
where $\eta_O \equiv 234~T^{1/2}/x_e~ {\rm cm}^2{\rm s}^{-1}$ is the Ohmic resistivity, and $x_e$ is the electron abundance relative to hydrogen. This region of the disk is traditionally called the dead zone. The change in turbulent $\alpha_{\rm turb}$ impacts both the direction of the net torques as well as the gap opening mass, but does not impact the global angular momentum transport through the disk. 


\section{Supplementary: Numerical Results}
\label{app:numerical-results}

\begin{figure*}
	\includegraphics[width=18cm]{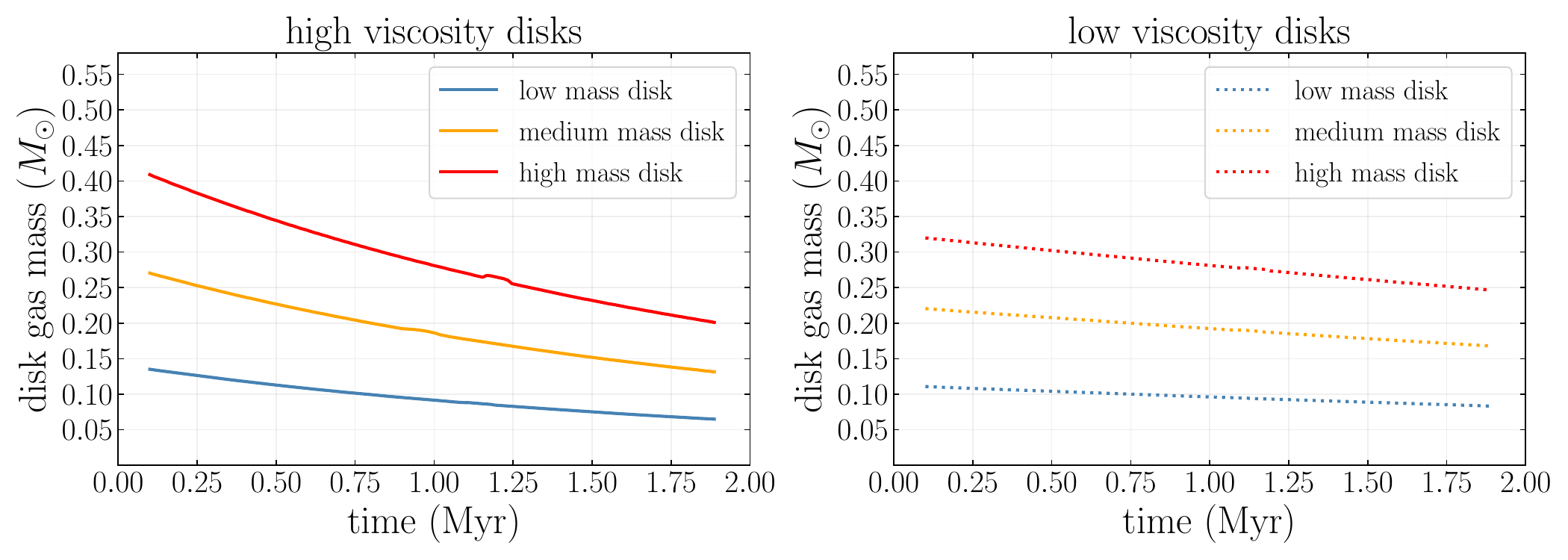}
    \caption{ \textbf{  \protect\hyperlink{hyp:fig:app:alldisks-masses-trunc93au}{Disk mass contained within 93 au.}} Disk masses are initialized so as to evolve to observational estimates at 1 Myr.}
    \label{fig:app:alldisks-masses-trunc93au}
\end{figure*}

\begin{figure*}
	\includegraphics[width=18cm]{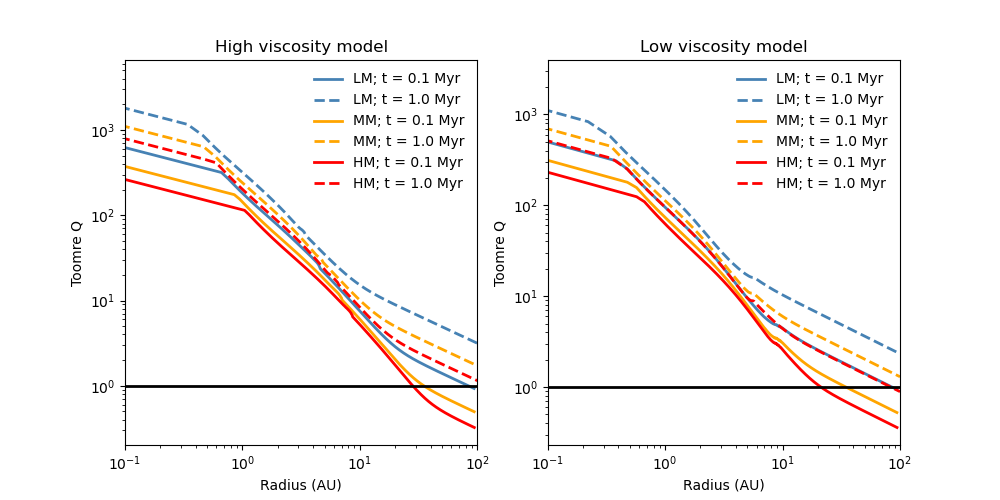}
    \caption{ \textbf{  \protect\hyperlink{hyp:fig:app:toomre}{Toomre-Q parameter}} at $t=0.1$ Myr (solid lines) and 1 Myr (dashed lines) for all 6 disk models.  A value of $Q < 1$ represents gravitationally unstable regions. } 
    \label{fig:app:toomre}
\end{figure*}

\hypertarget{hyp:fig:app:alldisks-masses-trunc93au}{Figure \ref{fig:app:alldisks-masses-trunc93au}} provides the disk mass inwards of 93 au over the course of our simulations, obtained by integrated surface density profiles shown in Figure \ref{fig:model_alphas_h_sigma}. The disk masses were initialized so as to be roughly equal to the observationally constrained values of \citet{carrasco2016vla} after 1 Myr. The careful reader may notice a tiny bump in the high-mass, high viscosity curve occurring between 1.167 and 1.260 Myr; the underlying cause of this bump is a small rise and fall in the disk surface density contained between 10 and 11 au at those times (see movie in $\Sigma$ column of Table \ref{tab:model-summary}), which is itself due to numerical artifacts arising from the resolution of our astrochemistry simulations. This brief perturbation does not have an effect on the planet migration trajectories (see movie of Fig. \ref{fig:planet_tracks}).

\hypertarget{hyp:fig:app:toomre}{Figure \ref{fig:app:toomre}} shows the gravitational instability of our disk models through the Toomre-Q parameter at two time snapshots: $t=0.1$ Myr and $t=1$ Myr. Initially, the high- and medium-mass disks are unstable outwards of $\sim 30$ au. The instability region begins at large radii and moves outward with time. By 1 Myr, all 6 models are fully gravitationally stable.

\begin{figure*}
	\includegraphics[width=18cm]{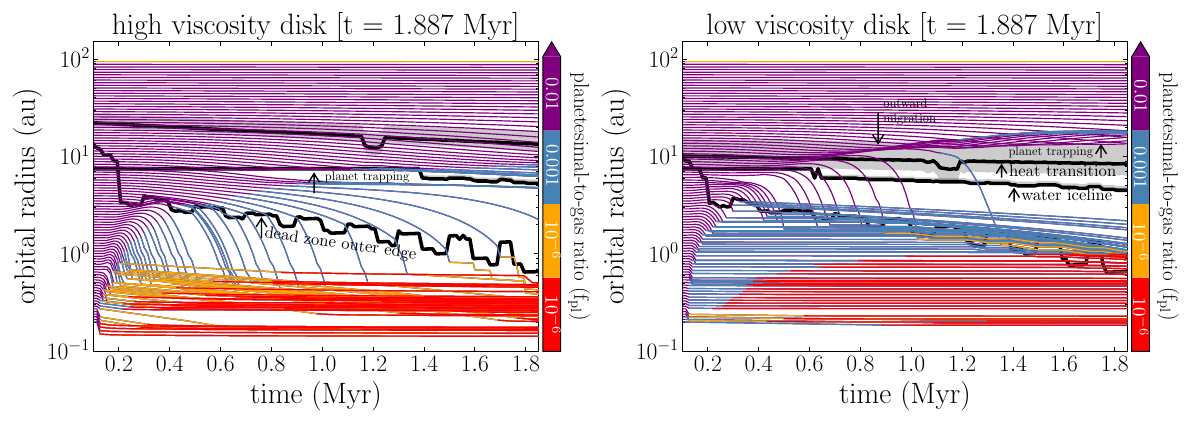}
	\includegraphics[width=18cm]{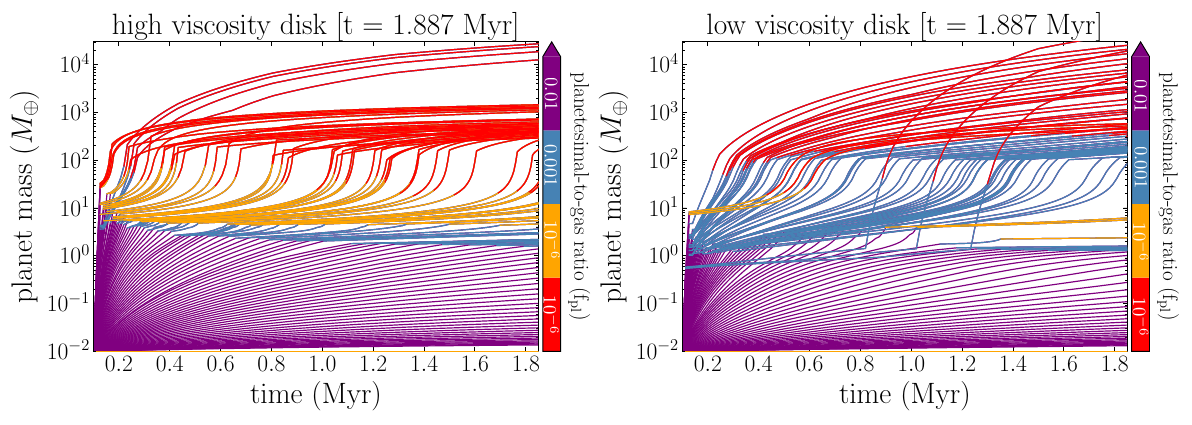}
    \caption{ \textbf{  \protect\hyperlink{hyp:fig:app:radius-vs-time}{Planetesimal-to-gas ratio ($\fpl$) ie. growth stage of planets over time }} (Sec. \ref{sec:formation-model}). [Shown for our high-mass disks]. Orange and red coloured lines correspond to the same $\fpl$, delineating instead when $\Mp > \Mcrit$ (Eqn \ref{eqn:Mcrit}), corresponding to when the viscous timescale is lengthened to $t^\prime_\nu$. \textbf{Top row:} Orbital radius vs. time. The thick grey fill-lines span ${r_{\rm IL}}_{-0.1}^{+0.5}$ au (${\rheat}_{-2}^{+5}$ au) to roughly account for the slight (large) difference between the formal location of the water iceline (heat transition) and the radial extent of the associated zero-torque contour. \textbf{Bottom row:} Same as above but planet mass vs. time.  }
    \label{fig:app:radius-vs-time}
\end{figure*}

\hypertarget{hyp:fig:app:radius-vs-time}{Figure \ref{fig:app:radius-vs-time}} shows the growth stages reached or not reached by our planetary cores (see Sec. \ref{sec:formation-model}). Each line represents a single planet, coloured according to the instantaneous value of $\fpl$, illustrating how the planetary core accretion rate $\Mcoredot \propto \fpl \Sigma$, at least for the first two stages (Eqns. \ref{eq:Mcoredot}-\ref{eqn:dust-surface-density}). The transition from orange to red (same $\fpl$) indicates $\Mp > \Mcrit$ (Eqn \ref{eqn:Mcrit}), when the viscous timescale is lengthened to $t^\prime_\nu$. The solid black lines in the background show the radial evolution of the three planet traps: the outer edge of the dead zone, the water iceline, and the heat transition (these lines correspond to the vertical dashed lines in Fig. \ref{fig:3trap_definitions}).

Note that the water iceline and heat transition planets remain in the first two stages, where $\fpl=0.01$ or 0.001, and so their growth depends on the local gas surface density $\Sigma$. Only planets that migrate to within $\sim 1$ au enter into the third and final growth state (which entails $\Mp > \Mgap$, Type II migration, $\fpl=10^{-6}$ and a reduced gas accretion rate).

\section{Supplementary: Theoretical Results}
\label{app:insights-planetmigration}

\begin{figure*}
	\includegraphics[width=\columnwidth]{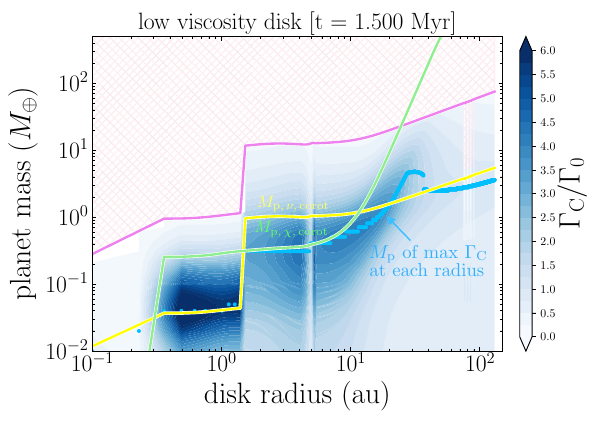}
	\includegraphics[width=\columnwidth]{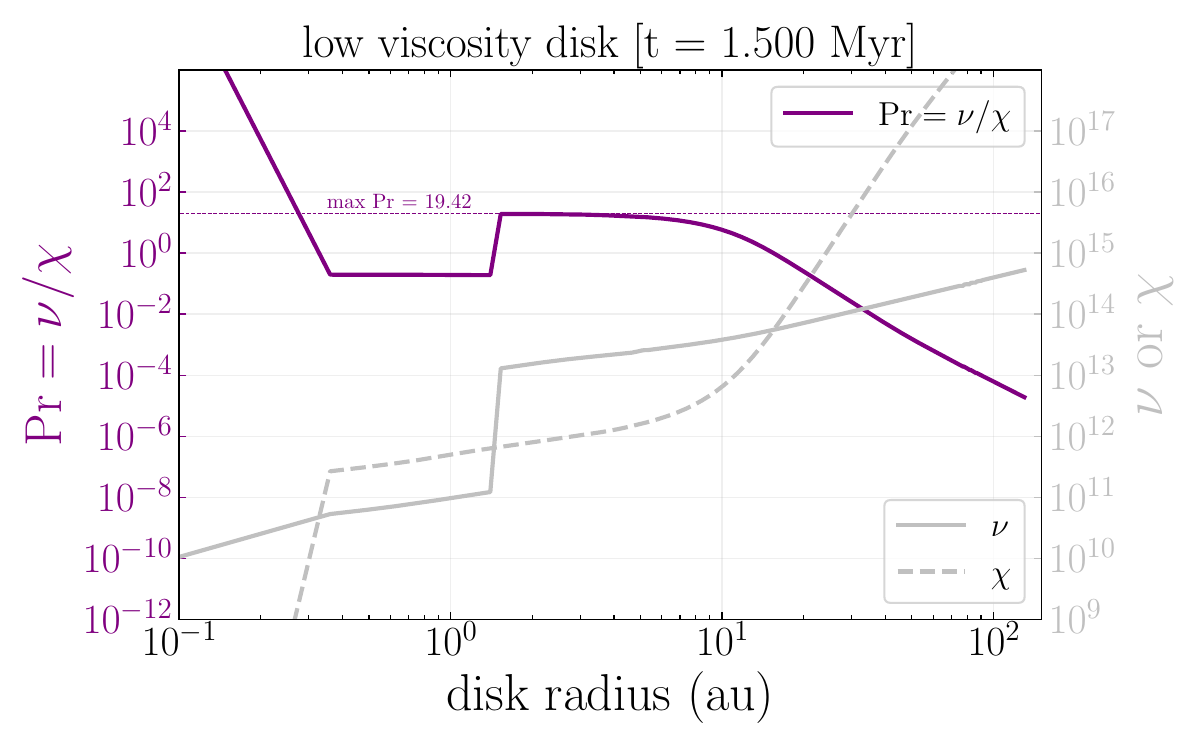}
    \caption{ \textbf{  \protect\hyperlink{hyp:fig:app:prandtl_vs_radius}{Alternative to Fig. \ref{fig:prandtl_fits}.}} \textbf{Left panel:} Like Fig. \ref{fig:prandtl_fits} but with overlaid atop the corotation torque map to illustrate the diverging bimodal $\Gamma_{\rm C}$ maxima at large disk radii, which $\Mpnucorot$ and $\Mpchicorot$ each follow. \textbf{Right panel:} Prandtl number (${\rm Pr}$), thermal diffusivity ($\chi$) and viscosity ($\nu$) vs. disk radius for the high-mass, low viscosity disk.}
    \label{fig:app:prandtl_vs_radius}
\end{figure*}

\hypertarget{hyp:fig:app:prandtl_vs_radius}{Figure \ref{fig:app:prandtl_vs_radius}} shows the thermal diffusivity $\chi=\chi(r)$, the level of viscosity $\nu=\nu(r)$ and therefore the Prandtl number $\Pr=\Pr(r)$ as a function of disk radius in our models. We note that numerical investigations by Paardekooper2011 showed that the lowest corotation torque was obtained for Prandtl numbers of order unity. The effect of thermal diffusivity is shown in modified curves in their Figure 6, where we see that while there is a slight adjustment in the peak value of the torque amplitude, the shape of the curves with thermal effects included is not too different from the purely viscous curves.

One important difference between the idealized adiabatic disks analyzed in the Paardekooper papers, and our own disk model is that we have included a treatment of disk viscous and radiative heating together with detailed disk astrochemistry.  This affects the thermal physics of the disk, and in particular the Prandtl number.  While in the former work the Prandtl number did not exceed unity and may go to low value $Pr \ll 1$. As already noted, in our models, the Prandtl number may take on values both much smaller, but also much greater than unity;  $ Pr \ll 1$; and  $ Pr \gg 1 $. We show that we can readily extend the theory to include these effects.

\bsp	
\label{lastpage}
\end{document}